\documentclass[a4paper,11pt]{revtex4}
\usepackage[a4paper, total={7in, 10in}]{geometry}
\usepackage{graphicx} 
\usepackage{amsmath} 
\usepackage{amssymb} 
\usepackage{bm} 
\usepackage{dcolumn}
\usepackage{color}
\usepackage{mathrsfs}
\usepackage{amsfonts}
\usepackage{varioref}
\usepackage{mathrsfs}
\usepackage{graphicx}
\usepackage{latexsym}
\usepackage{amsmath}
\usepackage{amssymb}
\usepackage{textcomp}
\usepackage{amsbsy}
\usepackage{graphics}
\usepackage{epstopdf}
\usepackage{color}
\usepackage[caption=false]{subfig}

\RequirePackage[colorlinks,citecolor=blue,urlcolor=magenta,linkcolor=blue]{hyperref}
\input epsf

\allowdisplaybreaks[4]
\begin{document}

\tolerance=5000

\title{Fitting NANOGrav 15-year data and ACT data with modified inflation in entropic cosmology}

\author{Simone~D'Onofrio$^{1}$\,\thanks{donofrio@ice.csic.es}
Sergei~Odintsov$^{1,2}$\,\thanks{odintsov@ieec.uab.es},
Tanmoy~Paul$^{3}$\,\thanks{tanmoy.paul@visva-bharati.ac.in}} \affiliation{
$^{1)}$ Institute of Space Sciences (ICE, CSIC) C. Can Magrans s/n, 08193 Barcelona, Spain\\
$^{2)}$ ICREA, Passeig Luis Companys, 23, 08010 Barcelona, Spain\\
$^{3)}$ Department of Physics, Visva-Bharati University, Santiniketan 731235}


\tolerance=5000

\begin{abstract}
Recent evidences of stochastic gravitational wave background (SGWB) through Pulsar Time Array (PTA) observations hint towards an alternative inflationary scenario, compared to the usual inflation, for describing the early stage of the universe in order to be compatible with the PTA data. Moreover, currently the Atacama Cosmology Telescope (combined with the Planck 2018 and BAO) refines the constraint on inflationary observables, compared to the only-Planck 2018 measurements. In the present work, we simultaneously address these two issues by incorporating certain modification during inflation over the usual inflationary scenario. Such modification amplifies the primordial tensor perturbation over the modes that are sensitive to the NANOGrav frequency region. For this purpose, we take the thermodynamic route of cosmology where the entropy of the apparent horizon is given by a generalized form of entropy that is able to generalize the other known form of horizon entropies for suitable representations. The constraints on the model parameters coming from the ACT data also fit the NANOGrav 15-year data (based on numerical analysis), which reveal the model's compatibility with both the ACT and the PTA data.
\end{abstract}

\maketitle

\section{Introduction}

Over the past few decades, the Cosmic Microwave Background (CMB) provides the most effective tool in probing the physics of the early universe \cite{Bennett:1996ce,Smoot:1998jt,Hinshaw_2013,Planck:2018jri}, as the CMB photons are non-interacting after the last scattering surface and thus carry the information of the early universe prior to the last scattering instant. On other hand, the recent Atacama Cosmology Telescope (ACT) data, combined with the Planck 2018 and the Baryon Acoustic Oscillation (BAO) from DESI, argue that the curvature perturbation should be less red tilted in comparison to the only-Planck measurement \cite{ACT:2025fju,ACT:2025tim}. In particular, the Data Release 6 (DR6) from the ACT, combined with the Planck 2018 and BAO (collectively it can be denoted by ACT-DR6+Planck18+BAO), indicates a $2\sigma$ deviation from the only-Planck 2018 measurements, on the constraints over the early universe observables like the spectral index for curvature perturbation and tensor-to-scalar ratio. Such new constraints on primordial perturbations from ACT data disfavors the models which previously lie at lower boundary value of spectral index. Based on these argument, a variety of inflationary models (which are viable from only-Planck 2018 result) are revisited from the angle of the recent ACT data \cite{Kallosh:2025rni,Aoki:2025wld,Odintsov:2025wai,Haque:2025uga,Haque:2025uis,Maity:2025czp,Dioguardi:2025vci,Salvio:2025izr,Antoniadis:2025pfa,Kim:2025dyi,Gao:2025onc,Drees:2025ngb,Odintsov:2025bmp,Peng:2025bws,Liu:2025qca,Mondal:2025kur,Gialamas:2025kef,Mohammadi:2025gbu}. However the constraints coming from the ACT-DR6, or only-Planck result, are effective mainly over the large scale modes, the corresponding constraints over the lower scale modes are comparatively weaker. For the lower scale modes, one may rely on gravitational waves (GWs).

During the last few years the detection of gravitational waves has opened a new window to view the physics operating at the early stage of universe, including over the lower scale modes \cite{Guzzetti:2016mkm,Caprini:2018mtu,Domenech:2021ztg,Roshan:2024qnv}. However the GWs observations made by Ligo-Virgo-Kagra (LVK) collaboration are of astrophysical origin, in particular, the detected GWs mainly originated from the mergers of binary black holes or neutron stars \cite{LIGOScientific:2016aoc,LIGOScientific:2017bnn,LIGOScientific:2017ycc,LIGOScientific:2017vox}; while if GWs have the primordial cosmological origin, then these would be stochastic in nature. However after a dedicated effort, the stochastic gravitational wave background (SGWB) remains undetected by the LVK collaboration over the frequency range $f=[1,100]  \mathrm{Hz}$ on which the LVK observatories are sensitive on \cite{KAGRA:2021kbb}. However the search of the stochastic GW background was lightened by the observations of Pulsar Timing Arrays (PTA)--- in particular by the joint collaboration of NANOGrav \cite{NANOGrav:2023gor,NANOGrav:2023hde,NANOGrav:2023hvm}, EPTA \cite{EPTA:2023sfo}, PPTA \cite{Zic:2023gta} and CPTA \cite{Xu:2023wog} --- over the frequency range $10^{-9} \mathrm{Hz} < f < 10^{-6} \mathrm{Hz}$. The detection of such SGWB by PTA $may~be$ favored to have a primordial origin by the NANOGrav 15-year data \cite{NANOGrav:2023gor,NANOGrav:2023hde,NANOGrav:2023hvm}; there can be other sources, for instance, the astrophysical sources from supermassive black hole binaries could also contribute to the GWs observation. This immediately invited different interesting proposals regarding the generation of GWs by cosmological origin, such as (but not limited to) --- models with blue tilted primordial tensor power spectra \cite{Vagnozzi:2023lwo,Borah:2023sbc,Datta:2023vbs,Choudhury:2023kam,Cai:2020qpu,Ye:2023tpz,Jiang:2023gfe,Zhu:2023lbf,Datta:2023xpr,Ben-Dayan:2023lwd,Liu:2023pau,Ashoorioon:2022raz}, by axion inflation \cite{Unal:2023srk,Niu:2023bsr,Murai:2023gkv}, phase transitions in the early universe \cite{Salvio:2023ynn,Gouttenoire:2023bqy,Ghosh:2023aum,An:2023jxf,Jiang:2023qbm}, models with enhanced formation of primordial black holes (a secondary source of SGWB) \cite{Inomata:2023zup,Franciolini:2023pbf,Cheung:2023ihl,Balaji:2023ehk,Firouzjahi:2023lzg}, mergers of primordial black holes \cite{Depta:2023uhy,Gouttenoire:2023nzr}, collapse of cosmic strings and domain walls \cite{Ellis:2023tsl,Lazarides:2023ksx,Zhang:2023nrs,Yamada:2023thl,Babichev:2023pbf}, by a secondary reheating era \cite{Haque:2021dha}, by primordial magnetic fields that in turn act as a secondary source of SGWB \cite{Maiti:2024nhv}, by pre Big-Bang scenario \cite{Tan:2024kuk,Conzinu:2024cwl,Conzinu:2025sot}, and some other interesting scenarios \cite{Ye:2023xyr,Chen:2024twp,Gangopadhyay:2023qjr}.

In the present work, we focus on the generation of SGWB through primordial tensor power spectra. For this purpose, we consider the thermodynamic route of cosmology where the thermodynamic laws are imposed on the apparent horizon given by: $r_\mathrm{A} = 1/H$ (with $H$ being the Hubble parameter) \cite{Cai:2005ra,Akbar:2006kj,Cai:2006rs,Akbar:2006er,Paranjape:2006ca}. Being a marginally trapped null surface, the apparent horizon sets the causal boundary between the observable and the unobservable universe. Now, with the cosmic evolution, there exists a matter flux from inside to outside of the horizon, which causes a decrease of the matter field's entropy inside the horizon \cite{Odintsov:2024ipb,Paul:2025rqe}. This in turn demands an entropy of gravitational field in order to validate the second law of thermodynamics for the combined system of ``matter field + gravitational field'' \cite{Odintsov:2024ipb,Paul:2025rqe}. Such entropy of the gravitational field is accounted by the entropy of the apparent horizon. Here it deserves mentioning that a microscopic origin of the entropy of the cosmic horizon still demands a proper justification (one may see \cite{Nojiri:2023bom} for some development in this regard). Based on the thermodynamics of the apparent horizon, the Bekenstein-Hawking entropy of the horizon leads to the usual Friedmann equations, while some variants of horizon entropy (compared to the Bekenstein-Hawking form) give rise to modified Friedmann equations and the modification comes from the effective entropic energy density. Such modified cosmological scenario turns out to have rich consequences from early inflation (or bounce) to the late dark energy era \cite{DAgostino:2019wko,Sanchez:2022xfh,Cognola:2005de,Nojiri:2022nmu,Jizba:2023fkp,Jizba:2024klq,Nojiri:2022aof,Nojiri:2022dkr,Odintsov:2022qnn,Odintsov:2023vpj,Housset:2023jcm,Bolotin:2023wiw,Lymperis:2023prf,Odintsov:2024sbo,Brevik:2024nzf,Okcu:2024tnw,Cruz:2023xjp,Cardenas:2023zmn,Adhikary:2025khr,Odintsov:2025sew,Adhikary:2021xym,Adhikary:2024sax,Luciano:2025hjn}, including the possible effects of thermodynamic-cosmology on BBN \cite{Sheykhi:2025kfw,Sheykhi:2025zre}. Some variants of Bekenstein-Hawking entropy that are frequently used in cosmology are the Tsallis entropy \cite{Tsallis:1987eu}, the R\'{e}nyi entropy \cite{Renyi}, the Barrow entropy \cite{Barrow_2020}, the Sharma-Mittal entropy \cite{Sayahian_Jahromi_2018}, the Kaniadakis entropy \cite{Kaniadakis_2005}, or more generally --- the few parameter dependent generalized form of entropy \cite{Nojiri:2022aof,Nojiri_2022,Odintsov:2022qnn}. The generalized entropy puts all the other entropies within a single umbrella for suitable representations of entropic parameters. Owing to such advantage, the generalized entropy proves to have considerable implications towards cosmology as well as black hole \cite{Nojiri:2022aof,Nojiri_2022,Odintsov:2022qnn,Odintsov_2023,Bolotin:2023wiw,Lymperis:2023prf,doi:10.1142/S0219887824503389,Odintsov:2024sbo,Tyagi:2025zov,Tariq:2025wiy,Adhikary:2025khr,Odintsov:2025sew}. Among different forms of generalized entropy, the 4-parameter dependent generalized entropy stands as the minimal generalized version that can generalize all the known entropies proposed so far.

In the present work, we concentrate on simultaneous validation of the ACT data and the NANOGrav 15-year data in the context of entropic cosmology where the entropy of the apparent horizon is given by the 4-parameter generalized entropy. As mentioned earlier that we focus on the generation of SGWB by a blue tilted primordial tensor power spectra regarding the validation of the NANOGrav 15-year data. Actually the detection of SGWB through PTA observations demand an alternative inflationary scenario, compared to the usual inflation, for describing the early stage of the universe in order to be compatible with the PTA data (as the usual inflation results to a nearly flat primordial tensor power spectrum with less amplitude than the PTA constraint). Thus we incorporate certain modification during the early inflation, compared to the usual inflation, which in turn helps to enhance the primordial tensor perturbation over the modes around the NANOGrav frequency rage. With such modified inflation, the current cosmological model proves to be compatible both with the ACT and the PTA constraints for a common range of entropic parameters.

\section{Thermodynamics of apparent horizon and cosmological field equations from the 4-parameter generalized entropy}\label{SecII}

In this section, we briefly review the derivation of the cosmological field equations from 4-parameter generalized entropy based on the thermodynamics of apparent horizon. The metric of spatially flat Friedmann-Lema\^{i}tre-Robertson-Walker (FLRW) spacetime suits our consideration, in particular,
\begin{equation}
    ds^2 = -dt^2 + a^2(t)\left(dr^2 + r^2 d\Omega^2\right) \ ,
\end{equation}
where $t$ is the cosmic time, $a(t)$ is the scale factor of universe and $d\Omega^2$ symbolizes the line element of a 2-sphere having unit radius. For the above metric, the radius of the apparent horizon takes the following form \cite{Cai_2005}
\begin{equation}\label{Horizon-Non-Flat}
    r_\mathrm{A} = \frac{1}{H} \ ,
\end{equation}
with $H = \dot{a}/a$ being the Hubble parameter of universe. As mentioned in the introduction, the apparent horizon needs to be associated with an entropy (in order to validate the second law of thermodynamics for the composite system ``apparent horizon + matter field''), and the corresponding first law of thermodynamics of the horizon is given by:
\begin{equation}
 TdS_\mathrm{h} = -d\left(\rho V\right) + WdV \ ,
 \label{n1}
\end{equation}
where $T = \frac{1}{2\pi r_\mathrm{A}} \left|1 - \frac{\dot{r_\mathrm{A}}}{2Hr_\mathrm{A}}\right|$ and $S_\mathrm{h}$ represent the temperature and the entropy, respectively, associated to the horizon. Moreover, the quantities in the rhs of Eq.~(\ref{n1}) are for the matter fields inside the horizon, in particular, $W = \frac{1}{2}\left(\rho - p\right)$ is known as the work density term where $\rho$ and $p$ are the energy density and the pressure of the matter fields inside the horizon. For a general form of horizon entropy: $S_\mathrm{h}=S_\mathrm{h}(S)$ (where $S = A/(4G)$ is the Bekenstein-Hawking entropy and $A = 4\pi r_\mathrm{A}^2$ is the area of the horizon), the first law of thermodynamics of the apparent horizon leads to \cite{odintsov2023entropicinflationpresencescalar,Nojiri:2024zdu},
\begin{equation}\label{Gen-Thermo-Eq}
   \frac{\Dot{r}_\mathrm{A}}{r_\mathrm{A}^3} \frac{\partial S_\mathrm{h}}{\partial S}  = - \frac{4 \pi G }{3}\Dot{\rho} \ .
\end{equation}
which, due to Eq.~(\ref{Horizon-Non-Flat}), yields
\begin{equation}\label{Gen-Thermo-Eq-with-Horizon}
     H  \Dot{H} \frac{\partial S_\mathrm{h}}{\partial S}  = \frac{4 \pi G }{3}\Dot{\rho} \ .
\end{equation}
From the local energy conservation of the matter fields, i.e. $\Dot{\rho}+3H(\rho+p)=0$, the above equation results to,
\begin{equation}\label{Gen_Fried_II}
     \Dot{H}\frac{\partial S_\mathrm{h}}{\partial S}  = -4\pi G(\rho+p) \ .
\end{equation}
Eq.~(\ref{Gen_Fried_II}) represents the second Friedmann equation and it arises from the thermodynamics of the apparent horizon where the entropy of the horizon is given by a general form: $S_\mathrm{h} = S_\mathrm{h}(S)$. The first Friedmann equation can be obtained by integrating both sides of Eq.~(\ref{Gen-Thermo-Eq-with-Horizon}), and as a result, one gets
\begin{equation}\label{Gen_Fried_I}
    \int d\left(\frac{1}{r_\mathrm{A}^2}\right)\frac{\partial S_\mathrm{h}}{\partial S}  = \frac{8 \pi G }{3}\rho + \frac{\Lambda}{3} \ ,
\end{equation}
where the cosmological constant $\Lambda$ naturally arises as an integrating constant. Thus, as a whole, Eq.~(\ref{Gen_Fried_I}) and Eq.~(\ref{Gen_Fried_II}) are the Friedmann equations derived from the thermodynamics of the apparent horizon where $S_\mathrm{h}$ represents the entropy of the horizon. In the present work, we consider the horizon entropy to be the 4-parameter generalized entropy given by \cite{Nojiri_2022},
\begin{equation}\label{gen-entropy}
    S_\mathrm{h} \equiv S_\mathrm{g}(\alpha_\mathrm{\pm},\beta,\gamma)= \frac{1}{\gamma}\left[\left(1+\frac{\alpha_+}{\beta}S\right)^\beta-\left(1+\frac{\alpha_-}{\beta}S\right)^{-\beta}\right] \ ,
\end{equation}
where $S = A/(4G)$ is the Bekenstein-Hawking like entropy (taking $\hbar =1$ and $c=1$, with $G$ being the Newton's gravitational constant); and $\alpha_\mathrm{\pm}$, $\beta$, and $\gamma$ are the entropic parameters which are assumed to be positive in order to have a monotonic increasing function of $S_\mathrm{g} = S_\mathrm{g}(S)$ (the suffix 'g' is for 'generalized'). For the 4-parameter generalized entropy, the Friedmann Eq.~(\ref{Gen_Fried_I}) and Eq.~(\ref{Gen_Fried_II}) take the following forms,
\begin{widetext}
\begin{equation}\label{Fried-Eq-1}
    \begin{aligned}
    \frac{\beta G H^4 }{\pi \gamma}  \Bigg[ &\frac{1}{2+\beta}\left(\frac{\beta G H^2}{\pi\alpha_-}\right)^{\beta} 2F_1\left(1+\beta,2+\beta,3+\beta,-\frac{\beta G H^2}{\pi\alpha_-}\right) \\ 
    & + \frac{1}{2-\beta}\left(\frac{\beta G H^2}{\pi\alpha_+}\right)^{-\beta} 2F_1\left(1-\beta,2-\beta,3-\beta,-\frac{\beta G H^2}{\pi\alpha_+}\right) \Bigg] = \frac{8 \pi G}{3}\rho + \frac{\Lambda}{3} \ , 
\end{aligned}
\end{equation}
and
\begin{equation}\label{Fried-Eq-2}
     \frac{1}{\gamma}\left[\alpha_+\left(1+\frac{\pi\alpha_+}{\beta G H^2}\right)^{\beta-1}+\alpha_-\left(1+\frac{\pi\alpha_-}{\beta G H^2}\right)^{-\beta-1}\right]\Dot{H}=-4\pi G(\rho+p) \ ,
\end{equation}
\end{widetext}

respectively. The above two equations can be equivalently expressed by,
\begin{align}\label{DE-FRW}
    H^2 &= \frac{8\pi G}{3}\left(\rho+\rho_\mathrm{g}\right) + \frac{\Lambda}{3} \\
    \Dot{H} &= -4\pi G \left[\left(\rho+\rho_\mathrm{g}\right)+\left(p+p_\mathrm{g}\right)\right] \ ,
\end{align}
where $\rho_\mathrm{g}$ and $p_\mathrm{g}$ have the following forms:
\begin{widetext}
\begin{equation}\label{entropy-energy}
    \begin{aligned}
        \rho_\mathrm{g} = \frac{3}{8\pi G} \Bigg\{ H^2 - \frac{\beta G H^4 }{\pi \gamma} \Bigg[& \frac{1}{2+\beta}\left(\frac{\beta G H^2}{\pi\alpha_-}\right)^{\beta} 2F_1\left(1+\beta,2+\beta,3+\beta,-\frac{\beta G H^2}{\pi\alpha_-}\right) \\
     & + \frac{1}{2-\beta}\left(\frac{\beta G H^2}{\pi\alpha_+}\right)^{-\beta} 2F_1\left(1-\beta,2-\beta,3-\beta,-\frac{G\beta G H^2}{\pi\alpha_+}\right) \Bigg] \Bigg\}
    \end{aligned}
\end{equation}
and
\begin{equation}\label{entropy-pressure}
     p_\mathrm{g} = \frac{\Dot{H}}{4\pi G}\left\{\frac{1}{\gamma}\left[\alpha_+\left(1+\frac{\pi\alpha_+}{\beta G H^2}\right)^{\beta-1}+\alpha_-\left(1+\frac{\pi\alpha_-}{\beta G H^2}\right)^{-\beta-1}\right] -1\right\} - \rho_\mathrm{g}\ ,
\end{equation}
\end{widetext}
respectively. Here $\rho_\mathrm{g}$ and $p_\mathrm{g}$ arises from the 4-parameter generalized entropy and thus they may be called as entropic energy density and entropic pressure. The presence of such $\rho_\mathrm{g}$ and $p_\mathrm{g}$ certainly modify the cosmological field equations compared to the usual ones, which may have rich cosmological consequences during different epochs of universe. Here we are interested to investigate the possible effects of $\rho_\mathrm{g}$ and $p_\mathrm{g}$ in simultaneous validation of ACT inflation and NANOGrav 15-year data from primordial tensor power spectrum.

\section{ACT inflation}\label{sec:ACT-Inflation}

We consider $\rho = p = \Lambda = 0$ during the early stage of the universe (as the normal matter gets diluted due to the fast expansion of the universe during its early stage), owing to which, the cosmological field equation from 4-parameter generalized entropy Eq.(\ref{Fried-Eq-2}) becomes,
\begin{align}
\frac{1}{\gamma}\left[\alpha_{+}\left(1 + \frac{\pi \alpha_+}{\beta GH^2}\right)^{\beta - 1}
+ \alpha_-\left(1 + \frac{\pi \alpha_-}{\beta GH^2}\right)^{-\beta-1}\right]\dot{H} = 0~~.
\label{FRW-2-inf}
\end{align}
This gives $H = \mathrm{constant}$. Thus in order to describe a quasi-dS inflationary scenario, here we will consider that the entropic parameters are not strictly constant, rather they slightly vary with the cosmic expansion of the universe. The entropic cosmology with varying exponents are well studied in \cite{Nojiri:2019skr,Odintsov:2023vpj}. In particular, here we consider only the parameter $\gamma$ varies by the following way and the other entropic parameters (i.e., $\alpha_+$, $\alpha_-$ and $\beta$) remain constant with $t$:
\begin{align}
\gamma(N) = \mathrm{exp}\left[\int_{N_\mathrm{f}}^{N}\sigma(N) dN\right]~~.
\label{gamma function}
\end{align}
Here $N$ denotes the e-folding number with $N_\mathrm{f}$ being the total e-folding number of the inflationary era. Therefore the FRW equation becomes,
\begin{align}
 -\left(\frac{2\pi}{G}\right)
\left[\frac{\alpha_+\left(1 + \frac{\alpha_+}{\beta}~S\right)^{\beta-1} + \alpha_-\left(1 + \frac{\alpha_-}{\beta}~S\right)^{-\beta-1}}
{\left(1 + \frac{\alpha_+}{\beta}~S\right)^{\beta} - \left(1 + \frac{\alpha_-}{\beta}~S\right)^{-\beta}}\right]\frac{1}{H^3}\frac{dH}{dN} = \sigma(N) \,,
\label{FRW-eq-viable-inf-main1}
\end{align}
on solving which for the Hubble parameter $H = H(N)$, we get,
\begin{align}
H(N) = 4\pi M_\mathrm{Pl}\sqrt{\frac{\alpha_+}{\beta}}
\left[\frac{2^{1/(2\beta)}\exp{\left[-\frac{1}{2\beta}\int^{N}\sigma(N)dN\right]}}
{\left\{1 + \sqrt{1 + 4\left(\alpha_+/\alpha_-\right)^{\beta}\exp{\left[-2\int^{N}\sigma(N)dN\right]}}\right\}^{1/(2\beta)}}\right] \,,
\label{solution-viable-inf-2}
\end{align}
where $M_\mathrm{Pl} = 1/\sqrt{16\pi G}$. The above two equations are valid for any form of $\sigma(N)$. In order to get a viable quasi dS inflation, let us consider that $\sigma(N)$ during inflation has the following form \cite{Nojiri_2022},
\begin{align}
\sigma(N) = \sigma_0 + \mathrm{e}^{-\left(N_\mathrm{f} - N\right)}~~;~~~~\mathrm{during~inflation} \,,
\label{sigma function}
\end{align}
where $\sigma_0$ is a constant. The above form of $\sigma(N)$ is important to understand, in particular --- the first term (i.e., $\sigma_0$ which is a constant) triggers a quasi dS inflation; while the second term (which remains subdominant during most of the inflationary period and becomes significant around $N \approx N_\mathrm{f}$) proves to be responsible for the exit of the inflation at $N=N_\mathrm{f}$. During inflation, $\sigma(N)$ is of the form of Eq.(\ref{sigma function}) and consequently we obtain $\int_0^{N}\sigma(N)dN = N\sigma_0 + \mathrm{e}^{-(N_\mathrm{f} - N)} - \mathrm{e}^{-N_\mathrm{f}}$. The above solution of $H(N)$ immediately leads to the slow roll parameter $\epsilon = -d\mathrm{ln}H/dN$ as follows:
\begin{align}
\epsilon(N) = \frac{\sigma_0 + \mathrm{e}^{-\left(N_\mathrm{f} - N\right)}}
{2\beta\sqrt{1 + 4\left(\alpha_+/\alpha_-\right)^{\beta}\exp{\left[-2\left(N\sigma_0 + \mathrm{e}^{N-N_\mathrm{f}}\right)\right]}}} \,.
\label{slow roll parameter}
\end{align}
With $\epsilon(N_\mathrm{f}) = 1$, we get
\begin{align}
\beta = \frac{(1 + \sigma_0)}
{2\sqrt{1 + 4\left(\alpha_+/\alpha_-\right)^{\beta}\exp{\left[-2\left(1 + \sigma_0N_\mathrm{f}\right)\right]}}} \,,
\label{end of inflation}
\end{align}
where we use $\int_0^{N_\mathrm{f}}\sigma(N)dN = 1+\sigma_0N_\mathrm{f}$.
\begin{figure}[!h]
\begin{center}
\centering
\includegraphics[width=3.5in,height=2.5in]{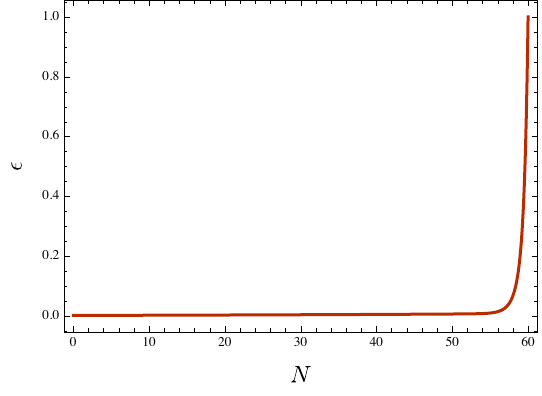}
\caption{$\epsilon(N)$ vs. $N$ for the considered form of $\sigma(N)$ in Eq.~(\ref{sigma function}) with $N_\mathrm{f} = 60$.}
 \label{plot-SRP-1}
\end{center}
\end{figure}
The important point to note here is that the slow roll parameter is flat during inflation (except at the end of inflation), which leads to a flat spectrum of tensor perturbation. The inflationary observables, in particular, the spectral tilt for the primordial curvature perturbation ($n_s$) and the tensor-to-scalar ratio ($r$) are defined by \cite{Schwarz:2001vv}:
\begin{eqnarray}
 n_s = 1- 2\epsilon - d\mathrm{ln}\epsilon/dN~~~~~~~~~~~\mathrm{and}~~~~~~~~~~~r=16\epsilon \,,
 \label{new-ns-r}
\end{eqnarray}
By using Eq.~(\ref{slow roll parameter}), the $n_s$ and $r$ from the above Eq.~(\ref{new-ns-r}) are evaluated as follows:
\begin{align}
n_s = 1 - \frac{2\sigma_0\sqrt{1 + 4\left(\alpha_+/\alpha_-\right)^{\beta}\exp{\left[-2\left(1 + \sigma_0N_\text{f}\right)\right]}}}
{(1+\sigma_0)\sqrt{1 + 4\left(\alpha_+/\alpha_-\right)^{\beta}}} - \frac{4\sigma_0\left(\alpha_+/\alpha_-\right)^{\beta}}
{1 + 4\left(\alpha_+/\alpha_-\right)^{\beta}} \,,
\label{ns final form}
\end{align}
and
\begin{align}
r = \frac{16\sigma_0\sqrt{1 + 4\left(\alpha_+/\alpha_-\right)^{\beta}\exp{\left[-2\left(1 + \sigma_0N_\text{f}\right)\right]}}}
{(1+\sigma_0)\sqrt{1 + 4\left(\alpha_+/\alpha_-\right)^{\beta}}}
\label{r final form}
\end{align}
respectively. To obtain such expressions of $n_s$ and $r$, we use Eq.(\ref{end of inflation}), i.e the above forms of $n_s$ and $r$ contain the information of $\epsilon(N_\mathrm{f}) = 1$. Confronting the theoretical expectations of $n_s$ and $r$ with the recent ACT-DR6+Planck18+BAO data ($n_s = 0.9743 \pm 0.0034$ and $r < 0.038$), we get the following viable ranges of the entropic parameters (see Table.~[\ref{Table-1}]):
\setlength{\arrayrulewidth}{0.1mm}
\setlength{\tabcolsep}{5pt}
\renewcommand{\arraystretch}{1.5}
\begin{table}[h]
  \centering
 {%
  \begin{tabular}{|c|c|c|c|}
   \hline
    Viable choices of $N_\mathrm{f}$ & Range of $\beta$ & Range of $\sigma_0$ & Range of $\left(\alpha_+/\alpha_-\right)^{\beta}$\\

   \hline
    (1) Set-1: $N_\mathrm{f} = 55$ & $\beta= (0,0.10]$ & $\sigma_0=[0.020,0.025]$ & $\left(\alpha_+/\alpha_-\right)^{\beta} \gtrsim 750$\\
   \hline
     (2) Set-2: $N_\mathrm{f} = 60$ & $\beta= (0,0.25]$ & $\sigma_0=[0.020,0.025]$ & $\left(\alpha_+/\alpha_-\right)^{\beta} \gtrsim 100$\\
   \hline
\end{tabular}%
 }
  \caption{Viable ranges on entropic parameters coming from the inflationary phenomenology for two different choices of $N_\mathrm{f}$.}
  \label{Table-1}
 \end{table}

 \begin{figure}[!h]
\begin{center}
\centering
\includegraphics[width=3.0in,height=2.0in]{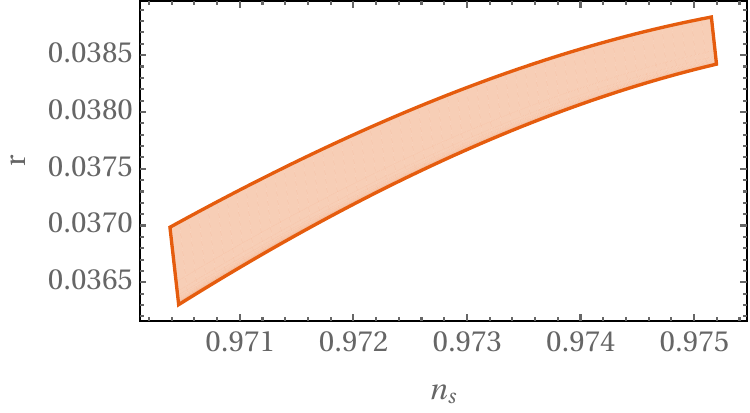}
\hskip0.6cm
\includegraphics[width=3.0in,height=2.0in]{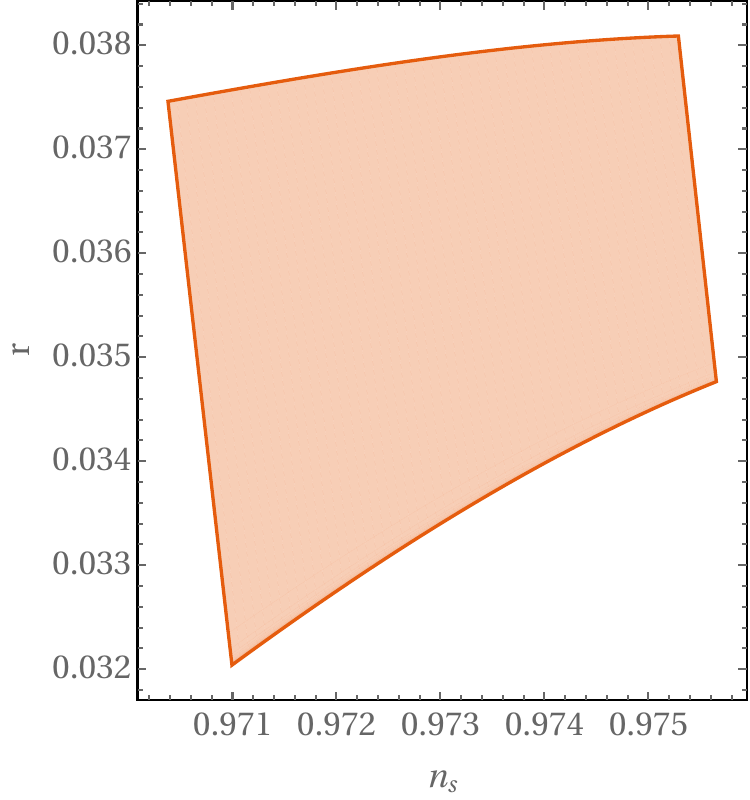}
\caption{Simultaneous compatibility of $n_s$ and $r$ according to Eq.~(\ref{ns final form}) and Eq.~(\ref{r final form}). The left and right plots correspond to $N_\mathrm{f}=55$ and $N_\mathrm{f}=60$ respectively.}
 \label{plot-observable}
\end{center}
\end{figure}

Therefore the considered form of $\sigma(N)$ in Eq.~(\ref{sigma function}) satisfies the ACT-DR6+Planck18+BAO data for suitable ranges of the entropic parameters given in the above Table.~[\ref{Table-1}]. Now we will investigate whether such inflationary scenario with the same range of entropic parameters (as shown in Table.~[\ref{Table-1}]) can satisfy the PTA data.

\section{Validation with NANOGrav}

It would be nice if the inflation, as showed in the previous section, is also compatible with the PTA data. However the problem is that the usual slow roll inflation gives rise to an almost flat tensor spectrum, which is way below than the GWs' amplitude observed by the PTA data. Therefore in order to have a simultaneous validation with the PTA as well as the ACT observations in the present context, the usual inflation needs to be accordingly modified in a way that it leads to a blue tensor spectrum over the modes around the NANOGrav frequency range, and at the same time, the validation with the ACT data (as ensured in the previous section) should not be harmed by that modification. This indicates that the usual slow roll parameter (as in Fig.~[\ref{plot-SRP-1}]) needs to be modified in such a way that:
\begin{itemize}
 \item the main modification on the slow roll parameter arises during the epoch when the NANOGrav modes exit the horizon, and

 \item the modification does not affect much during the initial phase of the inflation when the large CMB scales exit the horizon.
 \end{itemize}

 \subsection{Modified background dynamics}
Based on the above arguments, the modified slow roll parameter is considered as:
\begin{eqnarray}
 \epsilon_\mathrm{m}(N) = \epsilon_\mathrm{old}(N) + C\Bigg\{\mathrm{tanh}\left[D\left(N - N_\mathrm{1}\right)\right] + \mathrm{tanh}\left[D\left(N_\mathrm{2} - N\right)\right]\Bigg\} \, ,
 \label{Nano-1}
\end{eqnarray}
where the suffix 'm' stands for 'modified' form and $\epsilon_\mathrm{old}(N)$ is given in Eq.~(\ref{slow roll parameter}). The above expression clearly shows that the modified slow roll parameter gets corrected over the old one by a tanh function. Due to the tanh function, the correction term mainly affects the slow roll parameter only in-between $N_\mathrm{1} < N < N_\mathrm{2}$, otherwise $\epsilon_\mathrm{m}(N) \approx \epsilon_\mathrm{old}(N)$. For a better understanding, we give the plot of $\epsilon_\mathrm{m}(N)$ vs. $N$ in the following figure where we take $N_\mathrm{f} = 60$.

\begin{figure}[!h]
\begin{center}
\centering
\includegraphics[width=.5\linewidth]{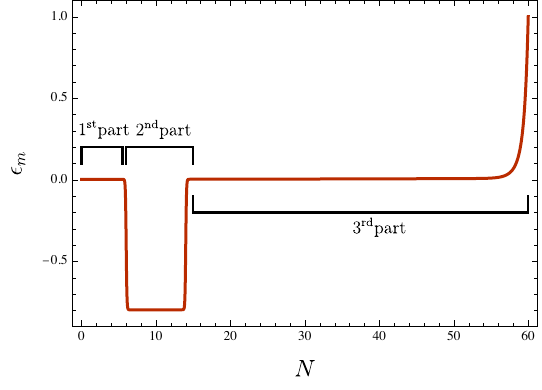}
\caption{$\epsilon_\mathrm{m}(N)$ vs. $N$. for $C=-0.4$, $D=10$, $\sigma_0 = 0.024$ and $N_\mathrm{f} = 60$. Moreover, we take $N_\mathrm{1}=6$ and $N_\mathrm{2}=14$ (later we will see that such values of $N_\mathrm{1}$ and $N_\mathrm{2}$ are compatible with the best fitted values of the parameters).}
 \label{plot-SRP-2}
\end{center}
\end{figure}

The Fig.~[\ref{plot-SRP-2}] clearly indicates that the entire duration of inflation ($0 \leq N \leq N_\mathrm{f}$) can be divided into three parts based on the behaviour of $\epsilon_\mathrm{m}(N)$:
\begin{enumerate}
 \item $\epsilon_\mathrm{m}(N) \approx \epsilon_\mathrm{old}(N)$ during the \textbf{1st~phase~of~inflation}, i.e. during $N < N_\mathrm{1}$, when the large scale modes exit the horizon;

 \item $\epsilon_\mathrm{m}$ gets a dip over the old one with $|\epsilon_\mathrm{m}| < 1$ in-between $N_\mathrm{1} < N < N_\mathrm{2}$ when, say, the modes around the NANOGrav exit the horizon (we consider $N_\mathrm{1}=6$ and $N_\mathrm{2}=14$ in the plot, later we will see that such values of $N_\mathrm{1}$ and $N_\mathrm{2}$ are compatible with the best fitted values of the parameters) --- this is termed as the \textbf{2nd phase (or the intermediate phase) of inflation}. Here it may be noted that a negative value of $C$ (sitting in front of the tanh function) makes a dip on the modified slow roll parameter (over the old one) while a positive $C$ leads to a bump on the same. In this regard, it is important to mention that the dip is crucial to get a blue tilted tensor power spectrum and validate the NANOGrav data, and thus we work with a negative value of $C$.

 \item Once again, $\epsilon_\mathrm{m}(N) \approx \epsilon_\mathrm{old}(N)$ during the \textbf{3rd (or final) phase of inflation} when the lower scale modes exit the horizon. This inflationary stage lasts for $N_\mathrm{2} < N < N_\mathrm{f}$, i.e. the entire inflation exits at $N = N_\mathrm{f}$ when $\epsilon_\mathrm{m}$ reaches to unity.
\end{enumerate}
The very first  property ensures the validation of ACT data even with this modified slow roll parameter, and we will show that the existence of the dip during the 2nd phase results to a blue tilted primordial tensor power spectrum over the modes that exit the horizon during $N_\mathrm{1} < N < N_\mathrm{2}$. Here it deserves mentioning that the modification of the slow roll parameter over $\epsilon_\mathrm{old}(N)$, in Eq.~(\ref{Nano-1}), is mainly considered from observational perspective based on the NANOGrav 15-year signal. As we will show that such $\epsilon_\mathrm{m}(N)$ results from four parameter generalized entropy with varying exponent, as also discussed around Eq.~(\ref{gamma function}). The microscopic origin of the four parameter generalized entropy with fixed exponent is described in \cite{Nojiri:2023bom}. However it will be interesting to extend such microscopic analysis of the generalized entropy in case of varying exponent. Since the main motivation of the present work is to examine the compatibility of generalized entropy with ACT and NANOGrav observations, the microscopic analysis of the same with varying exponent is expected to study in some future work.

By using $\epsilon_\mathrm{m}(N)$ from Eq.~(\ref{Nano-1}), we can solve the Hubble parameter as follows (by using $\epsilon_\mathrm{m}(N) = -\frac{d\ln{H_\mathrm{m}}}{dN}$):
\begin{eqnarray}
 H_\mathrm{m}(N) = H_\mathrm{old}(N)\mathrm{exp}\bigg[-C\int_{0}^{N}dN\, \left\{\mathrm{tanh}\left[D\left(N - N_\mathrm{1}\right)\right] + \mathrm{tanh}\left[D\left(N_\mathrm{2} - N\right)\right]\right\}\bigg] \, ,
 \label{reconstruct-1}
\end{eqnarray}
where $\int \mathrm{tanh}(x) dx= \ln{\left[\mathrm{cosh}(x)\right]}$ needs to be used, and $H_\mathrm{old}(N)$ is given in Eq.~(\ref{solution-viable-inf-2}). By plugging the above solution of $H(N)$ into Eq.~(\ref{FRW-eq-viable-inf-main1}), one can reconstruct the modified form $\sigma_\mathrm{m}(N)$. Similar to $\epsilon_\mathrm{m}$, the evolution of $H_\mathrm{m}(N)$ during inflation from Eq.~(\ref{reconstruct-1}) can also be decomposed into three parts as follows:
\begin{enumerate}
 \item During the \textbf{1st phase of inflation (with $N<N_\mathrm{1}$)},
 \begin{eqnarray}
  H_\mathrm{m}(N) = H_\mathrm{old}(N)\mathrm{exp}\bigg[-C\int_{0}^{N}dN\, \left\{\mathrm{tanh}\left[D\left(N - N_\mathrm{1}\right)\right] + \mathrm{tanh}\left[D\left(N_\mathrm{2} - N\right)\right]\right\}\bigg]\, .
 \label{H-0}
 \end{eqnarray}
Owing to the tanh function, the integrand within exponential almost vanishes for $N<N_\mathrm{1}$ and thus the Hubble parameter is given by,
 \begin{eqnarray}
  H_\mathrm{m}(N) \approx H_\mathrm{old}(N) \, .
  \label{H-1}
 \end{eqnarray}
Eq.~(\ref{H-1}) clearly indicates that $H_\mathrm{m}(N)$ remains almost constant during this phase.

 \item During the \textbf{2nd part of inflation (with $N_\mathrm{1}<N<N_\mathrm{2}$)},
 \begin{eqnarray}
  H_\mathrm{m}(N)&=&H_\mathrm{old}(N)\mathrm{exp}\bigg[-C\int_{N_\mathrm{1}}^{N}dN\, \left\{\mathrm{tanh}\left[D\left(N - N_\mathrm{1}\right)\right] + \mathrm{tanh}\left[D\left(N_\mathrm{2} - N\right)\right]\right\}\bigg]\nonumber\\
  &=&H_\mathrm{old}(N)\mathrm{exp}\bigg[-\frac{C}{D}\ln{\left\{\frac{\mathrm{cosh}[D(N - N_\mathrm{1})]~\mathrm{cosh}[D(N_\mathrm{2} - N_\mathrm{1})]}{\mathrm{cosh}[D(N_\mathrm{2} - N)]}\right\}}\bigg]\, .
 \label{H-2}
 \end{eqnarray}
where the lower limit is taken from $N_\mathrm{1}$, as the integrand does not contribute for $N<N_\mathrm{1}$. The above expression indicates that the Hubble parameter monotonically increases with $N$ during this phase (as $C<0$). Clearly such an increasing behaviour of the Hubble parameter is due to the modified slow roll parameter, otherwise the Hubble parameter follows $H_\mathrm{m}(N) = H_\mathrm{old}(N)$ during the entire inflation.


 \item During the \textbf{3rd part of inflation (with $N_\mathrm{2}<N<N_\mathrm{f}$)},
 \begin{eqnarray}
  H_\mathrm{m}(N)&=&H_\mathrm{old}(N)\mathrm{exp}\bigg[-C\int_{N_\mathrm{1}}^{N_\mathrm{2}}dN\, \left\{\mathrm{tanh}\left[D\left(N - N_\mathrm{1}\right)\right] + \mathrm{tanh}\left[D\left(N_\mathrm{2} - N\right)\right]\right\}\bigg]\nonumber\\
  &=&H_\mathrm{old}(N)\mathrm{exp}\bigg[-\frac{2C}{D} \ln{\left\{\mathrm{cosh}[D(N_\mathrm{2} - N_\mathrm{1})]\right\}}\bigg]\, .
 \label{H-3}
 \end{eqnarray}
Clearly the integration results to a constant factor which, due to $C < 0$, in turn acts as an amplification over $H_\mathrm{old}(N)$. Therefore the Hubble parameter during the 3rd phase is getting amplified compared to that of during the 1st inflationary phase.

\end{enumerate}

\begin{figure}[!h]
\begin{center}
\centering
\includegraphics[width=4.0in,height=2.5in]{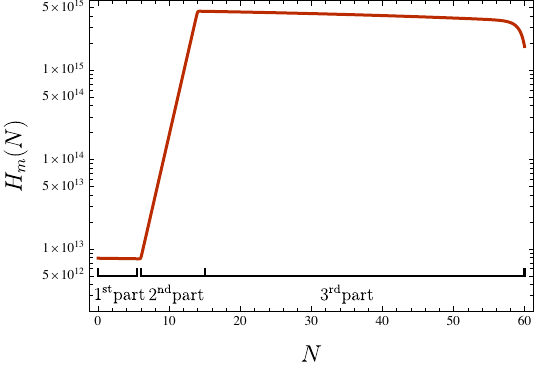}
\caption{$H_\mathrm{m}(N)$ (in the unit of $\mathrm{GeV}$) vs. $N$. for $C=-0.4$, $D=10$, $\sigma_0 = 0.024$ and $N_\mathrm{f} = 60$. Moreover, we take $N_\mathrm{1}=6$ and $N_\mathrm{2}=14$.}
 \label{plot-Hubble}
\end{center}
\end{figure}

For a better understanding, we give a plot of the Hubble dynamics during inflation in Fig.~[\ref{plot-Hubble}] by using Eq.~(\ref{reconstruct-1}). The figure is indeed compatible with the above arguments.

Here it may be mentioned that the above form of Hubble parameter (or equivalently, the slow roll parameter of Eq.~(\ref{Nano-1})) can also be obtained from a non-canonical scalar-tensor theory with suitable scalar potential and the non-canonical coupling function (see in the Appendix: Sec.~[\ref{sec-appendix-0}]). In this case, the scalar field (null) energy condition needs to be violated during the 2nd inflationary period when the Hubble parameter increases with the e-fold number, and thus, a canonical scalar-tensor theory can not lead to the above Hubble parameter. Therefore we may argue that the Hubble parameter of Eq.~(\ref{reconstruct-1}) can be triggered either from entropic cosmology (what we are interested in the present work) or from a non-canonical scalar-tensor theory (or, from some other gravity theories as well). This clearly hints a degeneracy of gravity theories in respect to the Hubble expansion as of Eq.~(\ref{reconstruct-1}). In this regard, we would like to mention that lifting of such degeneracy is not the motive of our present work, in particular, our motivation is two folded as follows --- (i) to examine possible $modification$ of the usual slow roll inflation in order to be compatible with the NANOGrav 15-year signal with primary gravitational waves, and (ii) to examine the compatibility of $entropic~cosmology$ from the perspective of the NANOGrav and the ACT data. Hence we take the thermodynamic route of cosmology, where the entropy of the apparent horizon is given by the four parameter generalized entropy, and will address its status (with proper constraints on entropic parameters) from both the NANOGrav and the ACT observations.\\

After the inflation ends, we consider an instantaneous reheating, during which, the entropic energy density corresponding to the 4-parameter generalized entropy (i.e. $\rho_\mathrm{g}$ in Eq.~(\ref{entropy-energy})) instantaneously decays to relativistic particles and starts the radiation dominated era (for the discussion of reheating in the context of 4-parameter generalized entropic cosmology, see \cite{Odintsov:2023vpj}). Therefore, during the radiation epoch, the entropy of the apparent horizon takes the form of Bekenstein-Hawking entropy like and radiation acts as the matter field inside of the horizon. Owing to this, the Friedmann equations in radiation era take the usual form, in particular,
\begin{eqnarray}
 H^2_\mathrm{m} = \frac{8\pi G}{3}\rho_\mathrm{rad}~~~~~\mathrm{and}~~~~~~\dot{H}_\mathrm{m} = -4\pi G\left(p_\mathrm{rad} + \rho_\mathrm{rad}\right)\, ,
 \label{H-rad}
\end{eqnarray}
(with $p_\mathrm{rad} = \rho_\mathrm{rad}/3$), and thus, $H_\mathrm{m} \propto a^{-2}$ during the post-inflation. With such background evolution in hand, we now analyze the primordial tensor perturbation generated during inflation from a Bunch-Davies vacuum state.

\subsection{Evolution and spectrum of primordial GWs}\label{subsec:PGWB}
At the very beginning of this section, let us clarify some symbols:
\begin{itemize}
 \item $a_\mathrm{1}$, $a_\mathrm{2}$ and $a_\mathrm{f}$ represent the scale factor at end instances of the 1st stage of inflation, the 2nd stage of inflation and the 3rd stage of inflation, respectively.

 \item Regarding the wave numbers of primordial tensor perturbation --- $k_\mathrm{1}$, $k_\mathrm{2}$ and $k_\mathrm{f}$ represent the modes that exit the horizon at end instances of the 1st stage of inflation, the 2nd stage of inflation and the 3rd stage of inflation, respectively.
\end{itemize}

Here we will analyze the evolution of the tensor perturbation (or gravitational waves) from the inflation to the radiation era with proper continuity conditions of the perturbation variable, and find the corresponding tensor spectrum at present epoch. Regarding the inflationary case, we have to consider all the three stages of the inflation as discussed in the previous section. The tensor perturbation, symbolized by $h_{ij}(t,\vec{x})$, over a spatially flat FRW spacetime is defined by
\begin{eqnarray}
 ds^2 = -dt^2 + a^2(t)\left[\left(\delta_{ij} + h_{ij}\right)dx^{i}dx^{j}\right]~~.
 \label{I-1}
\end{eqnarray}
The equation governing the dynamics of the tensor modes in the absence of anisotropic source is \cite{Odintsov:2024sbo},
\begin{eqnarray}
 \ddot{h}_{ij} + 3H_\mathrm{m}\dot{h}_{ij} - \nabla^2h_{ij} = 0~~,
 \label{I-2}
\end{eqnarray}
where $\nabla^2$ represents the Laplacian operator. On quantizing the tensor perturbation, $h_{ij}(t,\vec{x})$ can be decomposed in terms of its Fourier modes $h(k,t)$ as:
\begin{eqnarray}
 \hat{h}_{ij}(t,\vec{x}) = \sum_{\lambda=+,\times}\int \frac{d^3\vec{k}}{(2\pi)^{3/2}}\left[\hat{a}_k^{\lambda}\epsilon_{ij}^{\lambda}(\vec{k})h(k,t)\mathrm{e}^{i\vec{k}.\vec{x}} + c.c.\right]~~,
 \label{I-3}
\end{eqnarray}
where $\lambda = +,\times$ denotes two types of polarizations of the gravitational waves (GWs), $\epsilon_{ij}^{\lambda}(\vec{k})$ are the polarization tensor and $\hat{a}_k$ ($\hat{a}_k^{+}$) are the annihilation (creation) operators respectively that satisfy the usual commutation rules. Moreover, from the transverse condition of GWs, i.e. due to $\partial_{i}h^{ij} = 0$, one immediately gets $k^{i}\epsilon_{ij}^{\lambda}(\vec{k}) = 0$. Owing to Eq.~(\ref{I-2}), $h(k,t)$ obeys,
\begin{eqnarray}
 \ddot{h}(k,t) + 3H_\mathrm{m}\dot{h}(k,t) + \frac{k^2}{a^2}h(k,t) = 0~~.
 \label{I-4}
\end{eqnarray}
Passing to the conformal time $\eta$ (defined by $d\eta = dt/a(t)$) and introducing the Mukhanov-Sasaki variable as
\begin{equation}
    v(k,\eta) = \frac{a}{2} M_\mathrm{Pl}h(k,\eta) \ ,
    \label{I-5}
\end{equation}
Eq.~(\ref{I-4}) transforms to,
\begin{equation}\label{EoM_MS_variable}
    v''(k,\eta)+\left(k^2-\frac{a''}{a}\right)v(k,\eta)=0 \ ,
\end{equation}
where the overprime denotes ${d}/{d\eta}$. Clearly, the evolution of $v(k,\eta)$ depends on the background dynamics through the factor ${a''}/{a}$, which we consider in the following subsections.

\subsubsection{Evolution of GWs during 1st stage of inflation}
As mentioned earlier that during this phase the Hubble parameter is (almost) constant and leads to a dS kind of inflation, thus we may write $H_\mathrm{m}=H_\mathrm{1}$ (with $H_\mathrm{1}$ being a constant). Therefore the scale factor is given by $a(\eta) = -1/(H_\mathrm{1}\eta)$, in terms of the conformal time. With this form of $a(\eta)$, Eq.~(\ref{EoM_MS_variable}) becomes,
\begin{equation}\label{EoM_MS_variable_Inflation}
    v''(k,\eta)+\left(k^2-\frac{2}{\eta^2}\right)v(k,\eta)=0 \ ,
\end{equation}
on solving which, we get,
\begin{equation}\label{MS_variable_Inflation}
    v(k,\eta) = \frac{e^{-ik\eta}}{\sqrt{2k}}\left(1-\frac{i}{k\eta}\right) \ .
\end{equation}
This solution is compatible with the Bunch-Davies initial condition in the deep inside of the Hubble radius, given by,
\begin{equation}
    \lim_{|k\eta|\gg 1}v(k,\eta) = \frac{1}{\sqrt{2k}}e^{-ik\eta} \ .
    \label{I-6}
\end{equation}
Plugging the above solution of $v(k,\eta)$ into Eq.~(\ref{I-5}) yields the solution of $h(k,\eta)$ as
\begin{align}
    h(k,\eta) = -\sqrt{\frac{2}{k}}\left(\frac{H_\mathrm{1}}{M_\mathrm{Pl}}\right)\eta \,  e^{-ik\eta}\left(1-\frac{i}{k\eta}\right) \label{h_inflation} \ .
\end{align}
Consequently we calculate
\begin{equation}\label{I-7}
    h'(k,\eta)=i \, \sqrt{\frac{2}{k}} \left(\frac{H_\mathrm{1}}{M_\mathrm{Pl}}\right)(k\eta)e^{-ik\eta} \ ,
\end{equation}
which will be important later. Here it is convenient to define the transfer function as,
\begin{eqnarray}
 h(k,\eta) = i\sqrt{\frac{2}{k^3}}\left(\frac{H_\mathrm{1}}{M_\mathrm{Pl}}\right)\chi(k,\eta) \, ,
 \label{transfer function}
\end{eqnarray}
we get the evolution of the transfer function during the 1st stage of inflation as,
\begin{eqnarray}
 \chi^\mathrm{(1)}(k,\eta)&=&i(k\eta) \, e^{-ik\eta}\left(1-\frac{i}{k\eta}\right) \, ,\nonumber\\
 \frac{d\chi^\mathrm{(1)}(k,\eta)}{d\eta}&=&k(k\eta)e^{-ik\eta} \, ,
 \label{solution-transfer}
\end{eqnarray}
with the superfix number over $\chi$ denotes the 1st stage of inflation. At this stage, let us rescale the scale factor by: $A = a/a_\mathrm{1}$ --- such rescaled scale factor proves to be useful for subsequent calculations and thus we will use $A$ as the independent variable onwards. In such rescaled scale factor, $A=1$ denotes the end of 1st inflationary stage when the transfer function and its derivatives have the following forms:
\begin{eqnarray}
 \chi^\mathrm{(1)}(k,A=1)&=&1~~~~~~~\mathrm{and}~~~~~~~~\frac{d\chi^\mathrm{(1)}(k,A)}{dA}\big|_{A=1}=\mathcal{O}\left(\frac{k^2}{k_\mathrm{1}^2}\right) \approx 0~~~~\mathrm{for}~~k < k_\mathrm{1}\, ,\nonumber\\
 \chi^\mathrm{(1)}(k,A=1)&=&-i\left(\frac{k}{k_\mathrm{1}}\right)e^{ik/k_\mathrm{1}}~~~~~\mathrm{and}~~~~~\frac{d\chi^\mathrm{(1)}(k,A)}{dA}\big|_{A=1} = -\left(\frac{k^2}{k_\mathrm{1}^2}\right)e^{ik/k_\mathrm{1}}~~~~\mathrm{for}~~k > k_\mathrm{1}\, .
 \label{bc-1st inflation}
\end{eqnarray}
Here $|k_\mathrm{1}\eta_\mathrm{1}| = 1$; and the modes $k < k_\mathrm{1}$ exit the horizon during the 1st inflation, while all the modes with $k>k_\mathrm{1}$ exit the horizon after the end of the 1st inflation. The above Eq.~(\ref{bc-1st inflation}) provide the junction conditions of the transfer function in-between the 1st and the 2nd inflationary stage.

\subsubsection{Evolution of GWs during 2nd stage of inflation}
During the 2nd stage of inflation, the slow roll parameter is negative (with $|\epsilon_\mathrm{m} < 1|$) and almost constant (see Fig.~[\ref{plot-SRP-2}]). Owing to a $constant$ $\epsilon_\mathrm{m}$, the Hubble parameter can be parametrized by:
\begin{eqnarray}
 H_\mathrm{m}(A) = H_\mathrm{1}A^{-3(1+\omega)/2}\, ,
 \label{2I-1}
\end{eqnarray}
(in terms of rescaled scale factor) where $\omega = -1 - 2|\epsilon_\mathrm{m}|/3$ represents an effective equation of state during the 2nd inflation. As as consequence, the transfer function during this stage is governed by the following equation:
\begin{eqnarray}
 \frac{d^2\chi^\mathrm{(2)}}{dA^2} + \left(\frac{5-3\omega}{2A}\right)\frac{d\chi^\mathrm{(2)}}{dA} + \frac{(k/k_\mathrm{1})^2}{A^{1-3\omega}}\chi^\mathrm{(2)} = 0 \, ,
 \label{2I-2}
\end{eqnarray}
where the suffix '2' symbolizes the 2nd inflationary stage. The continuity conditions of the transfer function at $A=1$ (i.e. at the junction between 1st inflation-to-2nd inflation) are:
\begin{eqnarray}
 \chi^\mathrm{(2)}(k,1) = \chi^\mathrm{(1)}(k,1)~~~~~\mathrm{and}~~~~~\frac{d\chi^\mathrm{(2)}(k,1)}{dA} = \frac{d\chi^\mathrm{(1)}(k,1)}{dA}\, ,
 \label{2I-3}
\end{eqnarray}
respectively, where $\chi^\mathrm{(1)}(k,1)$ (and its derivatives) are shown in Eq.~(\ref{bc-1st inflation}) for different modes. With these junction conditions, we solve Eq.~(\ref{2I-2}) as follows:
\begin{equation}\label{2I-4}
    \chi^\mathrm{(2)}(k,A)=\frac{\mathscr{N}(k,A)}{\mathscr{D}(k)} \ .
\end{equation}
Here, along with
\begin{eqnarray}
 \nu \equiv \frac{3\omega-3}{2+6\omega}~~~~,~~~~~x=\frac{2k/k_\mathrm{1}}{1+3\omega}~~~~~,~~~~~y=\frac{2k/k_\mathrm{2}}{1+3\omega}
 \label{2I-5a}
\end{eqnarray}
the quantities $\mathscr{D}(k)$ and $\mathscr{N}(k,A)$ are given by
\begin{align}\label{2I-5}
    \mathscr{D}(k)\equiv2\left(\frac{k}{k_\mathrm{1}}\right)\Bigg\{J_{\nu}(x)\left[J_{-\nu-1}(x) - J_{-\nu+1}(x)\right] + J_{-\nu}(x)\left[J_{\nu+1}(x) - J_{\nu-1}(x)\right]\Bigg\}
\end{align}
and
\begin{equation}\label{2I-6}
    \mathscr{N}(k,A)\equiv A^{\frac{3\omega-3}{4}}\left[J_\nu\left(y\left({A}{/A_\mathrm{2}}\right)^\frac{1+3\omega}{2}\right) \mathscr{N}_1(k)+J_{-\nu}\left(y\left({A}/{A_\mathrm{2}}\right)^\frac{1+3\omega}{2}\right) \mathscr{N}_2(k) \right] \ ,
\end{equation}
respectively, with
\begin{equation}\label{2I-6a}
    \mathscr{N}_1(k)\equiv 2\left(\frac{k}{k_\mathrm{1}}\right)\chi^\mathrm{(1)}(k,1)\Big[J_{-\nu-1}(x)-J_{-\nu+1}(x)\Big]+\bigg((3\omega-3)\chi^\mathrm{(1)}(k,1)-4\frac{d\chi^\mathrm{(1)}(k,1)}{dA}\bigg)J_{-\nu}(x)
\end{equation}
and
\begin{equation}\label{2I-6b}
    \mathscr{N}_2(k)\equiv 2\left(\frac{k}{k_\mathrm{1}}\right)\chi^\mathrm{(1)}(k,1)\Big[J_{\nu+1}(x)-J_{\nu-1}(x)\Big] - \bigg((3\omega-3)\chi^\mathrm{(1)}(k,1)-4\frac{d\chi^\mathrm{(1)}(k,1)}{dA}\bigg)J_{\nu}(x) \ .
\end{equation}
Since the mode with $k = k_\mathrm{2}$ (present in Eq.~(\ref{2I-5a})) exits the horizon at the end of the 2nd inflation (i.e. at $A = A_\mathrm{2}$), the modes that exit the horizon during the 2nd stage of inflation are given by: $k_\mathrm{1} < k < k_\mathrm{2}$. In the present context, we choose $k_\mathrm{1}$ and $k_\mathrm{2}$ in such a way that the range of modes $k = [k_\mathrm{1}, k_\mathrm{2}]$ fall around the NANOGrav frequency range (see the numerical analysis in Sec.~[\ref{sec-numerics}]).

We will eventually use the above solution of $\chi(k,A)$ to arrive at the GWs spectrum at present epoch. At the moment, we also determine $d\chi^\mathrm{(2)}(k,A)/dA$ from Eq.~(\ref{2I-4}), which is given by,
\begin{eqnarray}
 \frac{d\chi^\mathrm{(2)}(k,A)}{dA} = \frac{1}{\mathscr{D}(k)}\frac{d\mathscr{N}(k,A)}{dA}~~,
 \label{2I-7}
\end{eqnarray}
(where one may use the property for the derivative of the Bessel function: $J'_{\nu}(x) = \frac{1}{2}\left[J_{\nu-1}(x) - J_{\nu+1}(x)\right]$). At the end of this subsection, we would like to mention that the above expressions of $\chi^\mathrm{(2)}(k,A)$ and $d\chi^\mathrm{(2)}(k,A)/dA$ at $A = A_\mathrm{2}$ provide the junction conditions of the transfer function in-between the 2nd and the 3rd inflationary stage.

\subsubsection{Evolution of GWs during 3rd inflation}

Recall that the 3rd stage of inflation can again be considered as a dS inflation (as $0 < \epsilon_\mathrm{m} \ll 1$), however, with a amplified Hubble parameter than that of in the 1st stage. Let the dS Hubble parameter during the 3rd stage of inflation is given by: $H_\mathrm{m}(N) = H_\mathrm{f}$. From the continuity of the Hubble parameter, we have the following relation between $H_\mathrm{1}$ and $H_\mathrm{f}$ as (by using Eq.~(\ref{2I-1})),
\begin{eqnarray}
H_\mathrm{f} = H_\mathrm{1}\left(\frac{a_\mathrm{2}}{a_\mathrm{1}}\right)^{-3(1+\omega)/2} \, ,
\label{continuity-Hubble}
\end{eqnarray}
(recall, $a_\mathrm{2}$ is the end instance of the 2nd inflation), and $H_\mathrm{f} > H_\mathrm{1}$. The transfer function in this stage follows:
\begin{eqnarray}
 \frac{d^2\chi^\mathrm{(3)}}{dA^2} + \frac{4}{A}\frac{d\chi^\mathrm{(3)}}{dA} + A_\mathrm{2}^2\left(\frac{k}{k_\mathrm{2}}\right)^2\frac{\chi^\mathrm{(3)}(k,A)}{A^4} = 0 \, ,
 \label{3I-1}
\end{eqnarray}
With the junction conditions at $A=A_\mathrm{2}$, in particular,
\begin{eqnarray}
 \chi^\mathrm{(3)}(k,A_\mathrm{2}) = \chi^\mathrm{(2)}(k,A_\mathrm{2})~~~~~\mathrm{and}~~~~~\frac{d\chi^\mathrm{(3)}(k,A_\mathrm{2})}{dA} = \frac{d\chi^\mathrm{(2)}(k,A_\mathrm{2})}{dA}\, ,
 \label{3I-n1}
\end{eqnarray}
(with $\chi^\mathrm{(2)}(k,A_\mathrm{2})$ and its derivative can be obtained from Eq.~(\ref{2I-4})), the solution of Eq.~(\ref{3I-1}) is given by:
\begin{eqnarray}
 \chi^\mathrm{(3)}(k,A)&=&\frac{A_\mathrm{2}(k_\mathrm{2}/k)^3}{2A}e^{-i\frac{k}{k_\mathrm{2}}\left(1 + \frac{A_\mathrm{2}}{A}\right)}\Bigg[e^{2i\frac{k}{k_\mathrm{2}}}\left(A_\mathrm{2}\frac{d\chi^\mathrm{(2)}(k,A_\mathrm{2})}{dA}\left(1 - i\frac{k}{k_\mathrm{2}}\right) + \chi^\mathrm{(2)}(k,A_\mathrm{2})\frac{k^2}{k_\mathrm{2}^2}\right)\left(\frac{k}{k_\mathrm{2}} - i\frac{A}{A_\mathrm{2}}\right)\nonumber\\
 &+&e^{2i\frac{k}{k_\mathrm{2}}\frac{A_\mathrm{2}}{A}}\left(A_\mathrm{2}\frac{d\chi^\mathrm{(2)}(k,A_\mathrm{2})}{dA}\left(1 + i\frac{k}{k_\mathrm{2}}\right) + \chi^\mathrm{(2)}(k,A_\mathrm{2})\frac{k^2}{k_\mathrm{2}^2}\right)\left(\frac{k}{k_\mathrm{2}} + i\frac{A}{A_\mathrm{2}}\right)\bigg]\, .
 \label{3I-2}
\end{eqnarray}
At the end instance of 3rd stage of inflation, the rescaled scale factor is given by: $A_\mathrm{f} = \frac{a_\mathrm{f}}{a_\mathrm{1}} = A_\mathrm{2}\left(\frac{k_\mathrm{f}}{k_\mathrm{2}}\right)$ (recall that $k_\mathrm{f}$ represents the mode that exits the horizon at $a = a_\mathrm{f}$).

\subsubsection{GWs' evolution during radiation era and its spectrum today}

During the radiation dominated era, the Hubble parameter behaves as $H \propto a^{-2}$ and thus we have $A^4H^2 = H_\mathrm{f}^2$ (where $H_\mathrm{f}$ is the Hubble parameter at the end of inflation), i.e.,
\begin{eqnarray}
 A^4H^2 = A_\mathrm{f}^4H_\mathrm{f}^2~~.
 \label{rd-1}
\end{eqnarray}
Using this, the equation of motion for $\chi(k,A)$ during RD takes the following form
\begin{equation}\label{EoM_chi_RD}
    \frac{d^2\chi}{dA^2}+\frac{2}{A}\frac{d\chi}{dA}+\frac{(k/k_\mathrm{f})^2}{A_\mathrm{f}^2}\chi(k,A)=0~~,
\end{equation}
where $k_\mathrm{f}=a_\mathrm{f}H_\mathrm{f}$ is the mode that re-enters the horizon at the end of the 3rd inflationary stage. On solving Eq.~(\ref{EoM_chi_RD}), we get the transfer function during RD as follows:
\begin{equation}\label{chi_RD}
    \chi^\mathrm{RD}(k,A) = \frac{1}{A}\left(c_\mathrm{1}\mathrm{e}^{-ibA} + c_\mathrm{2}\mathrm{e}^{ibA}\right) \ ,
\end{equation}
with $b = \left(\frac{k}{k_\mathrm{f}}\right)\frac{1}{A_\mathrm{f}}$. Moreover $c_\mathrm{1}$ and $c_\mathrm{2}$ are the integration constants that can be determined from the continuity conditions of the transfer function at the junction of 3rd inflation-to-radiation at $A=A_\mathrm{f}$, given by:
\begin{eqnarray}
 \chi^\mathrm{RD}(k,A_\mathrm{f}) = \chi^\mathrm{(3)}(k,A_\mathrm{f})~~~~~~~~~~~\mathrm{and}~~~~~~~~~~
 \frac{d\chi^\mathrm{RD}(k,A)}{dA}\bigg|_{A_\mathrm{f}} = \frac{d\chi^\mathrm{(3)}(k,A)}{dA}\bigg|_{A_\mathrm{f}}~~,
 \label{rd-2}
\end{eqnarray}
where the quantities in the r.h.s. can be determined from Eq.~(\ref{3I-2}). Such continuity conditions lead to the integration constants as,
\begin{align}\label{c_i}
    c_\mathrm{1,2}=\frac{e^{\pm ibA_\mathrm{f}}}{2}\left[\left(A_\mathrm{f} \mp \frac{1}{ib}\right)\chi^\mathrm{(3)}(k,A_\mathrm{f}) \mp \left(\frac{A_\mathrm{f}}{ib}\right)\frac{d\chi^\mathrm{(3)}(k,A_\mathrm{f})}{dA}\right] \ .
\end{align}
Consequently the final form of the transfer function during RD is given by,
\begin{eqnarray}\label{rd-3}
    \chi^\mathrm{RD}(k,A)&=&\frac{e^{-ib(A-A_\mathrm{f})}}{2A}\left[\left(A_\mathrm{f} - \frac{1}{ib}\right)\chi^\mathrm{(3)}(k,A_\mathrm{f}) - \left(\frac{A_\mathrm{f}}{ib}\right)\frac{d\chi^\mathrm{(3)}(k,A_\mathrm{f})}{dA}\right]\nonumber\\
    &+&\frac{e^{ib(A-A_\mathrm{f})}}{2A}\left[\left(A_\mathrm{f} + \frac{1}{ib}\right)\chi^\mathrm{(3)}(k,A_\mathrm{f}) + \left(\frac{A_\mathrm{f}}{ib}\right)\frac{d\chi^\mathrm{(3)}(k,A_\mathrm{f})}{dA}\right]  \ .
\end{eqnarray}

The dimensionless energy density of GWs, at an instant $A$, is defined by
\begin{align}\label{rd-4}
    \Omega_\mathrm{GW}(k,A)=&\frac{1}{3H^2M_\mathrm{Pl}^2}\rho_\mathrm{GW}(k,A) \\
    =&\frac{1}{6A^2a_\mathrm{f}^2H^2}\left(\frac{k^3}{2\pi^2}\right)\left\{a_\mathrm{f}^2A^4H^2\left|\frac{dh(k,A)}{dA}\right|^2 + k^2|h(k,A)|^2\right\} \ ,
\end{align}
where $\rho_\mathrm{GW}(k,A)$ represents the GWs energy density (at the instant $A$) per logarithmic interval of the modes. In terms of the transfer function, $\Omega_\mathrm{GW}(k,A)$ from Eq.~(\ref{rd-4}) becomes,
\begin{equation}\label{rd-5}
    \Omega_\mathrm{GW}(k,A)=\frac{1}{6\pi^2}\left(\frac{H_\mathrm{1}}{M_\mathrm{Pl}}\right)^2\left\{A^2\left|\frac{d\chi(k,A)}{dA} \right|^2 + \frac{k^2}{a^2H^2}|\chi(k,A)|^2\right\} \ ,
\end{equation}
with $H_\mathrm{1}$ being the Hubble parameter during the 1st dS inflation. One may simplify the term $k/(aH)$ (present in the r.h.s. of Eq.~(\ref{rd-5})) as,
\begin{eqnarray}
 \frac{k}{aH} = b\left(\frac{k_\mathrm{f}A_\mathrm{f}}{aH}\right) = b\left(\frac{A_\mathrm{f}^2H_\mathrm{f}}{AH}\right) = b\left(\frac{A^2H}{AH}\right) = bA~~,
 \label{rd-6}
\end{eqnarray}
where $b = \left(\frac{k}{k_\mathrm{f}}\right)\frac{1}{A_\mathrm{f}}$ (see after Eq.~(\ref{chi_RD})), and we have used $H\propto A^{-2}$ in the third equality. Therefore $\Omega_\mathrm{GW}(k,A)$ from Eq.~(\ref{rd-5}) can be expressed by,
\begin{equation}\label{rd-7}
    \Omega_\mathrm{GW}(k,A)=\frac{1}{6\pi^2}\left(\frac{H_\mathrm{1}}{M_\mathrm{Pl}}\right)^2\left\{A^2\left|\frac{d\chi^\mathrm{RD}(k,A)}{dA} \right|^2 + b^2A^2\left|\chi^\mathrm{RD}(k,A)\right|^2\right\} \ .
\end{equation}
The modes of interest are well inside the Hubble radius at late times (say, close to the radiation-matter equality) during radiation domination. Consequently the dimensionless energy density parameter $\Omega_\mathrm{GW}^\mathrm{(0)}(k)$ today (i.e. at present epoch) is given by,
\begin{eqnarray}
 \Omega_\mathrm{GW}^\mathrm{(0)}(k)h^2&\simeq&\left(\frac{g_{r,0}}{g_{r,eq}}\right)^{1/3}\Omega_\mathrm{R}h^2\Omega_\mathrm{GW}(k,A)\nonumber\\
 &=&\frac{1}{6\pi^2}\left(\frac{g_{r,0}}{g_{r,eq}}\right)^{1/3}\Omega_\mathrm{R}h^2\left(\frac{H_\mathrm{1}}{M_\mathrm{Pl}}\right)^2\left\{A^2\left|\frac{d\chi^\mathrm{RD}(k,A)}{dA}\right|^2 + b^2A^2\left|\chi^\mathrm{RD}(k,A)\right|^2\right\}~~,
 \label{rd-8}
\end{eqnarray}
where $\Omega_\mathrm{R}$ denotes the present day dimensionless energy density of radiation, $g_{r,eq}$ and $g_{r,0}$ represent the number of relativistic degrees of freedom at matter-radiation equality and today respectively. Furthermore by using Eq.~(\ref{rd-3}), we determine different terms present in the expression of $\Omega_\mathrm{GW}^\mathrm{(0)}(k)$, in particular,
\begin{eqnarray}\label{chi-0}
    \left|\chi^\mathrm{RD}(k,A)\right|^2 = \frac{1}{b^2A^2}&\Bigg|&bA_\mathrm{f}\, \chi^\mathrm{(3)}(k,A_\mathrm{f})\cos{(bA-bA_\mathrm{f})}+\chi^\mathrm{(3)}(k,A_\mathrm{f})\sin{(bA-bA_\mathrm{f})}\nonumber\\
    &+&A_\mathrm{f}\frac{d\chi^\mathrm{(3)}(k,A_\mathrm{f})}{dA}\sin{(bA-bA_\mathrm{f})}\Bigg|^2
\end{eqnarray}
and
\begin{eqnarray}\label{chi'_0}
    \left|\frac{d\chi^\mathrm{RD}(k,A)}{dA} \right|^2 = \frac{1}{b^2A^4}\Bigg|&\Big(&\chi^\mathrm{(3)}(k,A_\mathrm{f})+A_\mathrm{f}\frac{d\chi^\mathrm{(3)}(k,A_\mathrm{f})}{dA}\Big)\Big[\sin{(bA-bA_\mathrm{f})}-bA\cos{(bA-bA_\mathrm{f})}\Big]\nonumber\\
    &+&b^2AA_\mathrm{f}\, \chi^\mathrm{(3)}(k,A_\mathrm{f})\sin{(bA-bA_\mathrm{f})}\Bigg|^2 \ ,
\end{eqnarray}
respectively.  Thus as a whole, Eq.~(\ref{rd-8}) represents the dimensionless energy density of GWs at present epoch, symbolized by $\Omega_\mathrm{GW}^\mathrm{(0)}(k)$, where the respective quantities are determined above. Below we will determine the amplitude and the correspond spectral tilt (with respect to the wave number) of $\Omega_\mathrm{GW}^\mathrm{(0)}(k)$ for the modes $k<k_\mathrm{f}$. For this purpose, we need to understand that the explicit $k$ dependency on $\Omega_\mathrm{GW}^\mathrm{(0)}(k)$ comes through $\chi^{(3)}(k,A)$ (and its derivative as well) which follows Eq.~(\ref{3I-2}). Here it may be noted that the transfer function during the 3rd stage of inflation depends on the ratio $\frac{k}{k_\mathrm{1}}$ as well as on $\frac{k}{k_\mathrm{2}}$. Based on this fact, we will individually determine $\Omega_\mathrm{GW}^\mathrm{(0)}(k)$ for the following three cases: (I) $k<k_\mathrm{1}$, (II) $k_\mathrm{1} < k < k_\mathrm{2}$ and (III) $k_\mathrm{2} < k < k_\mathrm{f}$ respectively.

\subsection*{$\boxed{\text{Case (I): $k < k_\mathrm{1}$}}$}

The modes $k<k_\mathrm{1}$ exit the horizon during the 1st inflationary stage. Due to the condition $k < k_\mathrm{1}$, one may use the asymptotic expression for the Bessel function
\begin{equation}
    \lim_{x\ll 1} J_\nu(x) = \frac{x^\nu}{2^\nu\Gamma(\nu+1)}\ ,
\end{equation}
for all the Bessel functions containing $k/k_\mathrm{1}$ or $k/k_\mathrm{2}$ as arguments. The computation of the transfer function in this case is performed in Appendix \ref{sec-appendix-1}. For these modes, Eq.~(\ref{bc-1st inflation}) gives,
\begin{eqnarray}
 \chi^\mathrm{(1)}(k,A=1) = 1~~~~~~~~~\mathrm{and}~~~~~~~~\frac{d\chi^\mathrm{(1)}(k,1)}{dA} = 0 \, .
 \label{R1-1}
\end{eqnarray}
Using these, along with the asymptotic form of the Bessel functions, we get (see Eq.~\eqref{A1-11})
\begin{eqnarray}
 \chi^\mathrm{(2)}(k,A_\mathrm{2})&=&1 - \frac{1}{2(\nu+\nu^2)(1+3\omega)^2}\left(\frac{k^2}{k_\mathrm{1}^2}\right) \, ,\nonumber\\
 A_\mathrm{2}\, \frac{d\chi^\mathrm{(2)}(k,A_\mathrm{2})}{dA}&=&-\frac{1}{(5+3\omega)}\left(\frac{k^2}{k_\mathrm{1}^2}\right)\left[2\left(\frac{k_\mathrm{1}}{k_\mathrm{2}}\right)^2 - \left(\frac{k_\mathrm{1}}{k_\mathrm{2}}\right)^{2\nu}\right] \, .
 \label{R1-2}
\end{eqnarray}
The above expressions finally leads to (see Eq.~\eqref{A1-12})
\begin{eqnarray}
 \chi^\mathrm{(3)}(k,A_\mathrm{f})&=&1 + \mathcal{O}\left(\frac{k^2}{k_\mathrm{2}^2}\right) \, ,\nonumber\\
 A_\mathrm{f}\, \frac{d\chi^\mathrm{(3)}(k,A_\mathrm{f})}{dA}&=&-\frac{k_\mathrm{2}^2}{k_\mathrm{f}^2}\times\mathcal{O}\left(\frac{k^2}{k_\mathrm{2}^2}\right) \, .
 \label{R1-3}
\end{eqnarray}
In the limit of $\frac{k}{k_\mathrm{1}}<1$, the transfer function carries the same value (which is $\approx 1$, at leading order) from the end of the 1st inflation to the end of the 3rd stage of inflation. This is however expected as the modes $k<k_\mathrm{1}$ remain in the super-Hubble regime from $A=1$ to $A=A_\mathrm{f}$, where the tensor perturbation gets frozen. Plugging the expression from Eq.~(\ref{R1-3}) into the rhs of Eq.~(\ref{chi-0}), we get
\begin{eqnarray}
 \left|\chi^\mathrm{RD}(k,A)\right|^2 = \frac{1}{b^2A^2}&\Bigg\{&\left(\frac{k}{k_\mathrm{f}}\cos{(bA-bA_\mathrm{f})}+\sin{(bA-bA_\mathrm{f})}\right)\left(1 + \mathcal{O}\left(\frac{k^2}{k_\mathrm{2}^2}\right)\right)\nonumber\\
 &-&\frac{k_\mathrm{2}^2}{k_\mathrm{f}^2}\times\mathcal{O}\left(\frac{k^2}{k_\mathrm{2}^2}\right)\sin{(bA-bA_\mathrm{f})}\Bigg\}^2 \approx \frac{1}{b^2A^2}\sin^2{(bA)}~~.
    \label{CI-1}
\end{eqnarray}
where we keep the leading order term(s) (as $\frac{k}{k_\mathrm{2}}<1$ or $\frac{k}{k_\mathrm{f}}<1$) and use $bA_\mathrm{f} = k/k_\mathrm{f}$. Moreover we obtain,
\begin{eqnarray}
 \left|\frac{d\chi^\mathrm{RD}(k,A)}{dA} \right|^2&=&\frac{1}{A^2}\Bigg\{\left(1 + \mathcal{O}\left(\frac{k^2}{k_\mathrm{2}^2}\right) - \frac{k_\mathrm{2}^2}{k_\mathrm{f}^2}\, \mathcal{O}\left(\frac{k^2}{k_\mathrm{2}^2}\right)\right)\left[\frac{1}{bA}\sin{(bA-bA_\mathrm{f})}-\cos{(bA-bA_\mathrm{f})}\right]\nonumber\\
 &+&\frac{k}{k_\mathrm{f}}\left(1 + \mathcal{O}\left(\frac{k^2}{k_\mathrm{2}^2}\right)\right)\sin{(bA-bA_\mathrm{f})}\Bigg\}^2 \approx \frac{1}{A^2}\cos^2{(bA)}~~.
 \label{CI-2}
\end{eqnarray}
In arriving at the final result of Eq.~(\ref{CI-1}), we have used $bA \gg 1$ (as we are evaluating at late times during radiation domination) The above two expressions, due to Eq.~(\ref{rd-8}), immediately provide the dimensionless energy density of GWs today over the modes $k<k_\mathrm{1}$ as,
\begin{equation}\label{CI-3}
    \boxed{\ \Omega_\mathrm{GW}^\mathrm{(0)}(k)h^2 \simeq \left(\frac{1}{6\pi^2}\right)\Omega_\mathrm{R}h^2\left(\frac{H_\mathrm{1}}{M_\mathrm{Pl}}\right)^2 \ , \ }
\end{equation}
where we have assumed that $g_{r,0} = g_{r,eq}$. Therefore the GWs spectrum today for the modes $k<k_\mathrm{1}$ seems to be scale invariant and having the amplitude shown in Eq.~(\ref{CI-3}).

\subsection*{$\boxed{\text{Case (II): $k_\mathrm{1} < k < k_\mathrm{2}$}}$}

The modes $k_\mathrm{1} < k < k_\mathrm{2}$ exit the horizon during the 2nd inflationary stage. Therefore the Bessel function containing $k/k_\mathrm{1}$ has the following asymptotic form:
\begin{equation}
    \lim_{x\gg 1 }J_\nu(x) = \sqrt{\frac{2}{\pi x}}\cos\left(x-\frac{\pi}{2}\left(\nu+\frac{1}{2}\right)\right) \ . 
\end{equation}
while the Bessel function with $k/k_\mathrm{2}$ as the argument can be expressed by,
\begin{equation}
    \lim_{x\ll 1} J_\nu(x) = \frac{x^\nu}{2^\nu\Gamma(\nu+1)}\ .
\end{equation}
The computations in this case are reported in Appendix \ref{sec-appendix-2}.
For these modes, Eq.~(\ref{bc-1st inflation}) gives,
\begin{eqnarray}
 \chi^\mathrm{(1)}(k,A=1) = -i\left(\frac{k}{k_\mathrm{1}}\right)e^{ik/k_\mathrm{1}}~~~~~\mathrm{and}~~~~\frac{d\chi^\mathrm{(1)}(k,1)}{dA} = -\left(\frac{k^2}{k_\mathrm{1}^2}\right)e^{ik/k_\mathrm{1}} \, .
 \label{R2-1}
\end{eqnarray}
Using these, along with the asymptotic form of the Bessel functions, we get (see Eq. \eqref{A2-4})
\begin{eqnarray}
 \chi^\mathrm{(2)}(k,A_\mathrm{2})&=&\frac{\Gamma(\nu)(1+3\omega)^{\nu-\frac{1}{2}}}{\sqrt{\pi}}\left(\frac{k}{k_\mathrm{1}}\right)^{\frac{3}{2}-\nu}\mathrm{exp}\left[i\left(\frac{k}{k_\mathrm{1}} + x - \frac{\pi\nu}{2} - \frac{\pi}{4}\right)\right] \, ,\nonumber\\
 \frac{d\chi^\mathrm{(2)}(k,A_\mathrm{2})}{dA}&\sim&\mathrm{e}^{i\frac{k}{k_\mathrm{1}}}\left[\left(\frac{k}{k_\mathrm{2}}\right)^{\nu+\frac{3}{2}}\frac{\mathrm{e}^{i(x+\pi\nu/2)}}{(1+3\omega)^{\nu}\Gamma(\nu)} + \left(\frac{k}{k_\mathrm{2}}\right)^{-\nu+\frac{7}{2}}\frac{\mathrm{e}^{i(x-\pi\nu/2)}}{(1+3\omega)^{2-\nu}\Gamma(2-\nu)}\right] \, ,
 \label{R2-2}
\end{eqnarray}
where, due to $\nu > 0$, the derivative of the transfer function at $A=A_\mathrm{2}$ is less than unity, i.e. $\frac{d\chi^\mathrm{(2)}(k,A_\mathrm{2})}{dA} \ll 1$ which is expected as the modes $k_\mathrm{1} < k < k_\mathrm{2}$ lie in the super-Hubble regime at the end of the 2nd inflationary stage.
Using the above expressions (leading order terms) into Eq.~(\ref{3I-2}), we get (see Eq. \eqref{A2-5})
\begin{eqnarray}
 \chi^\mathrm{(3)}(k,A_\mathrm{f})&=&\frac{\Gamma(\nu)(1+3\omega)^{\nu-\frac{1}{2}}}{\sqrt{\pi}}\left(\frac{k}{k_\mathrm{1}}\right)^{\frac{3}{2}-\nu}\mathrm{exp}\left[i\left(\frac{k}{k_\mathrm{1}} + x - \frac{\pi\nu}{2} - \frac{\pi}{4}\right)\right]\left[1 + \mathcal{O}\left(\frac{k^2}{k_\mathrm{2}^2}\right)\right] \, ,\nonumber\\
 A_\mathrm{f}\, \frac{d\chi^\mathrm{(3)}(k,A_\mathrm{f})}{dA}&=&-\chi^\mathrm{(2)}(k,A_\mathrm{2})~\frac{k_\mathrm{2}^2}{k_\mathrm{f}^2}\times\mathcal{O}\left(\frac{k^2}{k_\mathrm{2}^2}\right) \, .
 \label{R2-3}
\end{eqnarray}
where $\chi^\mathrm{(2)}(k,A_\mathrm{2})$ is shown above in Eq.~(\ref{R2-2}). Here it deserves mentioning that the transfer function, for $k_\mathrm{1} < k < k_\mathrm{2}$, remains same (at the leading order) from the end of the 2nd inflationary stage to the end of the 3rd inflation --- this is due to the fact that such perturbation modes lie in the super-Hubble regime from $A=A_\mathrm{2}$ to $A=A_\mathrm{f}$. Plugging the expression of Eq.~(\ref{R2-3}) into Eq.~(\ref{chi-0}), we get
\begin{eqnarray}
 \left|\chi^\mathrm{RD}(k,A)\right|^2 = \frac{\left|\Gamma(\nu)(1+3\omega)^{\nu-\frac{1}{2}}\right|^2}{\pi b^2A^2}\left(\frac{k}{k_\mathrm{1}}\right)^{3-2\nu}\sin^2{(bA)}~~.
    \label{CII-1}
\end{eqnarray}
where we keep the leading order term(s) in the second equality. Moreover we obtain,
\begin{eqnarray}
 \left|\frac{d\chi^\mathrm{RD}(k,A)}{dA} \right|^2= \frac{\left|\Gamma(\nu)(1+3\omega)^{\nu-\frac{1}{2}}\right|^2}{\pi A^2}\left(\frac{k}{k_\mathrm{1}}\right)^{3-2\nu}\cos^2{(bA)}~~.
 \label{CII-2}
\end{eqnarray}
The above two expressions, due to Eq.~(\ref{rd-8}), immediately provide the dimensionless energy density of GWs today over the modes $k<k_\mathrm{1}$ as,
\begin{equation}\label{CII-3}
    \boxed{\ \Omega_\mathrm{GW}^\mathrm{(0)}(k)h^2 \simeq \left(\frac{1}{6\pi^2}\right)\Omega_\mathrm{R}h^2\left(\frac{H_\mathrm{1}}{M_\mathrm{Pl}}\right)^2\frac{\left|\Gamma(\nu)(1+3\omega)^{\nu-\frac{1}{2}}\right|^2}{\pi}\left(\frac{k}{k_\mathrm{1}}\right)^{3-2\nu} \ , \ }
\end{equation}
where we have assumed that $g_{r,0} = g_{r,eq}$. Therefore the GWs spectrum today for the modes $k_\mathrm{1} < k < k_\mathrm{2}$ seems to be scale dependent and having the amplitude shown in Eq.~(\ref{CII-3}). In particular, the spectral tilt is given by: $n_\mathrm{GW} = 3-2\nu$ which, due to Eq.~(\ref{2I-5a}), can be expressed as,
\begin{eqnarray}
 n_\mathrm{GW} = \frac{6(1+\omega)}{(1+3\omega)}\, ,
 \label{tilt-1}
\end{eqnarray}
recall that $\omega = -1-2|\epsilon_\mathrm{m}|/3$ is the effective EoS during the 2nd inflation stage. Eq.~(\ref{tilt-1}) indicates that the negative value of the slow roll parameter during the 2nd inflation (i.e.$\epsilon_\mathrm{m} < 0$, see Eq.~(\ref{Nano-1}), or Fig.~[\ref{plot-SRP-2}]) causes a blue tilted GWs' spectrum today ($n_\mathrm{GW}> 0$) over the modes that exit the horizon during the same.

\subsection*{$\boxed{\text{Case (III): $k_\mathrm{2} < k < k_\mathrm{f}$}}$}

The modes $k_\mathrm{2} < k < k_\mathrm{f}$ exit the horizon during the 3rd inflationary stage where $H = H_\mathrm{f}$. Therefore the Bessel function containing either $k/k_\mathrm{1}$ or $k/k_\mathrm{2}$ has the following asymptotic form:
\begin{equation}
    \lim_{x\gg 1 }J_\nu(x) = \sqrt{\frac{2}{\pi x}}\cos\left(x-\frac{\pi}{2}\left(\nu+\frac{1}{2}\right)\right) \ .
\end{equation}
For these modes, the transfer function at the end of 1st inflation takes the form as shown in Eq.~(\ref{bc-1st inflation}). Using these, along with the asymptotic form of the Bessel functions, we get (see the appendix)
\begin{eqnarray}
 \chi^\mathrm{(2)}(k,A_\mathrm{2})&=&\left(\frac{k_\mathrm{2}}{k_\mathrm{1}}\right)^{\frac{1}{2}-\nu}\left(\frac{k}{k_\mathrm{1}}\right)\mathrm{exp}\left[i\left(\frac{k}{k_\mathrm{1}} + x - y - \frac{\pi}{2}\right)\right] \, ,\nonumber\\
 A_\mathrm{2}\frac{d\chi^\mathrm{(2)}(k,A_\mathrm{2})}{dA}&=&-\left(\frac{k_\mathrm{1}}{k_\mathrm{2}}\right)^{\frac{1}{2}+\nu}\left(\frac{k}{k_\mathrm{1}}\right)^2\mathrm{exp}\left[i\left(\frac{k}{k_\mathrm{1}} + x - y\right)\right] \, ,
 \label{R3-1}
\end{eqnarray}
Using the above expressions into Eq.~(\ref{3I-2}), we get
\begin{eqnarray}
 \chi^\mathrm{(3)}(k,A_\mathrm{f})=\left(\frac{k_\mathrm{2}}{k_\mathrm{1}}\right)^{\frac{3}{2}-\nu}\mathrm{exp}\left[i\left(\frac{k}{k_\mathrm{1}} - \frac{k}{k_\mathrm{2}} + x - y\right)\right]\, .
 \label{R3-2}
\end{eqnarray}
Plugging the expression of Eq.~(\ref{R3-2}) into Eq.~(\ref{chi-0}), we get
\begin{eqnarray}
 \left|\chi^\mathrm{RD}(k,A)\right|^2 = \frac{1}{b^2A^2}\left(\frac{k_\mathrm{2}}{k_\mathrm{1}}\right)^{3-2\nu}\sin^2{(bA)}~~.
    \label{CIII-1}
\end{eqnarray}
where we keep the leading order term(s) in the second equality. Moreover we obtain,
\begin{eqnarray}
 \left|\frac{d\chi^\mathrm{RD}(k,A)}{dA} \right|^2= \frac{1}{A^2}\left(\frac{k_\mathrm{2}}{k_\mathrm{1}}\right)^{3-2\nu}\cos^2{(bA)}~~.
 \label{CIII-2}
\end{eqnarray}
The above two expressions, due to Eq.~(\ref{rd-8}), immediately provide the dimensionless energy density of GWs today over the modes $k_\mathrm{2}< k < k_\mathrm{f}$ as,
\begin{eqnarray}
 \Omega_\mathrm{GW}^\mathrm{(0)}(k)h^2 \simeq \left(\frac{1}{6\pi^2}\right)\Omega_\mathrm{R}h^2\left(\frac{H_\mathrm{1}}{M_\mathrm{Pl}}\right)^2\left(\frac{k_\mathrm{2}}{k_\mathrm{1}}\right)^{3-2\nu} \, .
 \label{CIII-3}
\end{eqnarray}
By utilizing $k_\mathrm{1} = a_\mathrm{1}H_\mathrm{1}$ and $k_\mathrm{2} = a_\mathrm{2}H_\mathrm{f}$ into Eq.~(\ref{continuity-Hubble}), we get $H_\mathrm{f} = H_\mathrm{1}\left(\frac{k_\mathrm{2}}{k_\mathrm{1}}\right)^{\frac{3}{2}-\nu}$, owing to which, the above expression finally leads to,
\begin{equation}\label{CIII-4}
    \boxed{\ \Omega_\mathrm{GW}^\mathrm{(0)}(k)h^2 \simeq \left(\frac{1}{6\pi^2}\right)\Omega_\mathrm{R}h^2\left(\frac{H_\mathrm{f}}{M_\mathrm{Pl}}\right)^2  \ } \ .
\end{equation}
Therefore the GWs spectrum today for the modes $k_\mathrm{2} < k < k_\mathrm{f}$ proves to be scale invariant and having the amplitude shown in Eq.~(\ref{CIII-4}). In particular, the amplitude is fixed by the Hubble scale of the 3rd stage of the inflation. Due to $H_\mathrm{f} > H_\mathrm{1}$, the comparison of Eqs.~(\ref{CI-3}) and (\ref{CIII-4}) show that the GWs spectrum over the modes $k_\mathrm{2} < k < k_\mathrm{f}$ has larger amplitude compared to that of over the modes $k < k_\mathrm{1}$. Regarding the spectral tilt, we may argue that the GWs spectrum today is flat for the modes that exit the horizon during the 1st and the 3rd stage of inflation, while the spectrum has a tilt over the modes exiting the horizon during the 2nd stage of the inflation and the amount of the tilt is fixed by the corresponding EoS parameter ($\omega$). In particular,
\begin{align}
\mathrm{Tilt~of~the~GWs~spectra} =
\begin{cases}
  0 \hspace{3.25cm} \text{for} \hspace{0.5cm} k < k_\mathrm{1} & \\
    n_\mathrm{GW} = \frac{6(1+\omega)}{(1+3\omega)}>0 \hspace{1.cm} \text{for} \hspace{0.5cm} k_\mathrm{1}<k<k_\mathrm{2} & \\
    0 \hspace{3.25cm} \text{for} \hspace{0.5cm} k_\mathrm{2} < k < k_\mathrm{f}~.
\end{cases}
\label{tilt-2}
\end{align}

\section{Pulsar Timing Arrays and the NANOGrav 15-year Detection}

The North American Nanohertz Observatory for Gravitational Waves (NANOGrav) collaboration has released its 15-year dataset, reporting compelling evidence for a low-frequency stochastic gravitational-wave background (SGWB) \cite{Agazie_2023}. The experiment exploits a \emph{pulsar timing array} (PTA): a network of millisecond pulsars whose remarkably stable radio pulses serve as galactic-scale clocks. The modeling of such pulsar is highly precise \cite{Zhu:2018etc} and their isotropic distribution in our galaxy and the relatively small period (around $1 \sim 10$ ms), make them perfect candidates for these detections. Gravitational waves passing between Earth and these pulsars induce characteristic fluctuations in the arrival times of the pulses, on the order of tens to hundreds of nanoseconds. By monitoring 68 millisecond pulsars over more than a decade, the PTA searched for the correlated timing deviations expected from a background of gravitational waves.
The detection hinges on the \emph{Hellings--Downs correlation function} $\Gamma(\zeta)$, which predicts the angular correlation of timing residuals between pairs of pulsars separated by angle $\zeta$ \cite{Hellings:1983fr}.
NANOGrav reports a strain spectrum consistent with a power-law of the form
\begin{equation}
h_c(f) = A_{\mathrm{GWB}} \left( \frac{f}{f_{\mathrm{yr}}} \right)^{-2/3},
\end{equation}
with amplitude 
\begin{equation}
A_{\mathrm{GWB}} \simeq 2.4^{+0.7}_{-0.6} \times 10^{-15}
\end{equation}
at the reference frequency $f_{\mathrm{yr}} = 1\,\mathrm{yr}^{-1}$. The spectral index is consistent with the inspiral of supermassive black hole binaries, but alternative cosmological mechanisms such as cosmic strings, first-order phase transitions, or relics from inflation remain possible and are actively constrained. The NANOGrav 15-year results therefore provide the first strong evidence of a nanohertz gravitational-wave background, opening a new observational window into both astrophysical populations and early-universe physics. The NANOGrav Collaboration explored models including inflationary relics, scalar-induced gravitational waves, first-order phase transitions, cosmic strings, and domain walls \cite{Afzal_2023}. Most models (except stable cosmic strings) can, in principle, replicate the observed spectrum, with Bayes factors ranging from $\mathcal{O}(10)$ to $\mathcal{O}(10^2)$ compared to the SMBHB model, but results remain model-sensitive and not conclusive. Several other cosmological models have been confronted with the NANOGrav 15-year (NG15), among which we have:
\begin{itemize}
\item \textit{String Cosmology Constraints:}
In \cite{Tan_2025} the authors constrain string cosmology parameters using NANOGrav 15-year. However, the current dataset lacks sensitivity to the post-string-phase Hubble parameter $H_r$, and Bayesian evidence modestly favors the string-cosmology scenario over a generic power-law, with a Bayes factor $\simeq 2.2$.
\item\textit{Pre-Big-Bang String Scenario:}
In contrast, the Pre-Big-Bang (PBB) scenario of string cosmology is strongly disfavored. As shown in \cite{tan2025prebigbangcosmologyexplainnanograv}, the key parameter
\begin{equation}
\beta_\text{PBB} = -0.12^{+0.06}_{-0.21} \ ,
\end{equation}
which lies outside its theoretically acceptable range ($0 \leq \beta_\text{PBB} < 3$) by more than $5\sigma$. A simple power-law spectrum is decisively favored, with a Bayes factor of $\sim 468$.
\item\textit{Massive Gravity and Early-Universe Models:}
Models invoking massive gravity, e.g. tensor perturbations from a time-varying graviton mass, demonstrate that certain parameter choices can reproduce the NG15 signal within $1\text{--}3\sigma$, though suppression mechanisms for high-frequency modes (to satisfy BBN constraints) are required \cite{Kenjale_2025}.
\item\textit{Axion-like Particle (ALP) Instabilities:}
The tachyonic instabilities induced by ALPs have been explored in \cite{geller2024challengesinterpretingnanograv15year}, resulting in dark gauge field production and associated gravitational-wave signals. They conclude that explaining the NG15 amplitude via this mechanism would grossly violate $\Delta N_{\rm eff}$ and dark matter relic-density constraints, rendering it highly unlikely to contribute significantly to the observed signal.
\end{itemize}
\section{Constraining with the NANOGrav 15-year data}\label{sec-numerics}
\begin{figure}[htb]
    \centering
    \begin{minipage}[b]{0.3\textwidth}
        \centering
    \includegraphics[width=1.\linewidth]{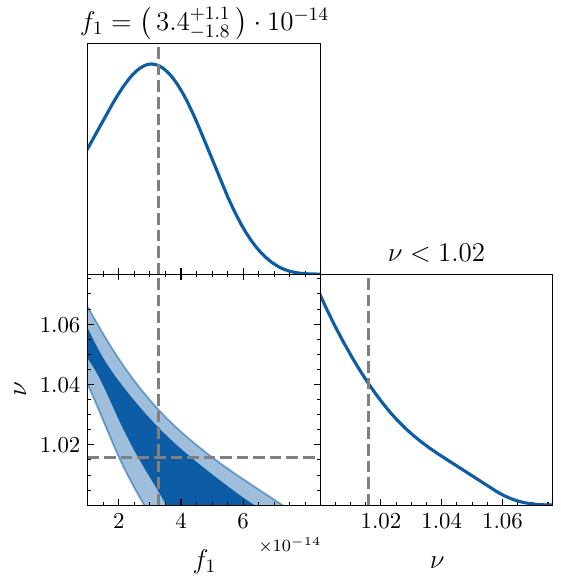}
    \end{minipage}%
    \hfill
    \begin{minipage}[b]{0.65\textwidth}
        \centering
    \includegraphics[width=1.\linewidth]{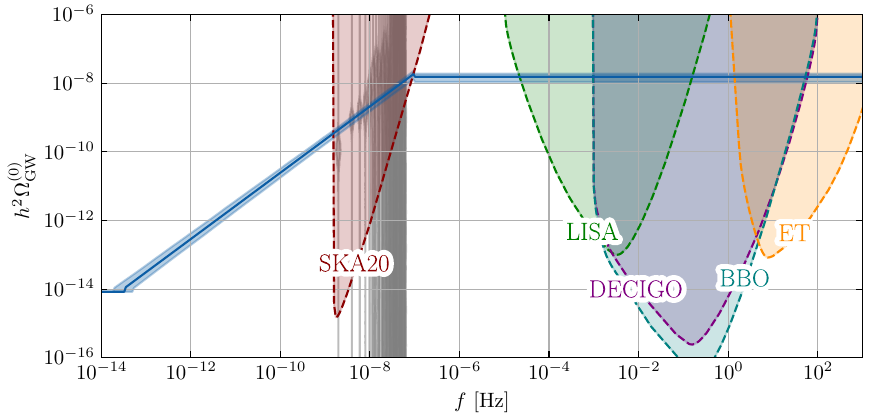}
    \end{minipage}%
    \par
    \begin{minipage}[t]{0.3\textwidth}
        \centering
    \caption{Corner plot for the entropic model sampling over the inflationary parameters $f_\mathrm{1}$ and $\nu$. The frequency $f_\mathrm{1}$ is given in Hz.}
    \label{fig:corner_plot}
    \end{minipage}%
    \hfill
    \begin{minipage}[t]{0.65\textwidth}
        \centering
    \caption{Plot of the best fit curve with its 1$\sigma$ region with the observational data represented by the gray violins. These represent the PDFs of the signal for each frequency bin. In the plot are also reported the sensibility curves for future GW observatories. We take $N_\mathrm{f} = 60$, $\beta = 0.25$, $\sigma_0 = 0.024$ and $f_\mathrm{2}= 10^{-7}$ Hz. The cosmological parameters have been fixed to $H_\mathrm{0} = 67.66 \text{ km s}^{-1}\text{Mpc}^{-1}$ and $\Omega_\mathrm{R} = 9.182\times 10 ^{-5}$ \cite{Planck2020}.}
    \label{fig:signal_sens}
    \end{minipage}%
\end{figure}

\begin{figure}[htb]
    \centering
    \includegraphics[width=0.8\linewidth]{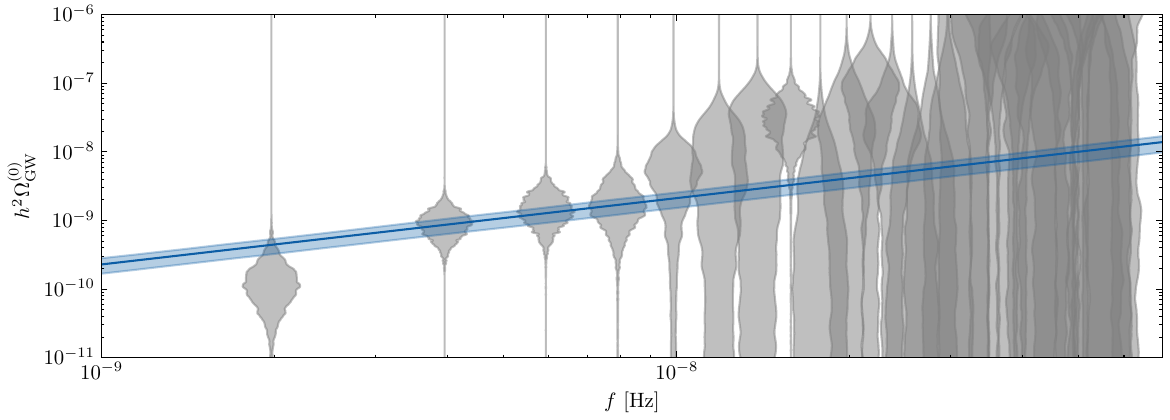}
    \caption{Plot of the best fit curve with its 1$\sigma$ region with the observational data represented by the gray violins. These represent the PDFs of the signal for each frequency bin. We fix $N_\mathrm{f} = 60$, $\beta = 0.25$, $\sigma_0 = 0.024$ and $f_\mathrm{2}= 10^{-7}$ Hz. The cosmological parameters have been fixed to $H_\mathrm{0} = 67.66 \text{ km s}^{-1}\text{Mpc}^{-1}$ and $\Omega_\mathrm{R} = 9.182\times 10 ^{-5}$ \cite{Planck2020}.}
    \label{fig:signal_violins}
\end{figure}

In this section we will constrain the parameters of our model using the detection of a background signal from the NANOGrav 15-year release. We will show the possibility of the entropic and inflationary parameters to be in agreement with both the ACT constraints of Sec.~[\ref{sec:ACT-Inflation}] and NANOGrav ones. \\
The NANOGrav 15-year signal \cite{thenanogravcollaboration202310344086} is  given as the probability densities from Kernel Density Estimators (KDEs) of free spectrum analyses of the dataset. In practice, we will have a density distribution for the signal for 30 frequency bins. The dataset is given as the time residual power distribution per bin $\rho(f)$. This is linked to the spectral density of the time residuals $S_r(f)$ as \cite{Lamb_2023}
\begin{equation}
    \rho(f) = \frac{S_r(f)}{T} \ ,
\end{equation}
where $T$ is the data time span ($T=(\text{width of the bin in Hz})^{-1}$). 
The $\rho(f)$ can then be linked to the characteristic strain $h_c(f)$ as 
\begin{equation}
    h_c(f)^2 = 12\pi^2f^3T\rho^2(f) \ .
\end{equation}
We can obtain the GW power spectrum as 
\begin{equation}
    \Omega_\mathrm{GW}^\mathrm{(0)} h^2 = \frac{2\pi^2}{3H_\mathrm{0}^2}f^2 h_c^2(f) = \frac{8\pi^4}{H_\mathrm{0}^2}f^5 T \rho(f)^2 \ .
\end{equation}
To have an easier comparison with the GWs observations, we will refer to frequencies instead of wave vector $k$ as we did in the previous sections. The frequency $f$ (in the unit of Hz.) is related to the wave vector $k$ (in $\text{Mpc}^{-1}$) by the relation
\begin{equation}\label{k_to_f}
    f = 1.55 \times 10^{-15}\left(\frac{k}{1 \, \text{Mpc}^{-1}}\right) \, \text{Hz} \ ,
\end{equation}
which can be used in converting $k_\mathrm{1,2}$ to $f_\mathrm{1,2}$, i.e. from the unit of $\text{Mpc}^{-1}$ to the unit of Hz.\\
At this point, we find the GWs signal to be a set of PDFs for each frequency bin. As done in the NANOGrav paper, we will perform a MCMC on the model's parameters, taking the likelihood to be given by these PDFs. As explained in the previous sections, the model consists of two entropic parameters $(\beta,\sigma_0)$ and four inflationary parameters $(N_\mathrm{f},f_\mathrm{1},f_\mathrm{2},\nu)$, and the corresponding GWs spectrum (today) is a piecewise continuous function defined in three frequency (equivalently wave vector) intervals. The analytical expression of the GWs spectrum is given by a flat initial ground base given by Eq.~(\ref{CI-3}) for $f<f_\mathrm{1}$, then a tilted portion of the spectrum as depicted in Eq.~(\ref{CII-3}) for $f_\mathrm{1}\leq f<f_\mathrm{2}$ and the final flat tail as Eq.~(\ref{CIII-4}) when $f\geq f_\mathrm{2}$. We will impose the ACT-DR6+Planck18+BAO data by using the values of the parameters in Table.~[\ref{Table-1}]. Performing the MCMC, we noted that the sampling over the entropic parameters was relevant in the sense that the maximum likelihood values were on the positive boundary of each validity region. For this reason, we decide to fix the entropic parameters, as well as the $N_\mathrm{f}$, to be $N_\mathrm{f} = 60$, $\beta = 0.25$ and $\sigma_0 = 0.024$ (that are well compatible with the ACT-DR6+Planck18+BAO constraints, see Sec.~[\ref{sec:ACT-Inflation}]). Regarding $f_\mathrm{2}$, recall that the $f_\mathrm{2}$ represents the mode that exits the Hubble radius at the end of the 2nd inflationary stage, and thus it locates the shift from a tilted nature of the GWs signal to a flat one at the junction of the 2nd inflation-to-3rd inflation phase. Now in order to be compatible with the NANOGrav 15-year data, we need to fix $f_\mathrm{1}$ and $f_\mathrm{2}$ in such a way, that the range of modes $f = [f_\mathrm{1},f_\mathrm{2}]$ fall around the NANOGrav frequency range. Therefore the only constraint on $f_\mathrm{2}$ is to be outside of the NANOGrav signal frequency domain, i.e. $f_\mathrm{2}$ needs to be: $f_\mathrm{2} \geq 10^{-7}$ Hz. Here, in particular, we consider $f_\mathrm{2}= 10^{-7}$ Hz. With such above considerations, we present the following plots. As can be seen from Fig.~[\ref{fig:signal_sens}] that the same range of entropic parameters, which are compatible with the ACT-DR6+Planck18+BAO data, also satisfy the NANOGrav 15-year data. This shows the simultaneous compatibility of the present model with the ACT as well as with the PTA data. Fig.~[\ref{fig:signal_sens}] also demonstrates that the GWs spectrum at higher frequencies will cross future GW observatories, this is due to the fact that GWs spectrum over the lower scale modes (i.e. for $f > f_\mathrm{2}$) is larger compared to that of for the larger scale modes (particularly for $f<f_\mathrm{1}$). At this point, we are left with the relevant parameters to sample on, which are the beginning of the tilt $f_1$, and $\nu$ which gives the inclination of it. We take the priors on these parameters to be $f_1\in[10^{-14},10^{-10}]$ in Hz and $\nu \in [1, 1.5]$. Furthermore, regarding $H_\mathrm{0}$ and $\Omega_\mathrm{R}$, the respective Planck values are considered, in particular: $H_\mathrm{0} = 67.66 \text{ km s}^{-1}\text{Mpc}^{-1}$ and $\Omega_\mathrm{R} = 9.182\times 10 ^{-5}$ \cite{Planck2020}. The results of such analysis are reported in Fig.~[\ref{fig:corner_plot}]. As evident that we have a tight constraint on $\nu$ as expected, since the value $\nu = 1$ is equivalent to $\epsilon_\mathrm{m}=-1$. We also have a gaussian fit on $k_\mathrm{1}$. From Fig.~[\ref{fig:signal_violins}] we see the tendency of the model to accommodate the frequencies in the middle, probably since the tilt is bounded by the prior on $\nu \geq 1$.


With the best fitted values, we can determine the duration of the 2nd phase of the inflation: by using $k_\mathrm{1} = a_\mathrm{1}H_\mathrm{1}$ and $k_\mathrm{2}=a_\mathrm{2}H_\mathrm{f}$, we obtain
\begin{eqnarray}
 \mathrm{Duration~of~the~2nd~inflationary~phase} = N_\mathrm{2} - N_\mathrm{1} = \left(\frac{2\nu - 1}{2}\right)\ln{\left[\frac{k_\mathrm{2}}{k_\mathrm{1}}\right]} = \left(\frac{2\nu - 1}{2}\right)\ln{\left[\frac{f_\mathrm{2}}{f_\mathrm{1}}\right]} \, ,
 \label{N1}
\end{eqnarray}
where we use Eq.~(\ref{2I-5a}). With $f_\mathrm{1} = 3.4\times10^{-14} \mathrm{Hz}$, $f_\mathrm{2} = 10^{-7} \mathrm{Hz}$ and $\nu \approx 1$, the above expression gives,
\begin{eqnarray}
 N_\mathrm{2} - N_\mathrm{1} \approx 8 \, ,
 \label{N2}
\end{eqnarray}
i.e., the 2nd inflationary phase lasts around $\sim 8$ e-fold. Moreover the 1st inflation lasts for $N_\mathrm{1} = \ln{\left[f_\mathrm{1}/f_\mathrm{CMB}\right]}$, with the consideration that the 1st inflation starts when the CMB scale exits the horizon. Taking $f_\mathrm{CMB} \approx 10^{-16} \mathrm{Hz}$ (or $k_\mathrm{CMB} = 0.05 \mathrm{Mpc}^{-1}$), we find: $N_\mathrm{1} \approx 6$. Thus as a whole, the duration of the individual three phases of the inflation are given by --- (i) 1st phase: $0 \leq N \leq N_\mathrm{1}$, (ii) 2nd phase: $N_\mathrm{1} \leq N \leq N_\mathrm{2}$, and (iii) 3rd phase: $N_\mathrm{2} \leq N \leq N_\mathrm{f}$, with $N_\mathrm{1} \approx 6$, $N_\mathrm{2} \approx 14$ and $N_\mathrm{f} = 60$.

\section{Conclusion}

Recent evidences of SGWB through PTA observations hint that the usual inflation is not the ultimate story of early universe in order to be compatible with the PTA data (as the usual inflation predicts a flat primordial tensor spectrum with less amplitude than the PTA constraint). Moreover the Atacama Cosmology Telescope, combined with the Planck-2018 and BAO, refines the inflationary observables (at a level of $2\sigma$), in comparison to the only Planck measurements. Based on these, here in the realm of entropic cosmology where the thermodynamics of apparent horizon is described by the 4-parameter generalized entropy, we show the simultaneous compatibility of the ACT data and the PTA data by incorporating certain modification during inflation over the usual inflationary scenario. The modification comes from the fact that the slow roll parameter becomes negative (with $|\epsilon_\mathrm{m}| < 1$) and violates the null energy condition during a short epoch of inflation when the modes, around the NANOGrav frequency range, cross the Hubble radius. As a whole, the entire duration of inflation is divided into three parts --- (i) during the initial phase of the inflation, the slow-roll parameter satisfies $0 < \epsilon_\mathrm{m} \ll 1$ (the usual case) and leads to an (almost) dS kind of inflation. This leads to a scale invariant primordial tensor power spectrum over the large scale modes (including the CMB scale), and the amplitude of the tensor perturbation is getting fixed by the corresponding Hubble parameter of the 1st phase of the inflation. The inflationary observables, like the spectral index of curvature perturbation and the tensor-to-scalar ratio corresponding to the CMB scale, are found to be compatible with the ACT-DR6+Planck-2018+BAO data for suitable choices of entropic parameters. (ii) During 2nd phase of the inflation (having e-fold duration of $6 \leq N \lesssim 14$), the slow roll parameter is negative with $|\epsilon_\mathrm{m}| < 1$, and moreover, the modes around the NANOGrav frequency range exit the Hubble radius during this short epoch. As a result, the primordial tensor power spectrum becomes blue tilted over the modes sensitive to the NANOGrav frequency range, and consequently, the GWs spectrum today passes the NANOGrav sensitivity curve. In order to confront the model with the observational data, we perform numerical analysis primarily focused on NANOGrav 15-year data. It turns out that the constraints on the entropic parameters, coming from the ACT-DR6+Planck-2018+BAO, also fit the NANOGrav 15-year data, which in turn shows the simultaneous compatibility of the model with the ACT and the PTA data. (iii) During 3rd stage of the inflation, the slow-roll parameter again satisfies $0 < \epsilon_\mathrm{m} \ll 1$ which leads to a scale invariant tensor power spectrum (over the lower scale modes) with the amplitude is getting fixed by the Hubble scale of the 3rd phase. Owing to violation of null energy condition during the 2nd phase of the inflation; the Hubble parameter of the 3rd inflationary stage acquires a larger value compared to that of during the 1st phase of the inflation. As a result, the GWs amplitude over the lower scale modes (in particular, that exit the Hubble radius during the 3rd phase of the inflation) is larger compared to that of for the large scale modes (particularly that exit the Hubble radius during the 1st phase of the inflation), see Fig.~[\ref{fig:signal_sens}]. As a consequence, beside the NANOGrav sensitivity curve, the GWs spectrum in the present context also crosses the sensitivity regions of other observatories like LISA, DECIGO, BBO etc.

Thus as a whole, the modified inflation in entropic cosmology with 4-parameter generalized entropy has the following advantages --- the 1st and the 2nd phases of the inflation are responsible for the compatibility of the model with the ACT-DR6+Planck-2018+BAO data and the PTA data respectively, for a common ranges of the entropic parameters; while the 3rd phase of the inflation is liable for the fact that the present GWs spectrum crosses the sensitivity regions of various other observatories like LISA, DECIGO, BBO etc. Therefore we may argue that if the future observatories can detect the signal of primordial GWs, then our theoretical expectation carried in the present work may provide a possible testbed for the generalized entropic cosmology.

\appendix
\section{Correspondence with non-canonical scalar-tensor theory}\label{sec-appendix-0}

\begin{figure}[!h]
\begin{center}
\centering
\includegraphics[width=3.0in,height=2.0in]{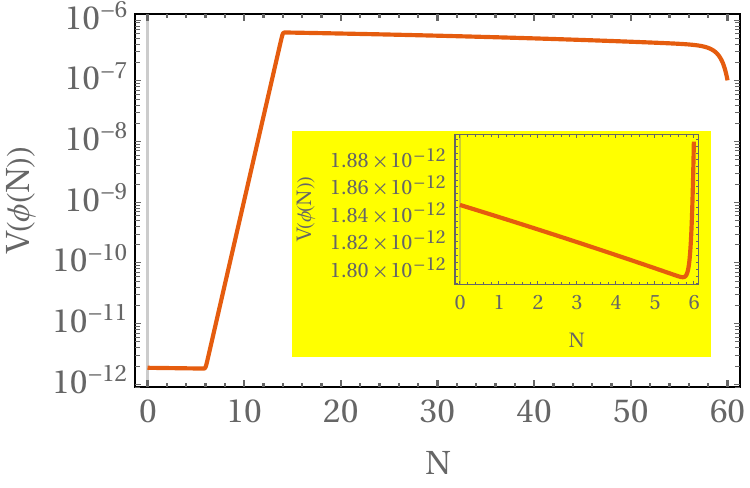}
\hskip0.5cm
\includegraphics[width=3.0in,height=2.0in]{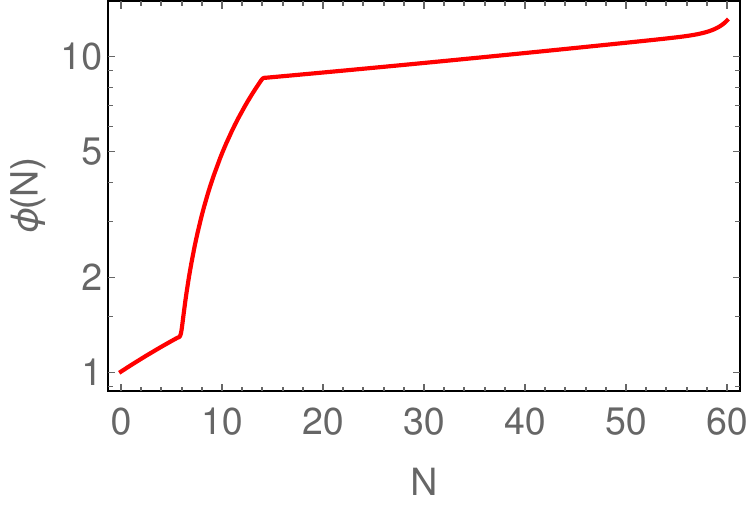}
\caption{{\color{blue}Left Plot}: $V(\phi(N))$ (in the unit of $M_\mathrm{Pl}^4$) vs. $N$; {\color{blue}Right Plot}: $\phi(N)$ (in the unit of $M_\mathrm{Pl}$) vs. $N$ for $N_\mathrm{f} = 60$. In both the plots, we consider $C=-0.4$, $D=10$, $\sigma_0 = 0.024$, $N_\mathrm{1}=6$ and $N_\mathrm{2}=14$ (which are compatible with the best fitted values of the parameters). The inset in the left plot is the zoomed-in version of scalar potential during the 1st inflationary stage (i.e. $N=[0,6]$).}
\label{plot-1-app}
\end{center}
\end{figure}

The action for a non-canonical scalar-tensor theory is given by:
\begin{eqnarray}
 S = \int d^4x \sqrt{-g}\left[-G(\phi)g^{\mu\nu}\partial_{\mu}\phi\partial_{\nu}\phi - V(\phi)\right] \, ,
 \label{app-0-1}
\end{eqnarray}
where $\phi$ is the scalar field under consideration, $G(\phi)$ is the non-canonical coupling function and $V(\phi)$ represents the scalar potential ($G(\phi) = \frac{1}{2}$ leads to the usual canonical scalar-tensor theory). In this case, the Friedmann equations (for homogeneous, isotropic and spatially flat universe) read as,
\begin{eqnarray}
 3M_\mathrm{Pl}^2~H_\mathrm{m}^2&= &G(\phi)\dot{\phi}^2 + V(\phi) \,,\nonumber\\
 M_\mathrm{Pl}^2~\dot{H}_\mathrm{m}&=&-G(\phi)\dot{\phi}^2 \,,
 \label{app-0-2}
\end{eqnarray}
respectively, where $H_\mathrm{m}$ is the Hubble parameter given in Eq.~(\ref{reconstruct-1}) and the scalar field is taken to be function of only the cosmic time due to the homogeneity of the spacetime. Moreover the scalar field equation comes as:
\begin{eqnarray}
 \ddot{\phi} + 3H_\mathrm{m}\dot{\phi} + \frac{V'(\phi)}{G(\phi)} + \frac{1}{2}\dot{\phi}^2\frac{G'(\phi)}{G(\phi)} = 0 \, .
 \label{app-0-3}
\end{eqnarray}
Eq.~(\ref{app-0-3}) depicts how the non-canonical coupling function affects the damping and the restoring forces acting on the scalar field, in particular, these are given by: $F_\mathrm{dam} = -3H_\mathrm{m}\dot{\phi} - \frac{1}{2}\dot{\phi}^2 G'(\phi)/G(\phi)$ and $F_\mathrm{res} = -V'(\phi)/G(\phi)$ respectively. The part of the damping force, coming from the coupling function, is proportional to $\dot{\phi}^2$, and may be negligible during the slow roll evolution; however the restoring force is significantly affected by $G(\phi)$ even during the slow roll motion. We assume the slow roll condition during the inflationary stage (due to $|\epsilon_\mathrm{m}| < 1$, see Fig.~[\ref{plot-SRP-2}]), owing to which, the Friedmann equations turn out to be,
\begin{eqnarray}
 3M_\mathrm{Pl}^2~H_\mathrm{m}^2&=&V(\phi) \,,\label{app-0-2a}\\
 M_\mathrm{Pl}^2~\dot{H}_\mathrm{m}&=&-G(\phi)\dot{\phi}^2 \,.
 \label{app-0-2b}
\end{eqnarray}
Eq.~(\ref{app-0-2b}) refers the importance of the non-canonical coupling function ($G(\phi)$) during the inflationary stage particularly when the energy condition gets violated with $\dot{H}_\mathrm{m} > 0$ (i.e., during the 2nd inflationary stage, see Fig.~[\ref{plot-Hubble}]). With the above set of equations, we now examine whether the solution of $H_\mathrm{m}(N)$ obtained in the generalized entropic scenario can be produced from a non-canonical scalar-tensor theory designated by $\{G(\phi),V(\phi)\}$. Plugging $H_\mathrm{m}(N)$ from Eq.~(\ref{reconstruct-1}) to Eq.~(\ref{app-0-2a}), we reconstruct the form of $V(\phi(N))$ as follows:
\begin{eqnarray}
 V(\phi(N)) = 3M_\mathrm{Pl}^2~H_\mathrm{old}^2(N)~\mathrm{exp}\bigg[-\frac{2C}{D}\ln{\left\{\frac{\mathrm{cosh}[D(N - N_\mathrm{1})]~\mathrm{cosh}[D N_\mathrm{2}]}{\mathrm{cosh}[D(N_\mathrm{2} - N)]~\mathrm{cosh}[D N_\mathrm{1}]}\right\}}\bigg] \, ,
\label{app-0-4}
\end{eqnarray}
where $N = [N_\mathrm{1}, N_\mathrm{2}]$ is the duration of the 2nd inflationary stage. To have a better understanding, we give the plot of $V(\phi(N))$ vs. $N$ for $N = [0,N_\mathrm{f}]$ (recall $N_\mathrm{f}$ is the total e-fold duration for the inflation in the entropic scenario), see the left plot of Fig.~[\ref{plot-1-app}]. This clearly demonstrates that, similar to the Hubble parameter, the scalar potential remains $almost$ constant during the 1st and the 3rd inflationary stage (with $V'(\phi) < 0$ which is clear from the inset plot), while it shows a rise during the intermediate stage. Regarding the evolution of the scalar field, Eq.~(\ref{app-0-2b}) can be framed in terms of e-fold number as follows:
\begin{eqnarray}
 \frac{1}{M_\mathrm{Pl}}\frac{d\phi}{dN} = \sqrt{\epsilon_\mathrm{m}(N)/G(\phi(N))} \,,
\label{app-0-5}
\end{eqnarray}
(we take only the positive mode of $d\phi/dN$), with $\epsilon_\mathrm{m}(N)$ being given in Eq.~(\ref{Nano-1}). Clearly the evolution of $\phi = \phi(N)$ demands an explicit form of $G(\phi)$, in particular, $G(\phi)$ should be positive (or negative) valued when $\epsilon_\mathrm{m} > 0$ (or, $\epsilon_\mathrm{m} < 0$) in order to have a real valued scalar field. For simplicity, we consider $G(\phi)$ as follows: $G(\phi) = 1$ during the 1st and the 3rd inflationary stage (when the $\epsilon_\mathrm{m}$ is positive), while $G(\phi)=-1$ during the 2nd inflationary stage (when the $\epsilon_\mathrm{m}$ is negative). Here it may be mentioned that some other forms of $G(\phi)$ can also do the job, however we take this choice for a simplified calculation. With the aforementioned choice of $G(\phi)$, Eq.~(\ref{app-0-5}) takes the following form:
\begin{eqnarray}
 \frac{1}{M_\mathrm{Pl}}\frac{d\phi}{dN} = \sqrt{\left|\epsilon_\mathrm{m}(N)\right|} \, .
\label{app-0-6}
\end{eqnarray}
Owing to the complicated nature, we numerically solve Eq.~(\ref{app-0-6}) for $\phi = \phi(N)$ which is shown in the right plot of the Fig.~[\ref{plot-1-app}]. This reveals that the scalar field monotonically increases with the e-fold number during the entire inflationary epoch. Such evolution of the scalar field is however expected from the pattern of $\{G(\phi),V(\phi)\}$ and can be demonstrated as follows --- (i) during the 1st and the 3rd inflationary stage, $V'(\phi) < 0$ and $G(\phi) > 0$ which leads to a positive restoring force, i.e. $F_\mathrm{res} = -V'(\phi)/G(\phi) > 0$; while (ii) during the 2nd inflationary epoch, $V'(\phi) > 0$ and $G(\phi) < 0$ that, once again, results to $F_\mathrm{res} > 0$. Thus the restoring force acting on the scalar field remains positive during the entire inflation, which actually causes the monotonic increasing behaviour (with time) of the scalar field.

\begin{figure}[!h]
\begin{center}
\centering
\includegraphics[width=3.0in,height=2.0in]{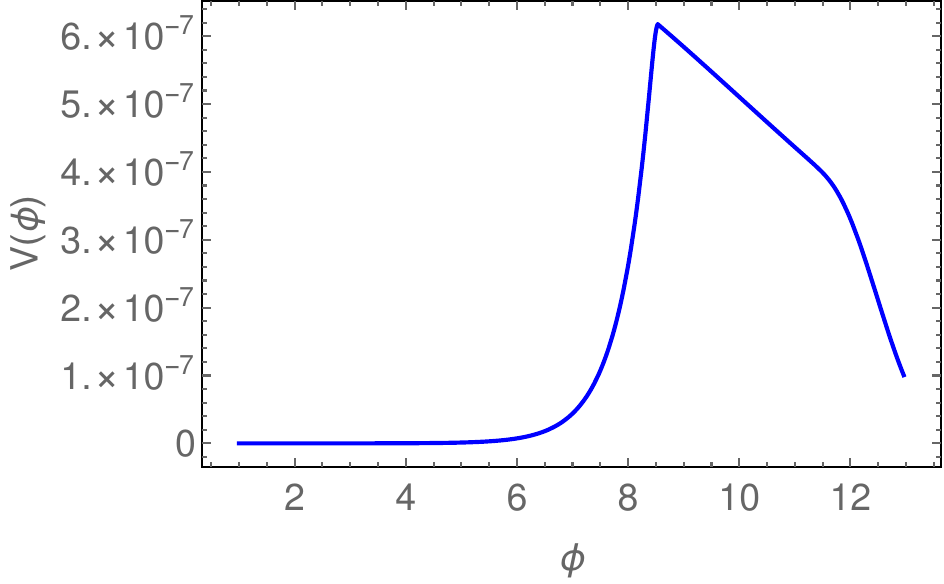}
\hskip0.7cm
\includegraphics[width=3.0in,height=2.0in]{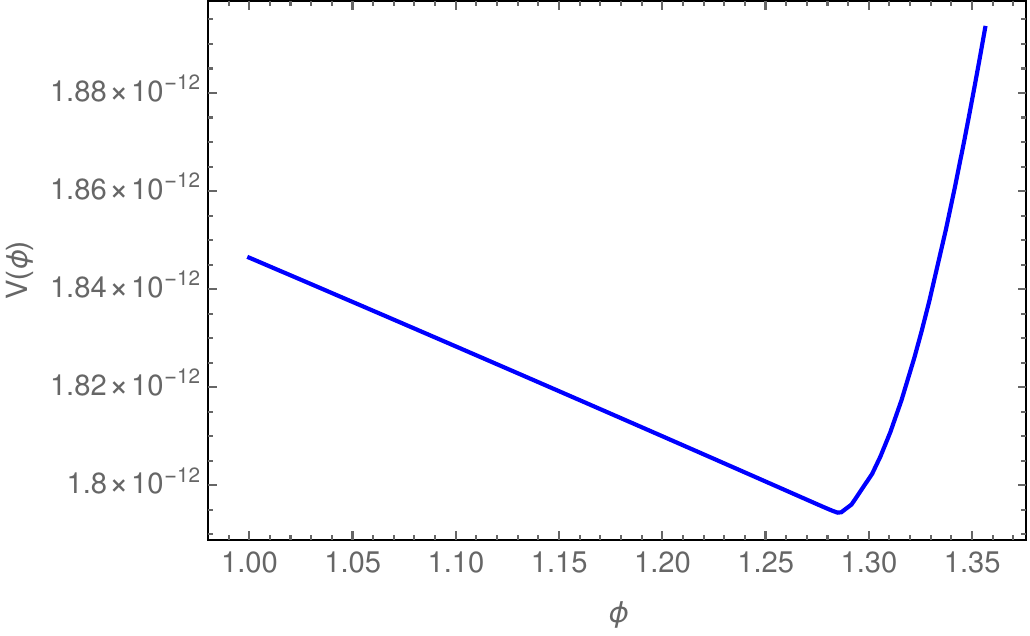}
\caption{{\color{blue}Left Plot}: $V(\phi)$ (in the unit of $M_\mathrm{Pl}^4$) vs. $\phi$ (in the unit of $M_\mathrm{Pl}$); {\color{blue}Right Plot}: Zoomed-in version of $V(\phi)$ vs. $\phi$ during $N=[0,6]$. The right plot indeed ensures that $V'(\phi) < 0$ during the 1st inflationary stage. In both the plots, we consider $C=-0.4$, $D=10$, $\sigma_0 = 0.024$, $N_\mathrm{1}=6$ and $N_\mathrm{2}=14$ (which are compatible with the best fitted values of the parameters).}
\label{plot-2-app}
\end{center}
\end{figure}

For completeness, we also present the variation of the scalar potential $V(\phi)$ with the scalar field $\phi$ by using $V = V(\phi(N))$ and $\phi = \phi(N)$ (by the ``ParametricPlot'' in M{\small{ATHEMATICA}}), see the left plot of Fig.~[\ref{plot-2-app}] and the right one is the zoomed-in version of $V(\phi)$ during $N=[0,6]$.

Thus as a whole, the solution of $H_\mathrm{m}(N)$ in the present context of generalized entropic cosmology can also be triggered from a non-canonical scalar-tensor theory where $V(\phi)$ and $\phi(N)$ are given by Fig.~[\ref{plot-1-app}].

\section{Transfer functions for the Case (I) $k < k_\mathrm{1}$}\label{sec-appendix-1}
In this Appendix we will report the computation of the transfer function and its derivative in the Case (I): $k < k_\mathrm{1}$. In this case $|x|<1$ and we can take the asymptotic expression of the Bessel function 
\begin{equation}
    \lim_{x\ll 1} J_\nu(x) = \frac{x^\nu}{2^\nu\Gamma(\nu+1)}\ .
\end{equation} 
We then need to compute the numerator Eq.~(\ref{2I-6}) and denominator Eq.~(\ref{2I-5}). In this case the denominator becomes
\begin{eqnarray}
 \mathscr{D}(k)=2\left(\frac{k}{k_\mathrm{1}}\right)\frac{x^{\nu}}{2^{\nu}\Gamma(\nu+1)}&\Bigg[&\frac{x^{-\nu-1}}{2^{-\nu-1}\Gamma(-\nu)} - \frac{x^{-\nu+1}}{2^{-\nu+1}\Gamma(-\nu+2)}\Bigg]\nonumber\\
 &+&2\left(\frac{k}{k_\mathrm{1}}\right)\frac{x^{-\nu}}{2^{-\nu}\Gamma(-\nu+1)}\left[\frac{x^{\nu+1}}{2^{\nu+1}\Gamma(\nu+2)} - \frac{x^{\nu-1}}{2^{\nu-1}\Gamma(\nu)}\right] \, ,
 \label{A1-1}
\end{eqnarray}
Owing to $|x| < 1$, the term with less power becomes the dominant one, thus the above expression is approximated as,
\begin{eqnarray}
 \mathscr{D}(k) \approx 4\left(\frac{k}{k_\mathrm{1}}\right)\frac{1}{x}\left[\frac{1}{\Gamma(-\nu)\Gamma(\nu+1)} - \frac{1}{\Gamma(\nu)\Gamma(-\nu+1)}\right] \, .
 \label{A1-2}
\end{eqnarray}
With $x=\frac{2(k/k_\mathrm{1})}{(1+3\omega)}$ along with the identity $1/(\Gamma(\nu)\Gamma(-\nu+1)) = \sin{(\pi\nu)}/\pi$, the above expression leads to the final form of $\mathscr{D}(k)$ as follows,
\begin{eqnarray}
 \mathscr{D}(k) = -\frac{4}{\pi}(1+3\omega)\sin{(\pi\nu)} \, .
 \label{A1-3}
\end{eqnarray}
At $A=A_\mathrm{2}$, the quantity $\mathcal{N}$ takes the form,
\begin{eqnarray}
 \mathscr{N}(k,A_\mathrm{2}) = A_\mathrm{2}^{\frac{3\omega-3}{4}}\left[\frac{y^{\nu}}{2^{\nu}\Gamma(\nu+1)}\mathscr{N}_\mathrm{1}(k) + \frac{y^{-\nu}}{2^{-\nu}\Gamma(-\nu+1)}\mathscr{N}_\mathrm{2}(k)\right] \, ,
 \label{A1-4}
\end{eqnarray}
Once again, due to $|y|< 1$, the term with less power is the dominant one, and thus the above expression becomes,
\begin{eqnarray}
 \mathscr{N}(k,A_\mathrm{2}) \approx A_\mathrm{2}^{\frac{3\omega-3}{4}}\left[\frac{y^{-\nu}}{2^{-\nu}\Gamma(-\nu+1)}\mathscr{N}_\mathrm{2}(k)\right] = A_\mathrm{2}^{\frac{3\omega-3}{4}}\left(\frac{k_\mathrm{2}}{k_\mathrm{1}}\right)^{\nu}\left[\frac{x^{-\nu}}{2^{-\nu}\Gamma(-\nu+1)}\mathscr{N}_\mathrm{2}(k)\right] \, ,
 \label{A1-5}
\end{eqnarray}
where we use $y=x k_\mathrm{1}/k_\mathrm{2}$ in the second equality. Due to the fact $A_\mathrm{2}^{\frac{3\omega-3}{4}} = (k_\mathrm{1}/k_\mathrm{2})^{\nu}$, the above expression is simplified to,
\begin{eqnarray}
 \mathscr{N}(k,A_\mathrm{2})&=&\frac{x^{-\nu}}{2^{-\nu}\Gamma(-\nu+1)}\mathscr{N}_\mathrm{2}(k)\nonumber\\
 &=&\frac{x^{-\nu}}{2^{-\nu}\Gamma(-\nu+1)}\left[2\frac{k}{k_\mathrm{1}}~J_\mathrm{\nu+1}(x) - 2\frac{k}{k_\mathrm{1}}~J_\mathrm{\nu-1}(x) - (3\omega-3)J_\mathrm{\nu}(x)\right] \, .
 \label{A1-6}
\end{eqnarray}
Once again, due to $|x|<1$, the Bessel functions inside the square bracket can be given by the asymptotic form, and thus the final form of $\mathscr{N}(k,A_\mathrm{2})$ comes as,
\begin{eqnarray}
 \mathscr{N}(k,A_\mathrm{2}) = -\frac{4}{\pi}(1+3\omega)\sin{(\pi\nu)} + \frac{2\sin{(\pi\nu)}}{\pi(\nu+\nu^2)(1+3\omega)}\left(\frac{k}{k_\mathrm{1}}\right)^2 \, .
 \label{A1-7}
\end{eqnarray}
By using Eq.~(\ref{A1-3}) and Eq.~(\ref{A1-7}), one gets the desired result of the transfer function at $A=A_\mathrm{2}$. In order to find the derivative of the transfer function at $A=A_\mathrm{2}$, for which, we first determine,
\begin{eqnarray}
 \frac{d\mathscr{N}(k,A)}{dA}\big|_{A_\mathrm{2}} = y\left(\frac{1+3\omega}{2A_\mathrm{2}}\right)A_\mathrm{2}^{\frac{3\omega-3}{4}}\left[\mathscr{N}_\mathrm{1}(k)J_\mathrm{\nu-1}(y) - \mathscr{N}_\mathrm{2}(k)J_\mathrm{-\nu+1}(y)\right] \, .
 \label{A1-9}
\end{eqnarray}
Using the asymptotic form of the Bessel functions, the first and second terms present in the above expression turns out to be,
\begin{eqnarray}
 \mathrm{1st~term~in~r.h.s.~of~Eq.~(\ref{A1-9})}&=&-\frac{1}{2A_\mathrm{2}}\left(\frac{k_\mathrm{1}}{k_\mathrm{2}}\right)^{2\nu}\frac{2(k/k_\mathrm{1})^2}{\Gamma(\nu)\Gamma(-\nu+2)} \, ,\nonumber\\
 \mathrm{2nd~term~in~r.h.s.~of~Eq.~(\ref{A1-9})}&=&\frac{1}{A_\mathrm{2}}\left(\frac{k_\mathrm{1}}{k_\mathrm{2}}\right)^{2}\frac{2(k/k_\mathrm{1})^2}{\Gamma(\nu)\Gamma(-\nu+2)} \, .
 \label{A1-10}
\end{eqnarray}
Therefore, Eq.~(\ref{A1-9}) leads to,
\begin{eqnarray}
 \frac{d\mathscr{N}(k,A)}{dA}\big|_{A_\mathrm{2}} = \frac{1}{2A_\mathrm{2}}\frac{2(k/k_\mathrm{1})^2}{\Gamma(\nu)\Gamma(-\nu+2)}\left\{2\left(\frac{k_\mathrm{1}}{k_\mathrm{2}}\right)^2 - \left(\frac{k_\mathrm{1}}{k_\mathrm{2}}\right)^{2\nu}\right\} \ll 1 \, .
 \label{A1-11}
\end{eqnarray}
The above equation along with the form of $\mathscr{D}(k)$ in Eq.~(\ref{A1-2}) immediately leads to the desired result of $\frac{d\chi^\mathrm{(2)}(k,A_\mathrm{2})}{dA}$ of Eq.~(\ref{R1-2}).

In order to obtain Eq.~(\ref{R1-3}), let us start by putting $\chi^\mathrm{(2)}(k,A_\mathrm{2})\approx 1$ and $\frac{d\chi^\mathrm{(2)}(k,A_\mathrm{2})}{dA} \approx 0$ into Eq.~(\ref{3I-2}) which simplifies the transfer function (and its derivative) during the 3rd inflationary stage as follows,
\begin{eqnarray}
 \chi^\mathrm{(3)}(k,A)&=&\frac{A_\mathrm{2}}{A}~\cos{\left[\frac{k}{k_\mathrm{2}}\left(1 - \frac{A_\mathrm{2}}{A}\right)\right]} + \frac{k_\mathrm{2}}{k}~\sin{\left[\frac{k}{k_\mathrm{2}}\left(1 - \frac{A_\mathrm{2}}{A}\right)\right]} \, ,\nonumber\\
 A_\mathrm{f}\frac{d\chi^\mathrm{(3)}(k,A)}{dA}&=&-\left(\frac{k}{k_\mathrm{2}}\right)\frac{A_\mathrm{f}A_\mathrm{2}^2}{A^3}~\sin{\left[\frac{k}{k_\mathrm{2}}\left(1 - \frac{A_\mathrm{2}}{A}\right)\right]} \, ,
 \label{A1-12}
\end{eqnarray}
respectively. Owing to $A_\mathrm{f} = A_\mathrm{2}(k_\mathrm{f}/k_\mathrm{2})$ along with the fact that $k<k_\mathrm{1}<k_\mathrm{2}<k_\mathrm{f}$, the sine and cosine functions present in Eq.~(\ref{A1-12}) can be expanded as Taylor series w.r.t. $\frac{k}{k_\mathrm{2}}$. As a result, we get the desired result of Eq.~(\ref{R1-3}).

\section{Transfer functions for the Case (II)   $k_\mathrm{1} < k < k_\mathrm{2}$}\label{sec-appendix-2}
In this Appendix we have reported the steps followed to obtain the transfer function in the Case (II):  $k_\mathrm{1} < k < k_\mathrm{2}$.
Under this hypothesis we can take the Bessel function containing $k/k_\mathrm{1}$ as
\begin{equation}
    \lim_{x\gg 1 }J_\nu(x) = \sqrt{\frac{2}{\pi x}}\cos\left(x-\frac{\pi}{2}\left(\nu+\frac{1}{2}\right)\right) \ .
\end{equation}
while the Bessel function with $k/k_\mathrm{2}$ becomes
\begin{equation}
    \lim_{x\ll 1} J_\nu(x) = \frac{x^\nu}{2^\nu\Gamma(\nu+1)}\ .
\end{equation}
For $k_\mathrm{1} < k < k_\mathrm{2}$ we have $|x| = \left|\frac{2k/k_\mathrm{1}}{1+3\omega}\right|> 1$ and $|y|= \left|\frac{2k/k_\mathrm{2}}{1+3\omega}\right| < 1$. 
Owing to this expansions, Eq.~(\ref{2I-5}) becomes
\begin{eqnarray}
 \mathscr{D}(k)&=&2\left(\frac{k}{k_\mathrm{1}}\right)\Bigg\{J_{\nu}(x)\left[J_{-\nu-1}(x) - J_{-\nu+1}(x)\right] + J_{-\nu}(x)\left[J_{\nu+1}(x) - J_{\nu-1}(x)\right]\Bigg\}\nonumber\\
 &=&\frac{4k/k_\mathrm{1}}{\pi x} \Bigg[\cos{\left(\phi - \frac{\pi}{4}\right)}\left\{\cos{\left(\theta + \frac{\pi}{4}\right)} - \cos{\left(\theta - \frac{3\pi}{4}\right)}\right\}\nonumber\\
 &+&\cos{\left(\theta - \frac{\pi}{4}\right)}\left\{\cos{\left(\phi - \frac{3\pi}{4}\right)} - \cos{\left(\phi + \frac{\pi}{4}\right)}\right\}\Bigg] \, ,
 \label{A2-1}
\end{eqnarray}
where we use the asymptotic form of the Bessel function having argument larger than unity, here $\theta = x + \pi\nu/2$ and $\phi = x - \pi\nu/2$. Using the identity: $\cos{(B\pm C)} = \cos{B}\cos{C} \mp \sin{B}\sin{C}$, the above equation can be simplified to get,
\begin{eqnarray}
 \mathscr{D}(k) = \frac{4(1+3\omega)}{\pi}~\sin{\left(\phi - \theta\right)} = -\frac{4(1+3\omega)}{\pi}~\sin{(\pi\nu)} \, .
 \label{A2-2}
\end{eqnarray}
Moreover, the numerator of Eq.~(\ref{2I-6}) becomes
\begin{eqnarray}
 \mathscr{N}(k,A_\mathrm{2})&=&A_\mathrm{2}^{\frac{3\omega-3}{4}}\left[J_\nu(y) \mathscr{N}_1(k)+J_{-\nu}(y) \mathscr{N}_2(k) \right]\nonumber\\
 &=&A_\mathrm{2}^{\frac{3\omega-3}{4}}\left[\frac{y^{\nu}}{2^{\nu}\Gamma(\nu+1)} \mathscr{N}_1(k)+ \frac{y^{-\nu}}{2^{-\nu}\Gamma(-\nu+1)} \mathscr{N}_2(k) \right]\nonumber\\
 &\approx&A_\mathrm{2}^{\frac{3\omega-3}{4}}\left[\frac{y^{-\nu}}{2^{-\nu}\Gamma(-\nu+1)} \mathscr{N}_2(k) \right] \, .
 \label{A2-3}
\end{eqnarray}
In the second equality, we use the asymptotic form of $J_\mathrm{\pm \nu}(y)$ with $|y| < 1$; and consequently, the term with $J_\mathrm{\nu}(y)$ becomes the sub-dominant one (w.r.t. the term containing $J_\mathrm{-\nu}(y)$) which we use in the third line of the above equation. By using the explicit form of $\mathscr{N}_2(k)$ (see Eq.~(\ref{2I-6b})) along with the expressions of $\chi^\mathrm{(1)}(k,A=1)$ and its derivative (Eq. \eqref{R2-1}), the above Eq.~(\ref{A2-3}) leads to,
\begin{eqnarray}
 \mathscr{N}(k,A_\mathrm{2}) = -\frac{4}{\sqrt{\pi}}\frac{x^{-\nu-1/2}}{2^{-\nu-1/2}\Gamma(-\nu+1)}\left(\frac{k^2}{k_\mathrm{1}^2}\right)\mathrm{exp}\left[i\left(\frac{k}{k_\mathrm{1}} + x - \frac{\pi\nu}{2} - \frac{\pi}{4}\right)\right] \, ,
\end{eqnarray}
which, due to $x= \frac{2k/k_\mathrm{1}}{1+3\omega}$, turns out to be
\begin{eqnarray}
 \mathscr{N}(k,A_\mathrm{2}) = -\frac{4}{\sqrt{\pi}}\frac{(1+3\omega)^{\nu+1/2}}{\Gamma(-\nu+1)}\left(\frac{k}{k_\mathrm{1}}\right)^{\frac{3}{2}-\nu}\mathrm{exp}\left[i\left(\frac{k}{k_\mathrm{1}} + x - \frac{\pi\nu}{2} - \frac{\pi}{4}\right)\right] \, .
 \label{A2-4}
\end{eqnarray}
From Eq.~(\ref{A2-2}) and Eq.~(\ref{A2-4}), one gets the expression of $\chi^\mathrm{(2)}(k,A_\mathrm{2}) = \mathscr{N}(k,A_\mathrm{2})/\mathscr{D}(k)$, present in Eq.~(\ref{R2-2}). In order to obtain Eq.~(\ref{R2-3}), we use $\chi^\mathrm{(2)}(k,A_\mathrm{2})$ (and its derivative) into Eq.~(\ref{3I-2}) which simplifies the transfer function (and its derivative) during the 3rd inflationary stage as follows,
\begin{eqnarray}
 \chi^\mathrm{(3)}(k,A)&=&\chi^\mathrm{(2)}(k,A_\mathrm{2})\left[\frac{A_\mathrm{2}}{A}~\cos{\left[\frac{k}{k_\mathrm{2}}\left(1 - \frac{A_\mathrm{2}}{A}\right)\right]} + \frac{k_\mathrm{2}}{k}~\sin{\left[\frac{k}{k_\mathrm{2}}\left(1 - \frac{A_\mathrm{2}}{A}\right)\right]}\right] \, ,\nonumber\\
 A_\mathrm{f}\frac{d\chi^\mathrm{(3)}(k,A)}{dA}&=&-\chi^\mathrm{(2)}(k,A_\mathrm{2})\left[\left(\frac{k}{k_\mathrm{2}}\right)\frac{A_\mathrm{f}A_\mathrm{2}^2}{A^3}~\sin{\left[\frac{k}{k_\mathrm{2}}\left(1 - \frac{A_\mathrm{2}}{A}\right)\right]}\right] \, ,
 \label{A2-5}
\end{eqnarray}
respectively. Owing to $A_\mathrm{f} = A_\mathrm{2}(k_\mathrm{f}/k_\mathrm{2})$ along with the fact that $k_\mathrm{1}<k<k_\mathrm{2}<k_\mathrm{f}$, the sine and cosine functions present in Eq.~(\ref{A2-5}) can be expanded as Taylor series w.r.t. $\frac{k}{k_\mathrm{2}}$. As a result, we get the desired result of Eq.~(\ref{R2-3}).

\section*{Acknowledgment}
T. Paul thanks M. R. Haque for some discussions.

\bibliography{bibliography}

@article{Cai_2005,
   title={First Law of Thermodynamics and Friedmann Equations of Friedmann–Robertson–Walker Universe},
   volume={2005},
   ISSN={1029-8479},
   url={http://dx.doi.org/10.1088/1126-6708/2005/02/050},
   DOI={10.1088/1126-6708/2005/02/050},
   number={02},
   journal={Journal of High Energy Physics},
   publisher={Springer Science and Business Media LLC},
   author={Cai, Rong-Gen and Kim, Sang Pyo},
   year={2005},
   month=feb, pages={050–050} 
}

@article{Gialamas:2025kef,
    author = "Gialamas, Ioannis D. and Karam, Alexandros and Racioppi, Antonio and Raidal, Martti",
    title = "{Has ACT measured radiative corrections to the tree-level Higgs-like inflation?}",
    eprint = "2504.06002",
    archivePrefix = "arXiv",
    primaryClass = "astro-ph.CO",
    doi = "10.1103/6fpc-67s1",
    journal = "Phys. Rev. D",
    volume = "112",
    number = "10",
    pages = "103544",
    year = "2025"
}

@article{Mohammadi:2025gbu,
    author = "Mohammadi, Abolhassan and Yogesh and Wang, Anzhong",
    title = "{Power Law Plateau Inflation and Primary Gravitational Waves in the light of ACT}",
    eprint = "2507.06544",
    archivePrefix = "arXiv",
    primaryClass = "astro-ph.CO",
    month = "7",
    year = "2025"
}

@article{Conzinu:2025sot,
    author = "Conzinu, PIetro and Gasperini, Maurizio and Pavone, Eliseo",
    title = "{A simple example of {\textquotedblleft}non-minimal{\textquotedblright} Pre-Big Bang scenario}",
    eprint = "2506.00995",
    archivePrefix = "arXiv",
    primaryClass = "gr-qc",
    reportNumber = "BA-TH/810-25",
    doi = "10.1088/1475-7516/2025/11/027",
    journal = "JCAP",
    volume = "11",
    pages = "027",
    year = "2025"
}

@article{Liu:2023pau,
    author = "Liu, Lang and Chen, Zu-Cheng and Huang, Qing-Guo",
    title = "{Probing the equation of state of the early Universe with pulsar timing arrays}",
    eprint = "2307.14911",
    archivePrefix = "arXiv",
    primaryClass = "astro-ph.CO",
    doi = "10.1088/1475-7516/2023/11/071",
    journal = "JCAP",
    volume = "11",
    pages = "071",
    year = "2023"
}

@article{Ashoorioon:2022raz,
    author = "Ashoorioon, Amjad and Rezazadeh, Kazem and Rostami, Abasalt",
    title = "{NANOGrav signal from the end of inflation and the LIGO mass and heavier primordial black holes}",
    eprint = "2202.01131",
    archivePrefix = "arXiv",
    primaryClass = "astro-ph.CO",
    reportNumber = "IPM/P-2022/06",
    doi = "10.1016/j.physletb.2022.137542",
    journal = "Phys. Lett. B",
    volume = "835",
    pages = "137542",
    year = "2022"
}

@misc{odintsov2023entropicinflationpresencescalar,
      title={Entropic Inflation in Presence of Scalar Field}, 
      author={Sergei D. Odintsov and Simone D'Onofrio and Tanmoy Paul},
      year={2023},
      eprint={2312.13587},
      archivePrefix={arXiv},
      primaryClass={gr-qc},
      url={https://arxiv.org/abs/2312.13587}, 
}

@article{Nojiri_2022,
   title={Early and late universe holographic cosmology from a new generalized entropy},
   volume={831},
   ISSN={0370-2693},
   url={http://dx.doi.org/10.1016/j.physletb.2022.137189},
   DOI={10.1016/j.physletb.2022.137189},
   journal={Physics Letters B},
   publisher={Elsevier BV},
   author={Nojiri, Shin’ichi and Odintsov, Sergei D. and Paul, Tanmoy},
   year={2022},
   month=aug, pages={137189} 
}

@article{Nojiri:2019skr,
    author = "Nojiri, Shin'ichi and Odintsov, Sergei D. and Saridakis, Emmanuel N.",
    title = "{Modified cosmology from extended entropy with varying exponent}",
    eprint = "1903.03098",
    archivePrefix = "arXiv",
    primaryClass = "gr-qc",
    doi = "10.1140/epjc/s10052-019-6740-5",
    journal = "Eur. Phys. J. C",
    volume = "79",
    number = "3",
    pages = "242",
    year = "2019"
}

@article{Tyagi:2025zov,
    author = "Tyagi, Udit K. and Haridasu, Sandeep and Basak, Soumen",
    title = "{Constraints on Generalized Gravity-Thermodynamic Cosmology from DESI DR2}",
    eprint = "2504.11308",
    archivePrefix = "arXiv",
    primaryClass = "astro-ph.CO",
    month = "4",
    year = "2025"
}

@article{Kaniadakis_2005,
   title={Statistical mechanics in the context of special relativity. II.},
   volume={72},
   ISSN={1550-2376},
   url={http://dx.doi.org/10.1103/PhysRevE.72.036108},
   DOI={10.1103/physreve.72.036108},
   number={3},
   journal={Physical Review E},
   publisher={American Physical Society (APS)},
   author={Kaniadakis, G.},
   year={2005},
   month=sep 
}

@article{Tariq:2025wiy,
    author = "Tariq, Hamza and Zafar, Usman and Chaudhary, Shahid and Bamba, Kazuharu and Jawad, Abdul and Shaymatov, Sanjar",
    title = "{Exploring the effects of generalized entropy onto Bardeen black hole surrounded by cloud of strings}",
    eprint = "2504.10528",
    archivePrefix = "arXiv",
    primaryClass = "gr-qc",
    reportNumber = "FU-PCG-147",
    doi = "10.1016/j.nuclphysb.2025.116906",
    journal = "Nucl. Phys. B",
    volume = "1016",
    pages = "116906",
    year = "2025"
}

@article{Sayahian_Jahromi_2018,
   title={Generalized entropy formalism and a new holographic dark energy model},
   volume={780},
   ISSN={0370-2693},
   url={http://dx.doi.org/10.1016/j.physletb.2018.02.052},
   DOI={10.1016/j.physletb.2018.02.052},
   journal={Physics Letters B},
   publisher={Elsevier BV},
   author={Sayahian Jahromi, A. and Moosavi, S.A. and Moradpour, H. and Morais Graça, J.P. and Lobo, I.P. and Salako, I.G. and Jawad, A.},
   year={2018},
   month=may, pages={21–24} 
}

@article{Hellings:1983fr,
    author = "Hellings, R. w. and Downs, G. s.",
    title = "{UPPER LIMITS ON THE ISOTROPIC GRAVITATIONAL RADIATION BACKGROUND FROM PULSAR TIMING ANALYSIS}",
    doi = "10.1086/183954",
    journal = "Astrophys. J. Lett.",
    volume = "265",
    pages = "L39--L42",
    year = "1983"
}

@article{Sheykhi:2025zre,
    author = "Sheykhi, Ahmad and Sooraki, Ava Shahbazi and Liravi, Leila",
    title = "{Big-Bang nucleosynthesis constraints on (dual) Kaniadakis cosmology}",
    eprint = "2504.21146",
    archivePrefix = "arXiv",
    primaryClass = "gr-qc",
    month = "4",
    year = "2025"
}

@article{Sheykhi:2025kfw,
    author = "Sheykhi, Ahmad and Sooraki, Ava Shahbazi",
    title = "{Constraints on R{\'e}nyi Entropy through Primordial Big-Bang Nucleosynthesis and Baryogenesis}",
    eprint = "2507.14250",
    archivePrefix = "arXiv",
    primaryClass = "physics.gen-ph",
    month = "7",
    year = "2025"
}

@article{Luciano:2025hjn,
    author = "Luciano, Giuseppe Gaetano and Paliathanasis, Andronikos and Saridakis, Emmanuel N.",
    title = "{Barrow and Tsallis entropies after the DESI DR2 BAO data}",
    eprint = "2504.12205",
    archivePrefix = "arXiv",
    primaryClass = "gr-qc",
    doi = "10.1088/1475-7516/2025/09/013",
    journal = "JCAP",
    volume = "09",
    pages = "013",
    year = "2025"
}

@article{Adhikary:2024sax,
    author = "Adhikary, Priyanka and Das, Sudipta",
    title = "{Interacting Barrow holographic dark energy in non-flat universe}",
    eprint = "2412.05577",
    archivePrefix = "arXiv",
    primaryClass = "gr-qc",
    doi = "10.1088/1475-7516/2025/04/027",
    journal = "JCAP",
    volume = "04",
    pages = "027",
    year = "2025"
}

@article{Adhikary:2021xym,
    author = "Adhikary, Priyanka and Das, Sudipta and Basilakos, Spyros and Saridakis, Emmanuel N.",
    title = "{Barrow holographic dark energy in a nonflat universe}",
    eprint = "2104.13118",
    archivePrefix = "arXiv",
    primaryClass = "gr-qc",
    doi = "10.1103/PhysRevD.104.123519",
    journal = "Phys. Rev. D",
    volume = "104",
    number = "12",
    pages = "123519",
    year = "2021"
}

@article{Odintsov:2025sew,
    author = "Odintsov, Sergei D. and D'Onofrio, Simone and Paul, Tanmoy",
    title = "{Generalized entropic dark energy with spatial curvature}",
    eprint = "2504.03470",
    archivePrefix = "arXiv",
    primaryClass = "gr-qc",
    doi = "10.1016/j.dark.2025.101920",
    journal = "Phys. Dark Univ.",
    volume = "48",
    pages = "101920",
    year = "2025"
}

@article{Adhikary:2025khr,
    author = "Adhikary, Priyanka and Das, Sudipta and Odintsov, Sergei D. and Paul, Tanmoy",
    title = "{Dark energy era with a resolution of Hubble tension in generalized entropic cosmology}",
    eprint = "2507.15273",
    archivePrefix = "arXiv",
    primaryClass = "gr-qc",
    doi = "10.1016/j.dark.2025.102037",
    journal = "Phys. Dark Univ.",
    volume = "49",
    pages = "102037",
    year = "2025"
}

@article{Odintsov:2025bmp,
    author = "Odintsov, Sergei D. and Paul, Tanmoy",
    title = "{ACT inflation and its influence on reheating era in Einstein-Gauss-Bonnet gravity}",
    eprint = "2508.11377",
    archivePrefix = "arXiv",
    primaryClass = "gr-qc",
    month = "8",
    year = "2025"
}

@article{Cardenas:2023zmn,
    author = "C{\'a}rdenas, V{\'\i}ctor H. and Cruz, Miguel and Lepe, Samuel",
    title = "{Generalized second law of thermodynamics for the matter creation scenario and emergence of phantom regime}",
    eprint = "2302.10155",
    archivePrefix = "arXiv",
    primaryClass = "gr-qc",
    doi = "10.1140/epjp/s13360-024-05447-x",
    journal = "Eur. Phys. J. Plus",
    volume = "139",
    number = "7",
    pages = "642",
    year = "2024"
}

@article{Cruz:2023xjp,
    author = "Cruz, Miguel and Lepe, Samuel and Saavedra, Joel",
    title = "{A new approach to P{\ensuremath{-}}V phase transitions: Einstein gravity and holographic type dark energy}",
    eprint = "2312.14257",
    archivePrefix = "arXiv",
    primaryClass = "gr-qc",
    doi = "10.1016/j.dark.2024.101580",
    journal = "Phys. Dark Univ.",
    volume = "46",
    pages = "101580",
    year = "2024"
}

@article{Okcu:2024tnw,
    author = {{\"O}kc{\"u}, {\"O}zg{\"u}r},
    title = "{Investigation of generalised uncertainty principle effects on FRW cosmology}",
    eprint = "2401.09477",
    archivePrefix = "arXiv",
    primaryClass = "gr-qc",
    doi = "10.1016/j.nuclphysb.2024.116551",
    journal = "Nucl. Phys. B",
    volume = "1004",
    pages = "116551",
    year = "2024"
}

@article{Brevik:2024nzf,
    author = "Brevik, I. and Timoshkin, A. V.",
    title = "{Little rip and pseudo rip cosmological models with coupled dark energy based on a new generalized entropy}",
    eprint = "2404.05597",
    archivePrefix = "arXiv",
    primaryClass = "gr-qc",
    doi = "10.1142/S0219887824501810",
    journal = "Int. J. Geom. Meth. Mod. Phys.",
    volume = "21",
    number = "11",
    pages = "2450181",
    year = "2024"
}

@article{Odintsov:2024sbo,
    author = "Odintsov, Sergei D. and D'Onofrio, Simone and Paul, Tanmoy",
    title = "{Primordial gravitational waves in horizon cosmology and constraints on entropic parameters}",
    eprint = "2407.05855",
    archivePrefix = "arXiv",
    primaryClass = "gr-qc",
    doi = "10.1103/PhysRevD.110.043539",
    journal = "Phys. Rev. D",
    volume = "110",
    number = "4",
    pages = "043539",
    year = "2024"
}

@article{Housset:2023jcm,
    author = "Housset, Joaquin and Saavedra, Joel F. and Tello-Ortiz, Francisco",
    title = "{Cosmological FLRW phase transitions and micro-structure under Kaniadakis statistics}",
    eprint = "2312.05683",
    archivePrefix = "arXiv",
    primaryClass = "gr-qc",
    doi = "10.1016/j.physletb.2024.138686",
    journal = "Phys. Lett. B",
    volume = "853",
    pages = "138686",
    year = "2024"
}

@article{Odintsov:2023vpj,
    author = "Odintsov, Sergei D. and D'Onofrio, Simone and Paul, Tanmoy",
    title = "{Holographic realization from inflation to reheating in generalized entropic cosmology}",
    eprint = "2306.15225",
    archivePrefix = "arXiv",
    primaryClass = "gr-qc",
    doi = "10.1016/j.dark.2023.101277",
    journal = "Phys. Dark Univ.",
    volume = "42",
    pages = "101277",
    year = "2023"
}

@article{Nojiri:2022aof,
    author = "Nojiri, Shin'ichi and Odintsov, Sergei D. and Faraoni, Valerio",
    title = "{From nonextensive statistics and black hole entropy to the holographic dark universe}",
    eprint = "2201.02424",
    archivePrefix = "arXiv",
    primaryClass = "gr-qc",
    doi = "10.1103/PhysRevD.105.044042",
    journal = "Phys. Rev. D",
    volume = "105",
    number = "4",
    pages = "044042",
    year = "2022"
}

@article{Jizba:2024klq,
    author = "Jizba, Petr and Lambiase, Gaetano and Luciano, Giuseppe Gaetano and Mastrototaro, Leonardo",
    title = "{Imprints of Barrow{\textendash}Tsallis cosmology in primordial gravitational waves}",
    eprint = "2403.09797",
    archivePrefix = "arXiv",
    primaryClass = "gr-qc",
    doi = "10.1140/epjc/s10052-024-13455-5",
    journal = "Eur. Phys. J. C",
    volume = "84",
    number = "10",
    pages = "1076",
    year = "2024"
}

@article{Jizba:2023fkp,
    author = "Jizba, Petr and Lambiase, Gaetano",
    title = "{Constraints on Tsallis Cosmology from Big Bang Nucleosynthesis and the Relic Abundance of Cold Dark Matter Particles}",
    eprint = "2310.19045",
    archivePrefix = "arXiv",
    primaryClass = "gr-qc",
    doi = "10.3390/e25111495",
    journal = "Entropy",
    volume = "25",
    number = "11",
    pages = "1495",
    year = "2023"
}

@article{Nojiri:2022nmu,
    author = "Nojiri, Shin'ichi and Odintsov, Sergei D. and Paul, Tanmoy",
    title = "{Modified cosmology from the thermodynamics of apparent horizon}",
    eprint = "2211.02822",
    archivePrefix = "arXiv",
    primaryClass = "gr-qc",
    doi = "10.1016/j.physletb.2022.137553",
    journal = "Phys. Lett. B",
    volume = "835",
    pages = "137553",
    year = "2022"
}

@article{Cognola:2005de,
    author = "Cognola, Guido and Elizalde, Emilio and Nojiri, Shin'ichi and Odintsov, Sergei D. and Zerbini, Sergio",
    title = "{One-loop f(R) gravity in de Sitter universe}",
    eprint = "hep-th/0501096",
    archivePrefix = "arXiv",
    doi = "10.1088/1475-7516/2005/02/010",
    journal = "JCAP",
    volume = "02",
    pages = "010",
    year = "2005"
}

@article{Sanchez:2022xfh,
    author = "Sanchez, Luis M. and Quevedo, Hernando",
    title = "{Thermodynamics of the FLRW apparent horizon}",
    eprint = "2208.05729",
    archivePrefix = "arXiv",
    primaryClass = "gr-qc",
    doi = "10.1016/j.physletb.2023.137778",
    journal = "Phys. Lett. B",
    volume = "839",
    pages = "137778",
    year = "2023"
}

@article{DAgostino:2019wko,
    author = "D'Agostino, Rocco",
    title = "{Holographic dark energy from nonadditive entropy: cosmological perturbations and observational constraints}",
    eprint = "1903.03836",
    archivePrefix = "arXiv",
    primaryClass = "gr-qc",
    doi = "10.1103/PhysRevD.99.103524",
    journal = "Phys. Rev. D",
    volume = "99",
    number = "10",
    pages = "103524",
    year = "2019"
}

@article{Nojiri:2023bom,
    author = "Nojiri, Shin'ichi and Odintsov, Sergei D. and Paul, Tanmoy",
    title = "{Microscopic interpretation of generalized entropy}",
    eprint = "2311.03848",
    archivePrefix = "arXiv",
    primaryClass = "gr-qc",
    doi = "10.1016/j.physletb.2023.138321",
    journal = "Phys. Lett. B",
    volume = "847",
    pages = "138321",
    year = "2023"
}

@article{Paul:2025rqe,
    author = "Paul, Tanmoy",
    title = "{Origin of bulk viscosity in cosmology and its thermodynamic implications}",
    eprint = "2504.00422",
    archivePrefix = "arXiv",
    primaryClass = "gr-qc",
    doi = "10.1103/PhysRevD.111.083540",
    journal = "Phys. Rev. D",
    volume = "111",
    number = "8",
    pages = "083540",
    year = "2025"
}

@article{Odintsov:2024ipb,
    author = "Odintsov, Sergei D. and Paul, Tanmoy and SenGupta, Soumitra",
    title = "{Natural validation of the second law of thermodynamics in cosmology}",
    eprint = "2409.05009",
    archivePrefix = "arXiv",
    primaryClass = "gr-qc",
    doi = "10.1103/PhysRevD.111.043544",
    journal = "Phys. Rev. D",
    volume = "111",
    number = "4",
    pages = "043544",
    year = "2025"
}

@article{Paranjape:2006ca,
    author = "Paranjape, Aseem and Sarkar, Sudipta and Padmanabhan, T.",
    title = "{Thermodynamic route to field equations in Lancos-Lovelock gravity}",
    eprint = "hep-th/0607240",
    archivePrefix = "arXiv",
    doi = "10.1103/PhysRevD.74.104015",
    journal = "Phys. Rev. D",
    volume = "74",
    pages = "104015",
    year = "2006"
}

@article{Cai:2006rs,
    author = "Cai, Rong-Gen and Cao, Li-Ming",
    title = "{Unified first law and thermodynamics of apparent horizon in FRW universe}",
    eprint = "gr-qc/0611071",
    archivePrefix = "arXiv",
    doi = "10.1103/PhysRevD.75.064008",
    journal = "Phys. Rev. D",
    volume = "75",
    pages = "064008",
    year = "2007"
}

@article{Akbar:2006kj,
    author = "Akbar, M. and Cai, Rong-Gen",
    title = "{Thermodynamic Behavior of Friedmann Equations at Apparent Horizon of FRW Universe}",
    eprint = "hep-th/0609128",
    archivePrefix = "arXiv",
    doi = "10.1103/PhysRevD.75.084003",
    journal = "Phys. Rev. D",
    volume = "75",
    pages = "084003",
    year = "2007"
}

@article{Agazie_2023,
   title={The NANOGrav 15 yr Data Set: Evidence for a Gravitational-wave Background},
   volume={951},
   ISSN={2041-8213},
   url={http://dx.doi.org/10.3847/2041-8213/acdac6},
   DOI={10.3847/2041-8213/acdac6},
   number={1},
   journal={The Astrophysical Journal Letters},
   publisher={American Astronomical Society},
   author={Agazie, Gabriella and Anumarlapudi, Akash and Archibald, Anne M. and Arzoumanian, Zaven and Baker, Paul T. and Bécsy, Bence and Blecha, Laura and Brazier, Adam and Brook and others},
   year={2023},
   month=jun, pages={L8} }

@article{Zhu:2018etc,
    author = "Zhu, W. W. and others",
    title = "{Tests of Gravitational Symmetries with Pulsar Binary J1713+0747}",
    eprint = "1802.09206",
    archivePrefix = "arXiv",
    primaryClass = "astro-ph.HE",
    doi = "10.1093/mnras/sty2905",
    journal = "Mon. Not. Roy. Astron. Soc.",
    volume = "482",
    number = "3",
    pages = "3249--3260",
    year = "2019"
}

@article{Afzal_2023,
   title={The NANOGrav 15 yr Data Set: Search for Signals from New Physics},
   volume={951},
   ISSN={2041-8213},
   url={http://dx.doi.org/10.3847/2041-8213/acdc91},
   DOI={10.3847/2041-8213/acdc91},
   number={1},
   journal={The Astrophysical Journal Letters},
   publisher={American Astronomical Society},
   author={Afzal, Adeela and Agazie, Gabriella and Anumarlapudi, Akash and Archibald, Anne M. and Arzoumanian, Zaven and Baker, Paul T. and Bécsy, Bence and Blanco-Pillado and others},
   year={2023},
   month=jun, pages={L11} }

@article{Tan_2025,
   title={Constraining string cosmology with the gravitational-wave background using the NANOGrav 15-year data set},
   volume={85},
   ISSN={1434-6052},
   url={http://dx.doi.org/10.1140/epjc/s10052-025-13998-1},
   DOI={10.1140/epjc/s10052-025-13998-1},
   number={3},
   journal={The European Physical Journal C},
   publisher={Springer Science and Business Media LLC},
   author={Tan, Qin and Wu, You and Liu, Lang},
   year={2025},
   month=mar }

@misc{tan2025prebigbangcosmologyexplainnanograv,
      title={Pre-Big-Bang Cosmology Cannot Explain NANOGrav 15-year Signal}, 
      author={Qin Tan and You Wu and Lang Liu},
      year={2025},
      eprint={2411.16505},
      archivePrefix={arXiv},
      primaryClass={astro-ph.CO},
      url={https://arxiv.org/abs/2411.16505}, 
}

@article{Kenjale_2025,
   title={Connecting inflation to the NANOGrav 15-year dataset via massive gravity},
   volume={111},
   ISSN={2470-0029},
   url={http://dx.doi.org/10.1103/PhysRevD.111.103515},
   DOI={10.1103/physrevd.111.103515},
   number={10},
   journal={Physical Review D},
   publisher={American Physical Society (APS)},
   author={Kenjale, Ved and Kahniashvili, Tina},
   year={2025},
   month=may }

@misc{geller2024challengesinterpretingnanograv15year,
      title={Challenges in Interpreting the NANOGrav 15-Year Data Set as Early Universe Gravitational Waves Produced by ALP Induced Instability}, 
      author={Michael Geller and Subhajit Ghosh and Sida Lu and Yuhsin Tsai},
      year={2024},
      eprint={2307.03724},
      archivePrefix={arXiv},
      primaryClass={hep-ph},
      url={https://arxiv.org/abs/2307.03724}, 
}

@article{Lamb_2023,
   title={Rapid refitting techniques for Bayesian spectral characterization of the gravitational wave background using pulsar timing arrays},
   volume={108},
   ISSN={2470-0029},
   url={http://dx.doi.org/10.1103/PhysRevD.108.103019},
   DOI={10.1103/physrevd.108.103019},
   number={10},
   journal={Physical Review D},
   publisher={American Physical Society (APS)},
   author={Lamb, William G. and Taylor, Stephen R. and van Haasteren, Rutger},
   year={2023},
   month=nov }

@article{Bennett:1996ce,
    author = "Bennett, C. L. and Banday, A. and Gorski, K. M. and Hinshaw, G. and Jackson, P. and Keegstra, P. and Kogut, A. and Smoot, George F. and Wilkinson, D. T. and Wright, E. L.",
    title = "{Four year COBE DMR cosmic microwave background observations: Maps and basic results}",
    eprint = "astro-ph/9601067",
    archivePrefix = "arXiv",
    reportNumber = "COBE-PREPRINT-96-01",
    doi = "10.1086/310075",
    journal = "Astrophys. J. Lett.",
    volume = "464",
    pages = "L1--L4",
    year = "1996"
}

@article{Smoot:1998jt,
    author = "Smoot, George F.",
    editor = "Maiani, L. and Melchiorri, F. and Vittorio, N.",
    title = "{COBE observations and results}",
    eprint = "astro-ph/9902027",
    archivePrefix = "arXiv",
    doi = "10.1063/1.59326",
    journal = "AIP Conf. Proc.",
    volume = "476",
    number = "1",
    pages = "1--10",
    year = "1999"
}

@article{Hinshaw_2013,
   title={NINE-YEAR
                    WILKINSON MICROWAVE ANISOTROPY PROBE
                    (
                    WMAP
                    ) OBSERVATIONS: COSMOLOGICAL PARAMETER RESULTS},
   volume={208},
   ISSN={1538-4365},
   url={http://dx.doi.org/10.1088/0067-0049/208/2/19},
   DOI={10.1088/0067-0049/208/2/19},
   number={2},
   journal={The Astrophysical Journal Supplement Series},
   publisher={American Astronomical Society},
   author={Hinshaw, G. and Larson, D. and Komatsu, E. and Spergel, D. N. and Bennett, C. L. and Dunkley, J. and Nolta, M. R. and Halpern, M. and Hill, R. S. and Odegard, N. and Page, L. and Smith, K. M. and Weiland, J. L. and Gold, B. and Jarosik, N. and Kogut, A. and Limon, M. and Meyer, S. S. and Tucker, G. S. and Wollack, E. and Wright, E. L.},
   year={2013},
   month=sep, pages={19} }

@article{Cai:2020qpu,
    author = "Cai, Yong and Piao, Yun-Song",
    title = "{Intermittent null energy condition violations during inflation and primordial gravitational waves}",
    eprint = "2012.11304",
    archivePrefix = "arXiv",
    primaryClass = "gr-qc",
    doi = "10.1103/PhysRevD.103.083521",
    journal = "Phys. Rev. D",
    volume = "103",
    number = "8",
    pages = "083521",
    year = "2021"
}

@article{Ye:2023tpz,
    author = "Ye, Gen and Zhu, Mian and Cai, Yong",
    title = "{Null energy condition violation during inflation and pulsar timing array observations}",
    eprint = "2312.10685",
    archivePrefix = "arXiv",
    primaryClass = "gr-qc",
    doi = "10.1007/JHEP02(2024)008",
    journal = "JHEP",
    volume = "02",
    pages = "008",
    year = "2024"
}

@article{Ben-Dayan:2023lwd,
    author = "Ben-Dayan, Ido and Kumar, Utkarsh and Thattarampilly, Udaykrishna and Verma, Amresh",
    title = "{Probing the early Universe cosmology with NANOGrav: Possibilities and limitations}",
    eprint = "2307.15123",
    archivePrefix = "arXiv",
    primaryClass = "astro-ph.CO",
    doi = "10.1103/PhysRevD.108.103507",
    journal = "Phys. Rev. D",
    volume = "108",
    number = "10",
    pages = "103507",
    year = "2023"
}

@article{Datta:2023xpr,
    author = "Datta, Satyabrata",
    title = "{Explaining PTA data with inflationary GWs in a PBH-dominated universe}",
    eprint = "2309.14238",
    archivePrefix = "arXiv",
    primaryClass = "hep-ph",
    doi = "10.1088/1475-7516/2024/10/009",
    journal = "JCAP",
    volume = "10",
    pages = "009",
    year = "2024"
}

@article{Zhu:2023lbf,
    author = "Zhu, Mian and Ye, Gen and Cai, Yong",
    title = "{Pulsar timing array observations as possible hints for nonsingular cosmology}",
    eprint = "2307.16211",
    archivePrefix = "arXiv",
    primaryClass = "astro-ph.CO",
    doi = "10.1140/epjc/s10052-023-11963-4",
    journal = "Eur. Phys. J. C",
    volume = "83",
    number = "9",
    pages = "816",
    year = "2023"
}

@article{Jiang:2023gfe,
    author = "Jiang, Jun-Qian and Cai, Yong and Ye, Gen and Piao, Yun-Song",
    title = "{Broken blue-tilted inflationary gravitational waves: a joint analysis of NANOGrav 15-year and BICEP/Keck 2018 data}",
    eprint = "2307.15547",
    archivePrefix = "arXiv",
    primaryClass = "astro-ph.CO",
    doi = "10.1088/1475-7516/2024/05/004",
    journal = "JCAP",
    volume = "05",
    pages = "004",
    year = "2024"
}

@article{Gouttenoire:2023nzr,
    author = "Gouttenoire, Yann and Trifinopoulos, Sokratis and Valogiannis, Georgios and Vanvlasselaer, Miguel",
    title = "{Scrutinizing the primordial black hole interpretation of PTA gravitational waves and JWST early galaxies}",
    eprint = "2307.01457",
    archivePrefix = "arXiv",
    primaryClass = "astro-ph.CO",
    doi = "10.1103/PhysRevD.109.123002",
    journal = "Phys. Rev. D",
    volume = "109",
    number = "12",
    pages = "123002",
    year = "2024"
}

@article{Depta:2023uhy,
    author = "Depta, Paul Frederik and Schmidt-Hoberg, Kai and Schwaller, Pedro and Tasillo, Carlo",
    title = "{Signals of merging supermassive black holes in pulsar timing arrays}",
    eprint = "2306.17836",
    archivePrefix = "arXiv",
    primaryClass = "astro-ph.CO",
    reportNumber = "DESY-23-093, MITP-23-036",
    doi = "10.1103/PhysRevResearch.7.013196",
    journal = "Phys. Rev. Res.",
    volume = "7",
    number = "1",
    pages = "013196",
    year = "2025"
}

@article{Babichev:2023pbf,
    author = "Babichev, E. and Gorbunov, D. and Ramazanov, S. and Samanta, R. and Vikman, A.",
    title = "{NANOGrav spectral index {\ensuremath{\gamma}}=3 from melting domain walls}",
    eprint = "2307.04582",
    archivePrefix = "arXiv",
    primaryClass = "hep-ph",
    doi = "10.1103/PhysRevD.108.123529",
    journal = "Phys. Rev. D",
    volume = "108",
    number = "12",
    pages = "123529",
    year = "2023"
}

@article{Yamada:2023thl,
    author = "Yamada, Masaki and Yonekura, Kazuya",
    title = "{Dark baryon from pure Yang-Mills theory and its GW signature from cosmic strings}",
    eprint = "2307.06586",
    archivePrefix = "arXiv",
    primaryClass = "hep-ph",
    reportNumber = "TU-1203",
    doi = "10.1007/JHEP09(2023)197",
    journal = "JHEP",
    volume = "09",
    pages = "197",
    year = "2023"
}

@article{Zhang:2023nrs,
    author = "Zhang, Zhao and Cai, Chengfeng and Su, Yu-Hang and Wang, Shiyu and Yu, Zhao-Huan and Zhang, Hong-Hao",
    title = "{Nano-Hertz gravitational waves from collapsing domain walls associated with freeze-in dark matter in light of pulsar timing array observations}",
    eprint = "2307.11495",
    archivePrefix = "arXiv",
    primaryClass = "hep-ph",
    doi = "10.1103/PhysRevD.108.095037",
    journal = "Phys. Rev. D",
    volume = "108",
    number = "9",
    pages = "095037",
    year = "2023"
}

@article{Lazarides:2023ksx,
    author = "Lazarides, George and Maji, Rinku and Shafi, Qaisar",
    title = "{Superheavy quasistable strings and walls bounded by strings in the light of NANOGrav 15~year data}",
    eprint = "2306.17788",
    archivePrefix = "arXiv",
    primaryClass = "hep-ph",
    doi = "10.1103/PhysRevD.108.095041",
    journal = "Phys. Rev. D",
    volume = "108",
    number = "9",
    pages = "095041",
    year = "2023"
}

@article{Ellis:2023tsl,
    author = "Ellis, John and Lewicki, Marek and Lin, Chunshan and Vaskonen, Ville",
    title = "{Cosmic superstrings revisited in light of NANOGrav 15-year data}",
    eprint = "2306.17147",
    archivePrefix = "arXiv",
    primaryClass = "astro-ph.CO",
    reportNumber = "KCL-PH-TH/2023-38, CERN-TH-2023-126, AION-REPORT/2023-07",
    doi = "10.1103/PhysRevD.108.103511",
    journal = "Phys. Rev. D",
    volume = "108",
    number = "10",
    pages = "103511",
    year = "2023"
}

@article{Murai:2023gkv,
    author = "Murai, Kai and Yin, Wen",
    title = "{A novel probe of supersymmetry in light of nanohertz gravitational waves}",
    eprint = "2307.00628",
    archivePrefix = "arXiv",
    primaryClass = "hep-ph",
    doi = "10.1007/JHEP10(2023)062",
    journal = "JHEP",
    volume = "10",
    pages = "062",
    year = "2023"
}

@article{Niu:2023bsr,
    author = "Niu, Xuce and Rahat, Moinul Hossain",
    title = "{NANOGrav signal from axion inflation}",
    eprint = "2307.01192",
    archivePrefix = "arXiv",
    primaryClass = "hep-ph",
    doi = "10.1103/PhysRevD.108.115023",
    journal = "Phys. Rev. D",
    volume = "108",
    number = "11",
    pages = "115023",
    year = "2023"
}

@article{Unal:2023srk,
    author = "Unal, Caner and Papageorgiou, Alexandros and Obata, Ippei",
    title = "{Axion-gauge dynamics during inflation as the origin of pulsar timing array signals and primordial black holes}",
    eprint = "2307.02322",
    archivePrefix = "arXiv",
    primaryClass = "astro-ph.CO",
    doi = "10.1016/j.physletb.2024.138873",
    journal = "Phys. Lett. B",
    volume = "856",
    pages = "138873",
    year = "2024"
}

@article{Haque:2021dha,
    author = "Haque, Md Riajul and Maity, Debaprasad and Paul, Tanmoy and Sriramkumar, L.",
    title = "{Decoding the phases of early and late time reheating through imprints on primordial gravitational waves}",
    eprint = "2105.09242",
    archivePrefix = "arXiv",
    primaryClass = "astro-ph.CO",
    doi = "10.1103/PhysRevD.104.063513",
    journal = "Phys. Rev. D",
    volume = "104",
    number = "6",
    pages = "063513",
    year = "2021"
}

@article{Jiang:2023qbm,
    author = "Jiang, Siyu and Yang, Aidi and Ma, Jiucheng and Huang, Fa Peng",
    title = "{Implication of nano-Hertz stochastic gravitational wave on dynamical dark matter through a dark first-order phase transition}",
    eprint = "2306.17827",
    archivePrefix = "arXiv",
    primaryClass = "hep-ph",
    doi = "10.1088/1361-6382/ad24c6",
    journal = "Class. Quant. Grav.",
    volume = "41",
    number = "6",
    pages = "065009",
    year = "2024"
}

@article{An:2023jxf,
    author = "An, Haipeng and Su, Boye and Tai, Hanwen and Wang, Lian-Tao and Yang, Chen",
    title = "{Phase transition during inflation and the gravitational wave signal at pulsar timing arrays}",
    eprint = "2308.00070",
    archivePrefix = "arXiv",
    primaryClass = "astro-ph.CO",
    doi = "10.1103/PhysRevD.109.L121304",
    journal = "Phys. Rev. D",
    volume = "109",
    number = "12",
    pages = "L121304",
    year = "2024"
}

@article{Ghosh:2023aum,
    author = "Ghosh, Tathagata and Ghoshal, Anish and Guo, Huai-Ke and Hajkarim, Fazlollah and King, Stephen F. and Sinha, Kuver and Wang, Xin and White, Graham",
    title = "{Did we hear the sound of the Universe boiling? Analysis using the full fluid velocity profiles and NANOGrav 15-year data}",
    eprint = "2307.02259",
    archivePrefix = "arXiv",
    primaryClass = "astro-ph.HE",
    doi = "10.1088/1475-7516/2024/05/100",
    journal = "JCAP",
    volume = "05",
    pages = "100",
    year = "2024"
}

@article{Gouttenoire:2023bqy,
    author = "Gouttenoire, Yann",
    title = "{First-Order Phase Transition Interpretation of Pulsar Timing Array Signal Is Consistent with Solar-Mass Black Holes}",
    eprint = "2307.04239",
    archivePrefix = "arXiv",
    primaryClass = "hep-ph",
    doi = "10.1103/PhysRevLett.131.171404",
    journal = "Phys. Rev. Lett.",
    volume = "131",
    number = "17",
    pages = "171404",
    year = "2023"
}

@article{Salvio:2023ynn,
    author = "Salvio, Alberto",
    title = "{Supercooling in radiative symmetry breaking: theory extensions, gravitational wave detection and primordial black holes}",
    eprint = "2307.04694",
    archivePrefix = "arXiv",
    primaryClass = "hep-ph",
    doi = "10.1088/1475-7516/2023/12/046",
    journal = "JCAP",
    volume = "12",
    pages = "046",
    year = "2023"
}

@article{Gangopadhyay:2023qjr,
    author = "Gangopadhyay, M. R. and Godithi, V. V. and Inui, R. and Ichiki, K. and Kajino, T. and Manusankar, A. and Mathews, G. J. and Yogesh",
    title = "{Is the NANOGrav detection evidence of resonant particle creation during inflation?}",
    eprint = "2309.03101",
    archivePrefix = "arXiv",
    primaryClass = "astro-ph.CO",
    doi = "10.1016/j.jheap.2025.100358",
    journal = "JHEAp",
    volume = "47",
    pages = "100358",
    year = "2025"
}

@article{Chen:2024twp,
    author = "Chen, Zu-Cheng and Liu, Lang",
    title = "{Can we distinguish between adiabatic and isocurvature fluctuations with pulsar timing arrays?}",
    eprint = "2402.16781",
    archivePrefix = "arXiv",
    primaryClass = "astro-ph.CO",
    doi = "10.1007/s11433-025-2614-0",
    journal = "Sci. China Phys. Mech. Astron.",
    volume = "68",
    number = "5",
    pages = "250412",
    year = "2025"
}

@article{Ye:2023xyr,
    author = "Ye, Gen and Silvestri, Alessandra",
    title = "{Can the Gravitational Wave Background Feel Wiggles in Spacetime?}",
    eprint = "2307.05455",
    archivePrefix = "arXiv",
    primaryClass = "astro-ph.CO",
    doi = "10.3847/2041-8213/ad2851",
    journal = "Astrophys. J. Lett.",
    volume = "963",
    number = "1",
    pages = "L15",
    year = "2024"
}

@article{Maiti:2024nhv,
    author = "Maiti, Subhasis and Maity, Debaprasad and Sriramkumar, L.",
    title = "{Constraining inflationary magnetogenesis and reheating via GWs in light of PTA data}",
    eprint = "2401.01864",
    archivePrefix = "arXiv",
    primaryClass = "gr-qc",
    month = "1",
    year = "2024"
}

@article{Choudhury:2023kam,
    author = "Choudhury, Sayantan",
    title = "{Single field inflation in the light of Pulsar Timing Array Data: quintessential interpretation of blue tilted tensor spectrum through Non-Bunch Davies initial condition}",
    eprint = "2307.03249",
    archivePrefix = "arXiv",
    primaryClass = "astro-ph.CO",
    doi = "10.1140/epjc/s10052-024-12625-9",
    journal = "Eur. Phys. J. C",
    volume = "84",
    number = "3",
    pages = "278",
    year = "2024"
}

@article{Datta:2023vbs,
    author = "Datta, Satyabrata and Samanta, Rome",
    title = "{Fingerprints of GeV scale right-handed neutrinos on inflationary gravitational waves and PTA data}",
    eprint = "2307.00646",
    archivePrefix = "arXiv",
    primaryClass = "hep-ph",
    doi = "10.1103/PhysRevD.108.L091706",
    journal = "Phys. Rev. D",
    volume = "108",
    number = "9",
    pages = "L091706",
    year = "2023"
}

@article{Borah:2023sbc,
    author = "Borah, Debasish and Jyoti Das, Suruj and Samanta, Rome",
    title = "{Imprint of inflationary gravitational waves and WIMP dark matter in pulsar timing array data}",
    eprint = "2307.00537",
    archivePrefix = "arXiv",
    primaryClass = "hep-ph",
    doi = "10.1088/1475-7516/2024/03/031",
    journal = "JCAP",
    volume = "03",
    pages = "031",
    year = "2024"
}

@article{Vagnozzi:2023lwo,
    author = "Vagnozzi, Sunny",
    title = "{Inflationary interpretation of the stochastic gravitational wave background signal detected by pulsar timing array experiments}",
    eprint = "2306.16912",
    archivePrefix = "arXiv",
    primaryClass = "astro-ph.CO",
    doi = "10.1016/j.jheap.2023.07.001",
    journal = "JHEAp",
    volume = "39",
    pages = "81--98",
    year = "2023"
}

@article{Firouzjahi:2023lzg,
    author = "Firouzjahi, Hassan and Talebian, Alireza",
    title = "{Induced gravitational waves from ultra slow-roll inflation and pulsar timing arrays observations}",
    eprint = "2307.03164",
    archivePrefix = "arXiv",
    primaryClass = "gr-qc",
    doi = "10.1088/1475-7516/2023/10/032",
    journal = "JCAP",
    volume = "10",
    pages = "032",
    year = "2023"
}

@article{Balaji:2023ehk,
    author = "Balaji, Shyam and Dom{\`e}nech, Guillem and Franciolini, Gabriele",
    title = "{Scalar-induced gravitational wave interpretation of PTA data: the role of scalar fluctuation propagation speed}",
    eprint = "2307.08552",
    archivePrefix = "arXiv",
    primaryClass = "gr-qc",
    doi = "10.1088/1475-7516/2023/10/041",
    journal = "JCAP",
    volume = "10",
    pages = "041",
    year = "2023"
}

@article{Cheung:2023ihl,
    author = "Cheung, Kingman and Ouseph, C. J. and Tseng, Po-Yan",
    title = "{NANOGrav and other PTA signals and PBH from the modified Higgs inflation}",
    eprint = "2307.08046",
    archivePrefix = "arXiv",
    primaryClass = "hep-ph",
    doi = "10.1140/epjc/s10052-024-13268-6",
    journal = "Eur. Phys. J. C",
    volume = "84",
    number = "9",
    pages = "906",
    year = "2024"
}

@article{Franciolini:2023pbf,
    author = "Franciolini, Gabriele and Iovino, Junior., Antonio and Vaskonen, Ville and Veermae, Hardi",
    title = "{Recent Gravitational Wave Observation by Pulsar Timing Arrays and Primordial Black Holes: The Importance of Non-Gaussianities}",
    eprint = "2306.17149",
    archivePrefix = "arXiv",
    primaryClass = "astro-ph.CO",
    doi = "10.1103/PhysRevLett.131.201401",
    journal = "Phys. Rev. Lett.",
    volume = "131",
    number = "20",
    pages = "201401",
    year = "2023"
}

@article{Inomata:2023zup,
    author = "Inomata, Keisuke and Kohri, Kazunori and Terada, Takahiro",
    title = "{Detected stochastic gravitational waves and subsolar-mass primordial black holes}",
    eprint = "2306.17834",
    archivePrefix = "arXiv",
    primaryClass = "astro-ph.CO",
    reportNumber = "KEK-TH-2535, KEK-Cosmo-0317, KEK-QUP-2023-0016, CTPU-PTC-23-28",
    doi = "10.1103/PhysRevD.109.063506",
    journal = "Phys. Rev. D",
    volume = "109",
    number = "6",
    pages = "063506",
    year = "2024"
}

@article{Xu:2023wog,
    author = "Xu, Heng and others",
    title = "{Searching for the Nano-Hertz Stochastic Gravitational Wave Background with the Chinese Pulsar Timing Array Data Release I}",
    eprint = "2306.16216",
    archivePrefix = "arXiv",
    primaryClass = "astro-ph.HE",
    doi = "10.1088/1674-4527/acdfa5",
    journal = "Res. Astron. Astrophys.",
    volume = "23",
    number = "7",
    pages = "075024",
    year = "2023"
}

@article{Zic:2023gta,
    author = "Zic, Andrew and others",
    title = "{The Parkes Pulsar Timing Array third data release}",
    eprint = "2306.16230",
    archivePrefix = "arXiv",
    primaryClass = "astro-ph.HE",
    doi = "10.1017/pasa.2023.36",
    journal = "Publ. Astron. Soc. Austral.",
    volume = "40",
    pages = "e049",
    year = "2023"
}

@article{EPTA:2023sfo,
    author = "Antoniadis, J. and others",
    collaboration = "EPTA",
    title = "{The second data release from the European Pulsar Timing Array - I. The dataset and timing analysis}",
    eprint = "2306.16224",
    archivePrefix = "arXiv",
    primaryClass = "astro-ph.HE",
    doi = "10.1051/0004-6361/202346841",
    journal = "Astron. Astrophys.",
    volume = "678",
    pages = "A48",
    year = "2023"
}

@article{NANOGrav:2023hde,
    author = "Agazie, Gabriella and others",
    collaboration = "NANOGrav",
    title = "{The NANOGrav 15 yr Data Set: Observations and Timing of 68 Millisecond Pulsars}",
    eprint = "2306.16217",
    archivePrefix = "arXiv",
    primaryClass = "astro-ph.HE",
    doi = "10.3847/2041-8213/acda9a",
    journal = "Astrophys. J. Lett.",
    volume = "951",
    number = "1",
    pages = "L9",
    year = "2023"
}

@article{NANOGrav:2023gor,
    author = "Agazie, Gabriella and others",
    collaboration = "NANOGrav",
    title = "{The NANOGrav 15 yr Data Set: Evidence for a Gravitational-wave Background}",
    eprint = "2306.16213",
    archivePrefix = "arXiv",
    primaryClass = "astro-ph.HE",
    doi = "10.3847/2041-8213/acdac6",
    journal = "Astrophys. J. Lett.",
    volume = "951",
    number = "1",
    pages = "L8",
    year = "2023"
}

@article{KAGRA:2021kbb,
    author = "Abbott, R. and others",
    collaboration = "KAGRA, Virgo, LIGO Scientific",
    title = "{Upper limits on the isotropic gravitational-wave background from Advanced LIGO and Advanced Virgo{\textquoteright}s third observing run}",
    eprint = "2101.12130",
    archivePrefix = "arXiv",
    primaryClass = "gr-qc",
    reportNumber = "LIGO-DCC-P2000314",
    doi = "10.1103/PhysRevD.104.022004",
    journal = "Phys. Rev. D",
    volume = "104",
    number = "2",
    pages = "022004",
    year = "2021"
}

@article{LIGOScientific:2017vox,
    author = "Abbott, B. . P. . and others",
    collaboration = "LIGO Scientific, Virgo",
    title = "{GW170608: Observation of a 19-solar-mass Binary Black Hole Coalescence}",
    eprint = "1711.05578",
    archivePrefix = "arXiv",
    primaryClass = "astro-ph.HE",
    reportNumber = "LIGO-DOCUMENT-P170608-V8",
    doi = "10.3847/2041-8213/aa9f0c",
    journal = "Astrophys. J. Lett.",
    volume = "851",
    pages = "L35",
    year = "2017"
}

@article{LIGOScientific:2017ycc,
    author = "Abbott, B. P. and others",
    collaboration = "LIGO Scientific, Virgo",
    title = "{GW170814: A Three-Detector Observation of Gravitational Waves from a Binary Black Hole Coalescence}",
    eprint = "1709.09660",
    archivePrefix = "arXiv",
    primaryClass = "gr-qc",
    doi = "10.1103/PhysRevLett.119.141101",
    journal = "Phys. Rev. Lett.",
    volume = "119",
    number = "14",
    pages = "141101",
    year = "2017"
}

@article{LIGOScientific:2017bnn,
    author = "Abbott, Benjamin P. and others",
    collaboration = "LIGO Scientific, VIRGO",
    title = "{GW170104: Observation of a 50-Solar-Mass Binary Black Hole Coalescence at Redshift 0.2}",
    eprint = "1706.01812",
    archivePrefix = "arXiv",
    primaryClass = "gr-qc",
    reportNumber = "LIGO-P170104",
    doi = "10.1103/PhysRevLett.118.221101",
    journal = "Phys. Rev. Lett.",
    volume = "118",
    number = "22",
    pages = "221101",
    year = "2017",
    note = "[Erratum: Phys.Rev.Lett. 121, 129901 (2018)]"
}

@article{LIGOScientific:2016aoc,
    author = "Abbott, B. P. and others",
    collaboration = "LIGO Scientific, Virgo",
    title = "{Observation of Gravitational Waves from a Binary Black Hole Merger}",
    eprint = "1602.03837",
    archivePrefix = "arXiv",
    primaryClass = "gr-qc",
    reportNumber = "LIGO-P150914",
    doi = "10.1103/PhysRevLett.116.061102",
    journal = "Phys. Rev. Lett.",
    volume = "116",
    number = "6",
    pages = "061102",
    year = "2016"
}

@article{Roshan:2024qnv,
    author = "Roshan, Rishav and White, Graham",
    title = "{Using gravitational waves to see the first second of the Universe}",
    eprint = "2401.04388",
    archivePrefix = "arXiv",
    primaryClass = "hep-ph",
    doi = "10.1103/RevModPhys.97.015001",
    journal = "Rev. Mod. Phys.",
    volume = "97",
    number = "1",
    pages = "015001",
    year = "2025"
}

@article{Domenech:2021ztg,
    author = "Dom{\`e}nech, Guillem",
    title = "{Scalar Induced Gravitational Waves Review}",
    eprint = "2109.01398",
    archivePrefix = "arXiv",
    primaryClass = "gr-qc",
    doi = "10.3390/universe7110398",
    journal = "Universe",
    volume = "7",
    number = "11",
    pages = "398",
    year = "2021"
}

@article{Caprini:2018mtu,
    author = "Caprini, Chiara and Figueroa, Daniel G.",
    title = "{Cosmological Backgrounds of Gravitational Waves}",
    eprint = "1801.04268",
    archivePrefix = "arXiv",
    primaryClass = "astro-ph.CO",
    doi = "10.1088/1361-6382/aac608",
    journal = "Class. Quant. Grav.",
    volume = "35",
    number = "16",
    pages = "163001",
    year = "2018"
}

@article{Guzzetti:2016mkm,
    author = "Guzzetti, M. C. and Bartolo, N. and Liguori, M. and Matarrese, S.",
    title = "{Gravitational waves from inflation}",
    eprint = "1605.01615",
    archivePrefix = "arXiv",
    primaryClass = "astro-ph.CO",
    doi = "10.1393/ncr/i2016-10127-1",
    journal = "Riv. Nuovo Cim.",
    volume = "39",
    number = "9",
    pages = "399--495",
    year = "2016"
}

@article{Planck:2018jri,
    author = "Akrami, Y. and others",
    collaboration = "Planck",
    title = "{Planck 2018 results. X. Constraints on inflation}",
    eprint = "1807.06211",
    archivePrefix = "arXiv",
    primaryClass = "astro-ph.CO",
    doi = "10.1051/0004-6361/201833887",
    journal = "Astron. Astrophys.",
    volume = "641",
    pages = "A10",
    year = "2020"
}

@article{Haque:2025uga,
    author = "Haque, Md Riajul and Maity, Debaprasad",
    title = "{Minimal Plateau Inflation in light of ACT DR6 Observations}",
    eprint = "2505.18267",
    archivePrefix = "arXiv",
    primaryClass = "astro-ph.CO",
    month = "5",
    year = "2025"
}

@article{Kallosh:2025rni,
    author = "Kallosh, Renata and Linde, Andrei and Roest, Diederik",
    title = "{A simple scenario for the last ACT}",
    eprint = "2503.21030",
    archivePrefix = "arXiv",
    primaryClass = "hep-th",
    month = "3",
    year = "2025"
}

@article{Odintsov:2025wai,
    author = "Odintsov, S. D. and Oikonomou, V. K.",
    title = "{GW170817 Viable Einstein-Gauss-Bonnet Inflation Compatible with the Atacama Cosmology Telescope Data}",
    eprint = "2506.08193",
    archivePrefix = "arXiv",
    primaryClass = "gr-qc",
    month = "6",
    year = "2025"
}

@article{Drees:2025ngb,
    author = "Drees, Manuel and Xu, Yong",
    title = "{Refined predictions for Starobinsky inflation and post-inflationary constraints in light of ACT}",
    eprint = "2504.20757",
    archivePrefix = "arXiv",
    primaryClass = "astro-ph.CO",
    reportNumber = "MITP-25-033",
    doi = "10.1016/j.physletb.2025.139612",
    journal = "Phys. Lett. B",
    volume = "867",
    pages = "139612",
    year = "2025"
}

@article{Gao:2025onc,
    author = "Gao, Qing and Gong, Yungui and Yi, Zhu and Zhang, Fengge",
    title = "{Non-minimal coupling in light of ACT}",
    eprint = "2504.15218",
    archivePrefix = "arXiv",
    primaryClass = "astro-ph.CO",
    month = "4",
    year = "2025"
}

@article{Kim:2025dyi,
    author = "Kim, Jinsu and Wang, Xinpeng and Zhang, Ying-li and Ren, Zhongzhou",
    title = "{Enhancement of primordial curvature perturbations in $R^3$-corrected Starobinsky-Higgs inflation}",
    eprint = "2504.12035",
    archivePrefix = "arXiv",
    primaryClass = "astro-ph.CO",
    month = "4",
    year = "2025"
}

@article{Antoniadis:2025pfa,
    author = "Antoniadis, Ignatios and Ellis, John and Ke, Wenqi and Nanopoulos, Dimitri V. and Olive, Keith A.",
    title = "{How Accidental was Inflation?}",
    eprint = "2504.12283",
    archivePrefix = "arXiv",
    primaryClass = "hep-ph",
    reportNumber = "UMN-TH-4418/25, FTPI-MINN-25/03, KCL-PH-TH/2025-09, CERN-TH-2025-076",
    month = "4",
    year = "2025"
}

@article{Salvio:2025izr,
    author = "Salvio, Alberto",
    title = "{Independent connection in ACTion during inflation}",
    eprint = "2504.10488",
    archivePrefix = "arXiv",
    primaryClass = "hep-ph",
    month = "4",
    year = "2025"
}

@article{Dioguardi:2025vci,
    author = "Dioguardi, Christian and Iovino, Antonio J. and Racioppi, Antonio",
    title = "{Fractional attractors in light of the latest ACT observations}",
    eprint = "2504.02809",
    archivePrefix = "arXiv",
    primaryClass = "gr-qc",
    doi = "10.1016/j.physletb.2025.139664",
    journal = "Phys. Lett. B",
    volume = "868",
    pages = "139664",
    year = "2025"
}

@article{Aoki:2025wld,
    author = "Aoki, Shuntaro and Otsuka, Hajime and Yanagita, Ryota",
    title = "{Higgs-Modular Inflation}",
    eprint = "2504.01622",
    archivePrefix = "arXiv",
    primaryClass = "hep-ph",
    reportNumber = "RIKEN-iTHEMS-Report-25, KYUSHU-HET-317",
    month = "4",
    year = "2025"
}

@article{ACT:2025fju,
    author = "Louis, Thibaut and others",
    collaboration = "ACT",
    title = "{The Atacama Cosmology Telescope: DR6 Power Spectra, Likelihoods and LCDM Parameters}",
    eprint = "2503.14452",
    archivePrefix = "arXiv",
    primaryClass = "astro-ph.CO",
    reportNumber = "FERMILAB-PUB-25-0071-PPD",
    month = "3",
    year = "2025"
}

@article{ACT:2025tim,
    author = "Calabrese, Erminia and others",
    collaboration = "ACT",
    title = "{The Atacama Cosmology Telescope: DR6 Constraints on Extended Cosmological Models}",
    eprint = "2503.14454",
    archivePrefix = "arXiv",
    primaryClass = "astro-ph.CO",
    reportNumber = "FERMILAB-PUB-25-0157-PPD",
    month = "3",
    year = "2025"
}

@article{Nojiri:2022dkr,
    author = "Nojiri, Shin'ichi and Odintsov, Sergei D. and Paul, Tanmoy",
    title = "{Early and late universe holographic cosmology from a new generalized entropy}",
    eprint = "2205.08876",
    archivePrefix = "arXiv",
    primaryClass = "gr-qc",
    doi = "10.1016/j.physletb.2022.137189",
    journal = "Phys. Lett. B",
    volume = "831",
    pages = "137189",
    year = "2022"
}

@article{Tsallis:1987eu,
    author = "Tsallis, Constantino",
    title = "{Possible Generalization of Boltzmann-Gibbs Statistics}",
    reportNumber = "CBPF-NF-062-87",
    doi = "10.1007/BF01016429",
    journal = "J. Statist. Phys.",
    volume = "52",
    pages = "479--487",
    year = "1988"
}

@article{Cai:2005ra,
    author = "Cai, Rong-Gen and Kim, Sang Pyo",
    title = "{First law of thermodynamics and Friedmann equations of Friedmann-Robertson-Walker universe}",
    eprint = "hep-th/0501055",
    archivePrefix = "arXiv",
    doi = "10.1088/1126-6708/2005/02/050",
    journal = "JHEP",
    volume = "02",
    pages = "050",
    year = "2005"
}

@article{Akbar:2006er,
    author = "Akbar, M. and Cai, Rong-Gen",
    title = "{Friedmann equations of FRW universe in scalar-tensor gravity, f(R) gravity and first law of thermodynamics}",
    eprint = "hep-th/0602156",
    archivePrefix = "arXiv",
    doi = "10.1016/j.physletb.2006.02.035",
    journal = "Phys. Lett. B",
    volume = "635",
    pages = "7--10",
    year = "2006"
}

@article{Tan:2024kuk,
    author = "Tan, Qin and Wu, You and Liu, Lang",
    title = "{Constraining string cosmology with the gravitational-wave background using the NANOGrav 15-year data set}",
    eprint = "2409.17846",
    archivePrefix = "arXiv",
    primaryClass = "gr-qc",
    doi = "10.1140/epjc/s10052-025-13998-1",
    journal = "Eur. Phys. J. C",
    volume = "85",
    number = "3",
    pages = "327",
    year = "2025"
}

@article{Conzinu:2024cwl,
    author = "Conzinu, P. and Fanizza, G. and Gasperini, M. and Pavone, E. and Tedesco, L. and Veneziano, G.",
    title = "{Constraints on the Pre-Big Bang scenario from a cosmological interpretation of the NANOGrav data}",
    eprint = "2412.01734",
    archivePrefix = "arXiv",
    primaryClass = "hep-th",
    reportNumber = "CERN-TH-2024-210, BA-TH/809-24",
    doi = "10.1088/1475-7516/2025/02/039",
    journal = "JCAP",
    volume = "02",
    pages = "039",
    year = "2025"
}

@article{NANOGrav:2023hvm,
    author = "Afzal, Adeela and others",
    collaboration = "NANOGrav",
    title = "{The NANOGrav 15 yr Data Set: Search for Signals from New Physics}",
    eprint = "2306.16219",
    archivePrefix = "arXiv",
    primaryClass = "astro-ph.HE",
    reportNumber = "FERMILAB-PUB-23-589-T",
    doi = "10.3847/2041-8213/acdc91",
    journal = "Astrophys. J. Lett.",
    volume = "951",
    number = "1",
    pages = "L11",
    year = "2023",
    note = "[Erratum: Astrophys.J.Lett. 971, L27 (2024), Erratum: Astrophys.J. 971, L27 (2024)]"
}

@article{Schwarz:2001vv,
    author = "Schwarz, Dominik J. and Terrero-Escalante, Cesar A. and Garcia, Alberto A.",
    title = "{Higher order corrections to primordial spectra from cosmological inflation}",
    eprint = "astro-ph/0106020",
    archivePrefix = "arXiv",
    doi = "10.1016/S0370-2693(01)01036-X",
    journal = "Phys. Lett. B",
    volume = "517",
    pages = "243--249",
    year = "2001"
}

@dataset{thenanogravcollaboration202310344086,
  author       = {The NANOGrav Collaboration},
  title        = {KDE Representations of the Gravitational Wave
                   Background Free Spectra Present in the NANOGrav
                   15-Year Dataset
                  },
  month        = jun,
  year         = 2023,
  publisher    = {Zenodo},
  doi          = {10.5281/zenodo.10344086},
  url          = {https://doi.org/10.5281/zenodo.10344086},
}

@article{Renyi,
    author = "Rényi, A.",
    title = "{On Measures of Entropy and Information}",
    journal = "Proceedings of the Fourth Berkeley Symposium on Mathematical Statistics and Probability, Volume 1: Contributions to the Theory of Statistics, University of California Press, Berkeley, California, 20 June-30 July 1960, 547-561",
    year = "1961"
}

@article{Odintsov:2022qnn,
    author = "Odintsov, Sergei D. and Paul, Tanmoy",
    title = "{A non-singular generalized entropy and its implications on bounce cosmology}",
    eprint = "2212.05531",
    archivePrefix = "arXiv",
    primaryClass = "gr-qc",
    doi = "10.1016/j.dark.2022.101159",
    journal = "Phys. Dark Univ.",
    volume = "39",
    pages = "101159",
    year = "2023"
}

@article{Odintsov_2023,
   title={Holographic realization from inflation to reheating in generalized entropic cosmology},
   volume={42},
   ISSN={2212-6864},
   url={http://dx.doi.org/10.1016/j.dark.2023.101277},
   DOI={10.1016/j.dark.2023.101277},
   journal={Physics of the Dark Universe},
   publisher={Elsevier BV},
   author={Odintsov, Sergei D. and D’Onofrio, Simone and Paul, Tanmoy},
   year={2023},
   month=dec, pages={101277} 
}

@article{Lymperis:2023prf,
    author = "Lymperis, Andreas",
    title = "{Holographic dark energy through Loop Quantum Gravity inspired entropy}",
    eprint = "2310.01050",
    archivePrefix = "arXiv",
    primaryClass = "gr-qc",
    month = "10",
    year = "2023"
}

@article{doi:10.1142/S0219887824503389,
author = {Elizalde, E. and Yurov, A. V. and Timoshkin, A. V.},
title = {Holographic bounce cosmological models induced by viscous dark fluid from a generalized non-singular entropy function},
journal = {International Journal of Geometric Methods in Modern Physics},
volume = {0},
number = {ja},
pages = {null},
year = {0},
doi = {10.1142/S0219887824503389},

URL = { 
    
        https://doi.org/10.1142/S0219887824503389
    
    

},
eprint = { 
    
        https://doi.org/10.1142/S0219887824503389
    
    

}

}

@article{Barrow_2020,
   title={The area of a rough black hole},
   volume={808},
   ISSN={0370-2693},
   url={http://dx.doi.org/10.1016/j.physletb.2020.135643},
   DOI={10.1016/j.physletb.2020.135643},
   journal={Physics Letters B},
   publisher={Elsevier BV},
   author={Barrow, John D.},
   year={2020},
   month=sep, pages={135643} 
}

@article{Nojiri:2024zdu,
    author = "Nojiri, Shin'ichi and Odintsov, Sergei D. and Paul, Tanmoy",
    title = "{Different Aspects of Entropic Cosmology}",
    eprint = "2409.01090",
    archivePrefix = "arXiv",
    primaryClass = "gr-qc",
    doi = "10.3390/universe10090352",
    journal = "Universe",
    volume = "10",
    number = "9",
    pages = "352",
    year = "2024"
}

@article{Bolotin:2023wiw,
    author = "Bolotin, Yu. L. and Yanovsky, V. V.",
    title = "{Cosmology based on entropy}",
    eprint = "2310.10144",
    archivePrefix = "arXiv",
    primaryClass = "gr-qc",
    month = "10",
    year = "2023"
}

@article{Planck2020,
    author = "Aghanim, N. and others",
    collaboration = "Planck",
    title = "{Planck 2018 results. VI. Cosmological parameters}",
    eprint = "1807.06209",
    archivePrefix = "arXiv",
    primaryClass = "astro-ph.CO",
    doi = "10.1051/0004-6361/201833910",
    journal = "Astron. Astrophys.",
    volume = "641",
    pages = "A6",
    year = "2020"
}

@article{Peng:2025bws,
    author = "Peng, Zhi-Zhang and Chen, Zu-Cheng and Liu, Lang",
    title = "{The polynomial potential inflation in light of ACT observations}",
    eprint = "2505.12816",
    archivePrefix = "arXiv",
    primaryClass = "astro-ph.CO",
    month = "5",
    year = "2025"
}

@article{Liu:2025qca,
    author = "Liu, Lang and Yi, Zhu and Gong, Yungui",
    title = "{Reconciling Higgs Inflation with ACT Observations through Reheating}",
    eprint = "2505.02407",
    archivePrefix = "arXiv",
    primaryClass = "astro-ph.CO",
    month = "5",
    year = "2025"
}

@article{Maity:2025czp,
    author = "Maity, Suvashis",
    title = "{ACT-ing on inflation: Implications of non bunch-Davies initial condition and reheating on single-field slow roll models}",
    eprint = "2505.10534",
    archivePrefix = "arXiv",
    primaryClass = "astro-ph.CO",
    doi = "10.1016/j.physletb.2025.139913",
    journal = "Phys. Lett. B",
    volume = "870",
    pages = "139913",
    year = "2025"
}

@article{Haque:2025uis,
    author = "Haque, Md Riajul and Pal, Sourav and Paul, Debarun",
    title = "{Improved predictions on Higgs-Starobinsky inflation and reheating with ACT DR6 and primordial gravitational waves}",
    eprint = "2505.04615",
    archivePrefix = "arXiv",
    primaryClass = "astro-ph.CO",
    doi = "10.1016/j.physletb.2025.139852",
    journal = "Phys. Lett. B",
    volume = "869",
    pages = "139852",
    year = "2025"
}

@article{Mondal:2025kur,
    author = "Mondal, Rajesh and Mondal, Sourav and Chakraborty, Ayan",
    title = "{Constraining Reheating Temperature, Inflaton-SM Coupling and Dark Matter Mass in Light of ACT DR6 Observations}",
    eprint = "2505.13387",
    archivePrefix = "arXiv",
    primaryClass = "hep-ph",
    month = "5",
    year = "2025"
}
\bibliographystyle{./utphys1}

\end{document}